\documentclass[reprint,sort&compress,longbibliography,superscriptaddress,amsmath,aps,showpacs,pra]{revtex4-2}
\usepackage[british]{babel}
\usepackage{graphicx}
\usepackage{epsfig}
\usepackage[hidelinks]{hyperref}
\usepackage{amsmath}
\usepackage{xcolor}
\usepackage{amssymb}
\usepackage{gensymb}
\usepackage{braket}
\usepackage{soul}
\usepackage{physics}

\hypersetup{ 
     colorlinks   = true,
     citecolor    = blue,
     linkcolor    = blue 
}

\renewcommand{\abs}[1]{|#1|}

\begin{document}
\title{Short vs. long range exchange interactions in twisted bilayer graphene}
\author{Alejandro Jimeno-Pozo}
\email{alejandro.jimeno@imdea.org}
\affiliation{Imdea Nanoscience, Faraday 9, 28049 Madrid, Spain}
\author{Zachary A. H. Goodwin}
\affiliation{Departments of Physics and Materials and the Thomas Young Centre for Theory and Simulation of Materials, Imperial College London, South Kensington Campus, London SW7 2AZ, UK}
\affiliation{John A. Paulson School of Engineering and Applied Sciences, Harvard University, Cambridge, MA 02138, USA}

\author{Pierre A. Pantale\'on}
\affiliation{Imdea Nanoscience, Faraday 9, 28049 Madrid, Spain}
\author{Valerio Vitale}
\affiliation{Departments of Physics and Materials and the Thomas Young Centre for Theory and Simulation of Materials, Imperial College London, South Kensington Campus, London SW7 2AZ, UK}
\affiliation{Dipartimento di Fisica, Universit\`a di Trieste, Strada Costiera 11, 34151, Trieste, Italy}
\author{Lennart Klebl}
\affiliation{I. Institute of Theoretical Physics, University of Hamburg, Notkestrasse 9, 22607 Hamburg, Germany}

\author{Dante M.~Kennes}
\affiliation{Institute for Theory of Statistical Physics, RWTH Aachen University, and JARA Fundamentals of Future Information Technology, 52062 Aachen, Germany}
\affiliation{Max Planck Institute for the Structure and Dynamics of Matter, Center for Free Electron Laser Science, 22761 Hamburg, Germany}

\author{Arash Mostofi}
\affiliation{Departments of Physics and Materials and the Thomas Young Centre for Theory and Simulation of Materials, Imperial College London, South Kensington Campus, London SW7 2AZ, UK}
\author{Johannes Lischner}
\affiliation{Departments of Physics and Materials and the Thomas Young Centre for Theory and Simulation of Materials, Imperial College London, South Kensington Campus, London SW7 2AZ, UK}
\author{Francisco Guinea}
\affiliation{Imdea Nanoscience, Faraday 9, 28049 Madrid, Spain}
\affiliation{Donostia International  Physics Center, Paseo Manuel de Lardizabal 4, 20018 San Sebastian, Spain}


\date{\today}

\begin{abstract}
We discuss the effect of long-range interactions within the self-consistent Hartree-Fock (HF) approximation in comparison to short-range atomic Hubbard interactions on the band structure of twisted bilayer graphene (TBG) at charge neutrality for various twist angles. Starting from atomistic calculations, we determine the quasi-particle band structure of TBG with Hubbard interactions for various magnetic orderings: modulated anti-ferromagnetic (MAFM), nodal anti-ferromagnetic (NAFM) and hexagonal anti-ferromagnetic (HAFM). Then, we develop an approach to incorporate these magnetic orderings along with the HF potential in the continuum approximation. Away from the magic angle, we observe a drastic effect of the magnetic order on the band structure of TBG compared to the influence of the HF potential. Near the magic angle, however, the HF potential seems to play a major role on the band structure compared to the magnetic order. These findings suggest that the spin-valley degenerate broken symmetry state often found in HF calculations of charge neutral TBG near the magic angle should favour magnetic order, since the atomistic Hubbard interaction will break this symmetry in favour of spin polarization.
\end{abstract}

\maketitle

\section{Introduction}

Magic-angle twisted bilayer graphene (TBG) has generated tremendous interest in twinstronics~\cite{Carr2020NatRev,TT,moiresim} since the discovery of correlated insulating states and superconductivity in the $\sim$1.1$\degree$ moir\'e supperlattice~\cite{NAT_I,NAT_S}. The initial reports~\cite{SLG,NAT_I,NAT_S} indicated strong electron-electron correlations in TBG which gives rise to unconventional superconductivity~\cite{NAT_S}. While this is not unanimously agreed upon~\cite{Balents2020,Andrei2020,Saito2020,Stepanov2020,Oh2021un}, TBG has also been found to host strange metallic behaviour~\cite{SMTBLG,Polshyn2019}, nematic order~\cite{Cao2020,NAT_MEI,NAT_CO,IEC}, Dirac revivals~\cite{Zondiner2020,Wong2020}, Pomeranchuk effect~\cite{Rozen2021,Saito2021} and Chern insulators~\cite{Chern_Wu,Chern_Nuckolls,Chern_Das,Xie2021frac,Youngjoon2021}, amongst other effects and phases~\cite{TSTBLG,NAT_SS,SOM,EFM,Serlin2020,Yoo2019,Uri2020}. 

To understand these phases, given that the magic-angle of TBG contains $\sim$12,000 atoms~\cite{AC,CRAC,LDE,NSCS}, it is typical for the low-energy continuum model~\cite{LopesDosSantos2007,LopesDosSantos2012}, based on that of Bistritzer and MacDonald~\cite{Bistritzer2011}, to be utilised. This theory couples states of the Dirac cones of each layer and valley at different moir\'e crystal momenta, which causes the onset of flat bands at $\sim$1.1$\degree$~\cite{Andrei2020}. The continuum model of TBG can naturally be extended to include long-ranged Hartree-Fock interactions~\cite{EE,Xie2020CorrelatedIns,Bultinck2020,Liu2021Insulating,Cea2019Pinning,Zhang2020InsulatingHF,CeaElectrostatic2022,CeaCoulomb2021,Cea2020}, since it is an expansion in the moir\'e crystal momenta (which are very small values, corresponding to large length scales). This interacting theory has provided some understanding into the phase diagram of TBG~\cite{Balents2020}, in terms of the superconducting phase, correlated insulating states, Dirac revivals, pinning of van Hove singularities~\cite{EE,Cea2019Pinning,CeaCoulomb2021}, for example. 





Including short-ranged interactions in this continuum model has proven more difficult, however. Short-ranged interactions can be included in the Wannier orbital Hamiltonians of the flat bands~\cite{Zou2018,Po2018,Kang2019StrongCoupling,Seo2019FerroMott,Bascones2020,Carr2019wannier,PHD_1,PHD_2,PHD_3}, which provides a reduced Hamiltonian matrix which can be solved with strongly correlated methods~\cite{Kennes2018,Roy2019Unconventional,Kang2019StrongCoupling,Hsu2020RGanalysis,Laksono2018}, but this also has long-ranged interactions. It is more natural, however, to include short-ranged interactions in atomistic models, such as DFT~\cite{AC,CRAC} or tight-binding (TB)~\cite{LDE,NSCS}, since the atomic-scale information is retained in such approaches. For example, Klebl and Honerkamp~\cite{LK_CH} studied the magnetic phase diagram of TBG based on on-site Hubbard interactions from RPA spin-susceptibility calculations~\cite{PHD_6,AF2020}. Moreover, these atomistic approaches can also handle long-ranged interactions, such as  self-consistent Hartree interactions~\cite{Rademaker2019,PHD_4,PHD_9,Vahedi2021,Stauber2021,Gonzalex2020SymmetryBreaking}. 

%
%
A significant limiting factor of self-consistent atomistic approaches for broken symmetry phases is their computational cost~\cite{Sboychakov2019ManyBodyEffects,Sboychakov2020,Vahedi2021}. Some examples exist in the literature~\cite{Chichinadze2020NematicSupercond,Gonzalex2020SymmetryBreaking,Stauber2021,Vahedi2021}, but a full phase diagram - over twist angle and doping level, amongst other experimental variables - has not yet been achieved. For example, Stauber and Gonz\'alez ~\cite{Stauber2021,Gonzalex2020SymmetryBreaking} developed a theory based on Green's functions, where only some of the states were retained. They were able to study long-ranged interactions, and also the interplay between long and short ranged interactions, but only at 1.16$\degree$ and either at charge neutrality or $-2$ electrons per moir\'e unit cell. Moreover, Vahedi \textit{et al.}~\cite{Vahedi2021} investigated several twist angles (1.08$\degree$, 1.30$\degree$ and 1.47$\degree$) but only focused on charge neutrality. Usually, this has been overcome from re-scaling the TB parameters~\cite{GonzalezArraga2017,Vahedi2021,Ramires2019,Wolf2019} or applying hydrostatic pressure~\cite{PDTBLG,Bezanilla2019,Chen2022first}, such that flat bands can be created with unit cells only containing a few hundred to a few thousand atoms. 





%
%



Here we develop an approach which can include short-ranged interactions, such as the on-site Hubbard interaction of the p$_z$ orbitals, in the continuum model. Starting from the RPA spin susceptibility calculations of Klebl and Honerkamp~\cite{LK_CH}, we perform self-consistent atomistic Hubbard calculations to obtain the mean-field magnetic order parameters for different ordering tendencies (at a large twist angle and charge neutrality). We develop analytical forms for the magnetic ordering in real space, which we are able to include in the continuum model as an effective a scalar sublattice potential. In order to elucidate the angle dependence of the interplay between the magnetic orderings and the Hartree-Fock potential, we perform self-consistent Hartree-Fock calculations at charge neutrality, to which we later add the effective magnetic potential at different twist angles. Overall, it is found that the long range contribution dominates at the magic angle, but away from the magic-angle, these magnetic orders become more significant. We discuss the competition between these long and short range exchange interactions in detail, and finish with a discussion of future directions.

\section{Methods}

\subsection{Atomistic Calculations}
\label{sec:methods_atom}

We study commensurate moir\'e unit cells of TBG~\cite{LDE}, starting from AA stacked bilayers and rotating the top layer anticlockwise about an axis perpendicular to the graphene sheets that passes through a carbon atom in each layer. The moir\'e lattice vectors of the commensurate structures are $\textbf{R}_{1} = n\textbf{a}_{1} + m \textbf{a}_{2}$ and $\textbf{R}_{2} = -m\textbf{a}_{1} + (n + m) \textbf{a}_{2}$, where $n$ and $m$ are integers which define the commensurate TBG structure, and $\textbf{a}_{1} = (\sqrt{3}/2, -1/2)a_{0}$ and $\textbf{a}_{2} = (\sqrt{3}/2, 1/2)a_{0}$ are the lattice vectors of graphene, where $a_{0} = 2.46~\textrm{\AA}$ is the lattice constant of graphene.

At small twist angles, TBG undergoes significant atomic relaxations~\cite{AC,LREBM,CRAC,STBBG,SETLA,LDLE,Xia2020}. We calculate these relaxations using a classical force field implemented in the LAMMPS software package~\cite{LAMMPS}. The interlayer interactions are modelled using the AIREBO-Morse potential~\cite{AIREBO}, while intralyer interactions are described with the Kolmogorov-Crespi potential~\cite{KC}.

To investigate the electronic structure of TBG, we use the Hamiltonian 
\begin{equation}
\mathcal{\hat{H}} = \sum_{i\sigma}\varepsilon_{i\sigma}\hat{c}^{\dagger}_{i\sigma}\hat{c}_{i\sigma} + \sum_{ij\sigma}[t(\textbf{r}_{i} - \textbf{r}_{j})\hat{c}^{\dagger}_{j\sigma}\hat{c}_{i\sigma} + \text{H.c.}],
\label{eq:H}
\end{equation}
where $\varepsilon_{i\sigma}$ and $\hat{c}^{\dagger}_{i\sigma}$ ($\hat{c}_{i\sigma}$) denote the on-site energy of atom $i$ with spin $\sigma$ and the electron creation (annihilation) operator associated with atom $i$ and spin $\sigma$, respectively. The hopping parameters between atoms $i$ and $j$, $t(\textbf{r}_{i} - \textbf{r}_{j})$, are calculated using the Slater-Koster rules~\cite{SK}
\begin{equation}
t(\textbf{r}) = V_{pp\sigma}(\textbf{r})\bigg(\dfrac{\textbf{r}\cdot\textbf{e}_{z}}{|\textbf{r}|}\bigg)^{2} + V_{pp\pi}(\textbf{r})\bigg(1 - \dfrac{\textbf{r}\cdot\textbf{e}_{z}}{|\textbf{r}|}\bigg)^{2},
\end{equation}
where $V_{pp\sigma}(\textbf{r}) = V_{pp\sigma}^{0}\exp\{q_{\sigma}(1 - |\textbf{r}|/d_{AB})\}\Theta(R_c-|\mathbf{r}|)$ and $V_{pp\pi}(\textbf{r}) = V_{pp\pi}^{0}\exp\{q_{\pi}(1 - |\textbf{r}|/a)\}\Theta(R_c-|\mathbf{r}|)$. We take the pre-factor for the $pp\sigma$-hopping and $pp\pi$-hopping to be $V_{pp\sigma}^{0} = 0.48$ eV and $V_{pp\pi}^{0} = -2.7$ eV, respectively. The carbon-carbon bond length is given by $a = a_{0}/\sqrt{3}$ and the interlayer separation is taken to be $d_{\textrm{AB}} = 3.35~\textrm{\AA}$. We take the decay parameters of the Slater-Koster rules to be $q_{\sigma} = d_{\textrm{AB}}/0.184a_0$ and $q_{\pi} = 1/0.184\sqrt{3}$~\cite{LDE,NSCS}. Hoppings between carbon atoms separated by more than $R_c=10~\textrm{\AA}$ are neglected~\cite{EDS}.

To include the effects of short-range Hubbard interactions, the on-site energy is determined by the mean-field Hubbard interaction
\begin{equation}
\varepsilon_{i\sigma} = U[n_{i\sigma^{\prime}} - \textrm{const.}],
\end{equation}
where $U$ is the Hubbard parameter of the carbon p$_z$ orbital, and $n_{i\sigma^{\prime}}$ is the mean-field electron density on atom $i$ with the spin $\sigma^{\prime}$ being the opposite to $\sigma$. 

The electron density can be determined from the Bloch eigenstates $\psi_{n\mathbf{k}\sigma}(\mathbf{r})$ (with subscripts $n$ and $\mathbf{k}$ denoting a band index and the crystal momentum, respectively) according to
\begin{equation}
\begin{split}
n_{\sigma}(\textbf{r})& = \sum_{n\textbf{k}} f_{n\textbf{k}\sigma} |\psi_{n\textbf{k}\sigma}(\textbf{r})|^2 \\  &=\sum_{j}n_{j\sigma}\chi_j(\textbf{r}),
\end{split}
\end{equation}

\noindent where $f_{n\textbf{k}\sigma}=\Theta(\varepsilon_{F}-\varepsilon_{n\mathbf{k}\sigma})$ is the occupancy of state $\psi_{n\mathbf{k}\sigma}$ with eigenvalue $\varepsilon_{n\mathbf{k}\sigma}$ (where $\varepsilon_{F}$ is the Fermi energy), $\chi_j(\textbf{r}) = \sum_{\textbf{R}} \phi_{z}^2(\textbf{r}-\textbf{t}_j-\textbf{R})$ (with $\mathbf{R}$ denoting the moir\'e lattice vectors) and $n_{j\sigma}$ is the total number of electrons in the $j$-th orbital with spin $\sigma$. To characterizes the magnetic ordering, we calculate the spin polarisation
\begin{equation}
\zeta = \dfrac{n_{\uparrow} - n_{\downarrow}}{n_{\uparrow} + n_{\downarrow}}.
\end{equation}

To start the mean-field calculations, instead of guessing random magnetic configurations, we perform RPA spin-susceptibility calculations, following the methods outlined in Refs.~\citenum{LK_CH,PHD_6}. The eigenvalues of these calculations provide the critical interaction strength of an instability ($U_c$) and the eignevector is the form of the magnetic order. By using these eigenvectors as an initial on-site interaction, we can induce spin polarisation, which can then be used to perform self-consistent calculations. 

To obtain a self-consistent solution of the atomistic Hubbard equation, we use a simple mixing scheme with a mixing parameter of 0.1 typically (0.1 of the new electron spin density is mixed into the same spin density). When determining the Fermi energy, the total electron number is again forced to be $N + \nu$, but this does not restrict the spin densities to be the same. We mix the up and down spin density by the same amount, instead of choosing to work with the total electron density and magnetic order parameter, as we find it is sometimes more stable. 

Only the leading instability was able to be stabilised with this method~\cite{Goodwin2022thesis}. Therefore, we also perform constrained calculations. The constrained calculations require an analytical form for the magnetic order to be specified (which is outlined in Section \ref{sec:results_atom}), say $\zeta^{(j)}$, and we self-consistently determine the magnitude of this magnetic order through projecting onto the magnetic order parameter, $\zeta$. The on-site energy term is then given by
\begin{equation}
\varepsilon_{i\sigma} = \pm \dfrac{U}{2}\zeta_i^{(j)},
\end{equation}

\noindent where the sign depends on the spin. A self-consistent solution is obtained through a linear mixing of the spin densities. 

\subsection{Continuum Model Calculations}
\label{sec:methods_cont}

The mini-Brillouin Zone (mBZ) of the continuum model is spanned by the two reciprocal lattice vectors given by, $\vb{G}_{1} = \frac{2\pi}{L_{m}}\left( \frac{1}{\sqrt{3}}, 1\right)$ and $\vb{G}_{2} = \frac{4\pi}{L_{m}}\left( -\frac{1}{\sqrt{3}}, 0\right)$, where $L_{m} = \frac{a_{0}}{2 \sin{(\theta/2)}}$ is the moir\'e period and $a_{0}$ is the lattice constant of graphene. These vectors form the basis to define any reciprocal lattice vector, $\vb{G}_{i} = n \vb{G}_{1} + m \vb{G}_{2}$ with $n,m\in \mathbb{Z}$, $i\in \mathbb{N}$. The first star of reciprocal lattice vectors, i.e. the six first $\vb{G}_{i}$, are defined by $n,m\in[-1,1]$. 



The non-interacting continuum model is 4-fold degenerate, since it accounts for valley and spin quantum numbers, with the Hamiltonian at crystal momentum $\vb{k}$ being written as
 \begin{equation}
     \mathcal{\hat{H}}_{TBG, \xi}\left(\vb{k}\right) = \mqty( \hat{H}_{1, \xi}\left(\vb{k}\right) & \hat{T} \\ \hat{T}^{\dagger} & \hat{H}_{2,\xi}\left(\vb{k}\right) ),
 \end{equation}
where $\hat{H}_{l, \xi}$ is the continuum single layer graphene Hamiltonian of valley $\xi$, given by
\begin{equation}
    \hat{H}_{l,\xi}\left(\vb{k}\right) = \xi \hbar v_{F} \left( \vb{k} - \xi \vb{K}_{l} \right)\tau_{\theta, l}.
\end{equation}
with $v_{F} = (\sqrt{3} V_{pp\pi}^{0} a)/(2 \hbar)$ denoting the Fermi velocity, $\vb{K}_{l}$ is the position of the Dirac point of layer $l$, $\tau_{\theta, l} = e^{\mathrm{i}\xi\tau_{z} \theta / 2 } \mqty(\tau_{x}, \xi\tau_{y}) e^{-\mathrm{i}\xi\tau_{z} \theta / 2 }$, with $\tau_{i}$ being the Pauli matrices acting on the sublattice degree of freedom. The matrix $\hat{T}$ is a periodic function in the moir\'e unit cell that hybridises layers. For small angles, the main contribution comes from the first three reciprocal lattice vectors, $\vb{G} = (0,0)$, $\vb{G} = \vb{G}_{1}$ and $\vb{G} = \vb{G}_{1} + \vb{G}_{2}$~\cite{LopesDosSantos2007}
\begin{equation}
\begin{split}
    \hat{T} = \sum_{\vb{G}}\hat{T}(\vb{G}) &= \mqty(u_{1} & u_{2} \\ u_{2} & u_{1} ) + \mqty(u_{1} & u_{2}e^{-2\mathrm{i}\theta\pi / 3} \\ u_{2}e^{2\mathrm{i}\theta\pi / 3} & u_{1} ) \\ &+ \mqty(u_{1} & u_{2}e^{2\mathrm{i}\theta\pi / 3} \\ u_{2}e^{-2\mathrm{i}\theta\pi / 3} & u_{1} ),
\end{split}
\end{equation}
where $u_{1} = 0.0797$~eV and $u_{2} = 0.0975$~eV~\cite{Koshino2018} are, respectively, the hopping amplitudes between AB/BA and AA stacking, which takes into account the atomic relaxation in the continuum model.


To account for electron-electron interactions, we include the mean-field Hartree-Fock terms to the Hamiltonian. The Hartree contribution to the Hamiltonian is given by
\begin{equation}
    \label{eq:Hartree_Hamiltonian_Def}
     \mathcal{\hat{H}}_{H} = \sum_{i,\xi, \sigma} \int_{\Omega}\dd^{2}{\vb{r}} \psi_{\xi, \sigma}^{i,\dagger}\left( \vb{r} \right) \psi_{\xi, \sigma}^{i}\left( \vb{r} \right) V_{H}\left( \vb{r} \right),
\end{equation}
where $i\in[1,4]$ labels the combined sublattice and layer degree of freedom, $\sigma$ accounts for the spin, $\Omega$ is the area of the mBZ, and the local Hartree potential is given by
\begin{equation}
    \label{eq:Hartree_Potential_Def}
    V_{H}\left( \vb{r} \right) = \int_{\Omega} \dd^{2}{\vb{r'}} v_{C}\left( \vb{r} - \vb{r'} \right) \expval{ \delta\rho\left( \vb{r'} \right) }.
\end{equation}
Here $\delta\rho\left( \vb{r} \right) \equiv \rho\left( \vb{r} \right) - \rho_{CN}\left( \vb{r} \right)$ denotes the fluctuation in charge density, with $\rho\left( \vb{r} \right) = \sum_{\xi, \sigma}{\psi_{\xi, \sigma}^{\dagger}\left( \vb{r} \right) \psi_{\xi, \sigma}\left( \vb{r} \right) }$ corresponding to the charge density and $\rho_{CN}\left( \vb{r} \right)$ is the average density corresponding to the non-interacting TBG at charge neutrality point. We assume that the Coulomb interaction is screened by a double-metallic gate~\cite{Cea2019Pinning}
\begin{equation}
   v_{C}\left( \vb{q} \right) = \frac{2 \pi e^{2}}{\epsilon} \frac{\tanh{\left(d \abs{\vb{q}}\right)}}{\abs{\vb{q}}},
\end{equation}
where $d = 40$~nm is the distance to the metallic gates and $\varepsilon=10$ is the dielectric constant~\cite{Cea2020,PHD_3,PHD_4}. 


The Fock contribution to the Hamiltonian is given by
\begin{equation}
    \label{eq:Fock_Hamiltonian_Def}
     \mathcal{\hat{H}}_{F} = \sum_{i,j,\xi, \sigma} \int_{\Omega}\dd^{2}{\vb{r}}\dd^{2}{\vb{r'}}  \psi_{\xi, \sigma}^{i,\dagger}\left( \vb{r} \right) V_{F}^{ij}\left( \vb{r},\vb{r'} \right) \psi_{\xi, \sigma}^{j}\left( \vb{r'} \right),
\end{equation}
where $i,j$ run over the sublattice and layer indexes. The non-local Fock potential is described by,
\begin{equation}
    \label{eq:Fock_Potential_Def}
    V_{F}^{ij} = - \expval{\psi_{\xi, \sigma}^{j,\dagger}\left( \vb{r'} \right) \psi_{\xi, \sigma}^{i}\left( \vb{r} \right) }  v_{C}\left( \vb{r} - \vb{r'} \right),
\end{equation}
As we want to express the matrix elements of $\mathcal{\hat{H}}_{F}$, defined in Eq.~\eqref{eq:Fock_Hamiltonian_Def}, in the reciprocal space for which we must transform the non-local Fock potential into the Fourier space. By this procedure we compute the Fock matrix elements as
\begin{equation}
    \begin{split}
        &\mel{\vb{k}+\vb{G},\xi, \sigma,i}{\mathcal{H}_{F}}{\vb{k'}+\vb{G'},\xi', \sigma',i'} = \\
         &-\sum_{i,j,\xi, \sigma, n} \sum_{\vb{k''},\vb{G''}}  \psi_{n,\xi, \sigma}^{i}\left(\vb{k}+  \vb{G'} + \vb{G''} \right) \psi_{n,\xi, \sigma}^{j,*}\left(\vb{k'} + \vb{G} + \vb{G''} \right) \\ &\qquad \times v_{C}\left( \vb{k} - \vb{k''} + \vb{G''} \right),
    \end{split}
\end{equation}
where the index $n$ runs over the occupied bands at a given Fermi energy.
For the Hartree-Fock calculations we work with a continuum model of TBG expanded up to the third star. We use a density of points between $2-6\times 10^{5}~\textrm{\AA}^{2}$ in the mBZ, depending on the twisting angle. The convergence of the Hartree-Fock potential is normally reached after $5-6$ self-consistency steps and it has been proved for increasing density of points. 

To include the $\alpha$-magnetic potential ($\alpha=\text{M, N, H}$) in the continuum model we use a scalar sublattice and spin dependent potential expressed through its harmonic decomposition in the first star reciprocal lattice vectors
\begin{equation}
    \hat{H}^{\alpha}(\vb{G}_{i},\vb{G}_{j}) = \sum_{i,j=0}^{6} U_{\delta}^{\alpha}(\vb{G}_{i}-\vb{G}_{j}),
\end{equation}
The full form of $U_{\delta}^{\alpha}(\vb{G}_{i}-\vb{G}_{j})$ is discussed in the Section \ref{sec:results_cont_short}. The final Hamiltonian that combines both the effective magnetic potential derived from atomistic calculations and the Hartree-Fock potential at half-filling is given by, 

\begin{equation}
    \mathcal{\hat{H}}\left(\vb{k}\right) = \mathcal{\hat{H}}_{TBG}\left(\vb{k}\right) + \mathcal{\hat{H}}_{F}\left(\vb{k}\right) + \mathcal{\hat{H}}^{\alpha}.
\end{equation}
Note that this final Hamiltonian is not treated in a self-consistent way since the effective magnetic Hamiltonian is just added to the self-consistent Hartree-Fock Hamiltonian, which would be equivalent to a first-order approximation of the magnetic orderings in perturbation theory.


\section{Results}

\subsection{Short Range Atomistic Hubbard Interactions}
\label{sec:results_atom}

From the RPA ($\textbf{q} = 0$) spin-susceptibility calculations~\cite{LK_CH,PHD_6}, a number of leading antiferromagnetic instabilities are found: modulated antiferrogmagnetic order (MAFM), nodal antiferromagnetic order (NAFM) and hexagonal antiferromagnetic order (HAFM). To find which instability is the ground state at a given twist angle and doping level, (mean-field) atomistic Hubbard calculations must be performed. As these atomistic calculations are extremely computationally expensive at the magic angle, we focus on 1.54$\degree$ at charge neutrality. At this twist angle and doping level, TBGs leading instability was found to be MAFM (with a critical Hubbard interaction of $U_c \approx 5.1$~eV), with NAFM and HAFM having slightly larger critical interaction strengths (of $U_c \approx 5.4$~eV). For these latter instabilities, we perform unconstrained and constrained atomistic Hubbard calculations, as outlined in Section \ref{sec:methods_atom}.

\begin{figure*}[ht]
\includegraphics[width=\textwidth]{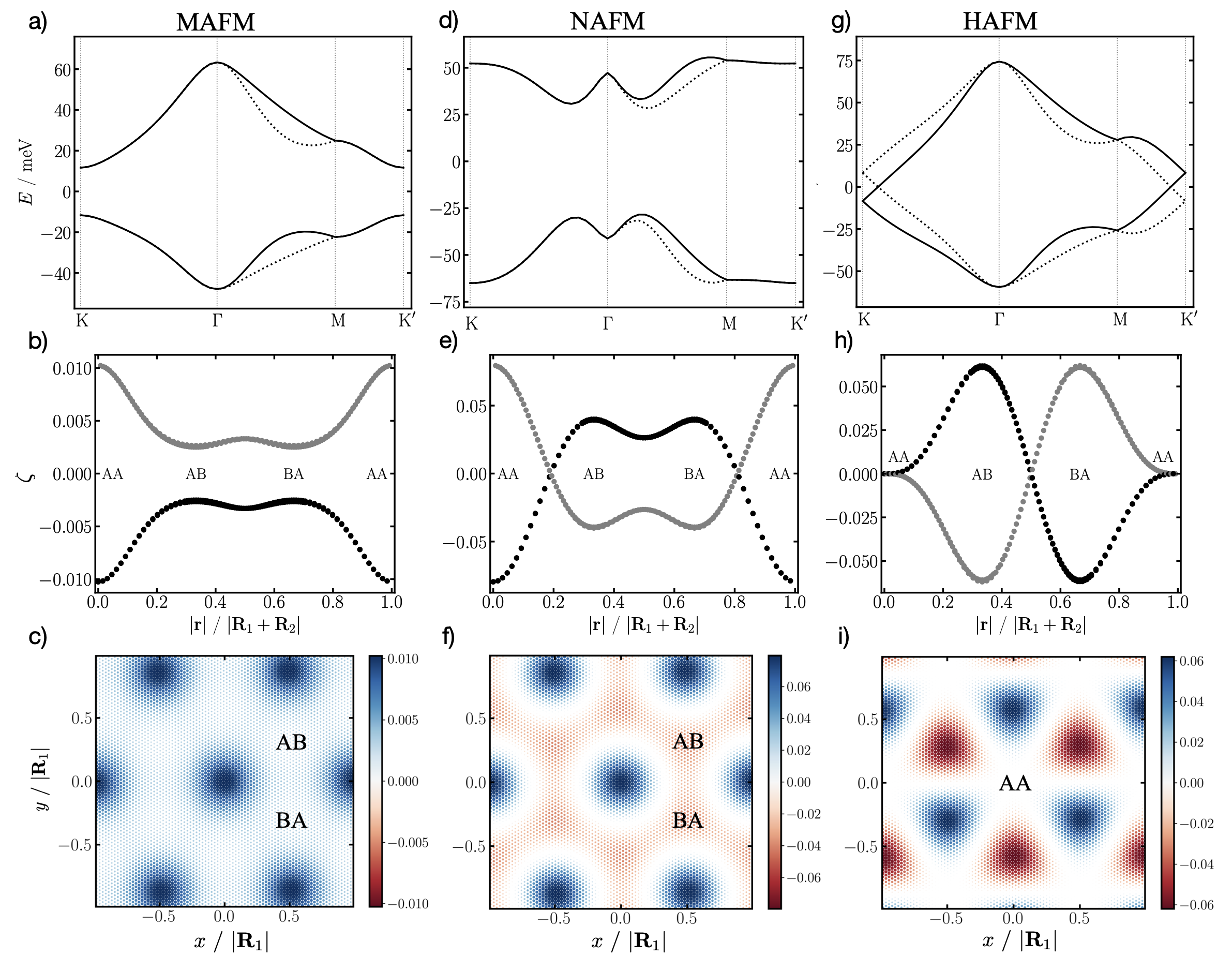}
\caption{(a,d,g) Self-consistent quasi-particle band structures for the studied magnetic orders of TBG at 1.54$\degree$ and charge neutrality along the high symmetry path. For MAFM, $U=5.4$~eV and unconstrained calculations were performed; whereas, for NAFM and HAFM, $U=5.94$~eV and constrained calculations were performed. (b,e,h) Corresponding self-consistent magnetic order parameter plotted along the diagonal of the moir\'e supperlattice, where $\textbf{R}_1$ and $\textbf{R}_2$ are the moir\'e lattice vectors. Sublattice A is shown in black and sublattice B is shown in grey for the top graphene layer (bottom layer not shown). (c,f,i) Corresponding plots in real space for a single layer and sublattice, where only sublattice B of the top layer is shown. Note, the $U$'s were chosen to be slightly above the critical values. For MAFM, if $U = 5.94$~eV is used, the gap at the Dirac point is extremely large~\cite{Goodwin2022thesis}.}
\label{fig:SPXA_test}
\end{figure*}

The MAFM instability, as seen in Fig.~\ref{fig:SPXA_test}(b) and (c), is characterised by a sub-lattice oscillation in the magnetic order parameter $\zeta = (n_{\uparrow} - n_{\downarrow})/(n_{\uparrow} + n_{\downarrow})$ that is modulated throughout the moir\'e supercell~\cite{LK_CH}. In Fig.~\ref{fig:SPXA_test}(b), which plots the magnetic structure along the diagonal of the moir\'e supperlattice, sublattice A is shown in black and sublattice B is shown in grey for the top graphene layer. To show its real-space structure more clearly, Fig.~\ref{fig:SPXA_test}(c) plots the magnetic order parameter on sublattice B of the top layer over the moir\'e superlattice. The constant contribution of the MAFM order is larger than the moir\'e-scale variation, which means the sub-lattice polarisation is the same throughout the moir\'e unit cell. The moir\'e scale variation of the magnetic order peaks in the AA regions, as might be expected from the LDOS of TBG peaking peaked in the AA regions~\cite{LDE}. The MAFM order can be approximated with the following analytical form
\begin{equation}
\zeta^{M}(\vb{r}) \approx \zeta_s' + \dfrac{\zeta_s}{6}\sum_{i = 1}^6\cos(\textbf{G}_{i}\cdot \textbf{r}),
\end{equation}


\noindent where $\vb{G}_i$ are reciprocal lattice vectors, and the sublattice oscillation $\zeta_s^{\prime}$ is the constant sublattice polarisation and $\zeta_s$ describes how this sublattice polarisation changes on the mor\'e scale. For sublattice $A_{l}$ ($B_{l}$), where $l=1,2$ is the layer index, the sign of the polarisation is $-$ ($+$), or vice versa. Note this equation assumes the AA region is located at the origin of the moir\'e unit cell.



In Fig.~\ref{fig:SPXA_test}(b) and (c) the unconstrained, self-consistent $\zeta$ for 1.54$\degree$ at charge neutrality and $U = 5.4$~eV is shown. The peaks of $\zeta \approx 0.1$ reside in the AA regions. If one electron is delocalized across all atoms in the moir\'e unit cell, which contains 5548 atoms at 1.54$\degree$, then $\zeta \approx 10^{-4}$. Therefore, the magnetic structures from these atomistic calculations are involving many more electrons than just the 4 flat band electrons, which is in agreement with other works~\cite{Stauber2021,Gonzalex2020SymmetryBreaking,Vahedi2021}.

In Fig.~\ref{fig:SPXA_test}(a), we show the corresponding self-consistent quasi-particle band structure. The different valleys, K and K', have been identified by applying the valley operator to the states (see Refs.~\cite{Alejandro2020,Wolf2019,Ramires2019} for details of this calculation), and shown in solid black and dotted grey, respectively. Since the MAFM order breaks C$_2$ symmetry, it causes a gap to open at the Dirac cones at the K/K' points of the moir\'e Brillouin zone of TBG. This instability was not able to be stabilised at $U = 5.1$~eV, but for $U$'s larger than $5.4$~eV, the constant contribution dominates and it becomes graphene-like ($\zeta_s^{'}$ $\gg$ $\zeta_s$, such that there is only a constant sub-lattice polarisation in each graphene sheet), with the gap at the K/K' points becoming very large ($100$'s of meV)~\cite{Goodwin2022thesis}.

Similarly, NAFM also has a moir\'e-scale peak in the magnetic order in the AA regions, but it does not possess a constant contribution to $\zeta$, as seen in the self-consistent values plotted in Fig.~\ref{fig:SPXA_test}(e). The corresponding real-space structure is shown in Fig.~\ref{fig:SPXA_test}(f), where nodes in the magnetic order around the AA region separate the regions of opposite signs of $\zeta$ spin polarisation. This magnetic order is referred to as nodal anti-ferromagnetic order because $\zeta$ goes through $0$ between the AA and AB/BA regions, causing the sign of $\zeta$ to change on each sub-lattice between these types of stacking~\cite{LK_CH}. Therefore, it can be described by 
\begin{equation}
\zeta^{N}(\vb{r}) \approx \dfrac{\zeta_s}{6}\sum_{i = 1}^6\cos(\textbf{G}_{i}\cdot \textbf{r}),
\end{equation}

\noindent This instability was not the leading instability~\cite{LK_CH}, and we found the unconstrained calculations could never stabilise this order, as it would always eventually revert to MAFM order. Therefore, we performed constrained mean-field Hubbard calculations to find the ``excited'' magnetic order, as explained in Section \ref{sec:methods_atom}. The quasi-particle band structure is shown in Fig.~\ref{fig:SPXA_test}(d), and again it is a Mott insulator. A large gap is obtained in the Dirac cone because of the Hubbard interaction parameter being taken as $U = 5.94$~eV. This instability could be stabilised with $U = 5.67$~eV and also larger $U$'s in the constrained method. 

Finally, the HAFM instability is similar to NAFM, but where the magnetic order parameter peaks on the AB/BA regions instead of the AA regions. The self-consistent HAFM magnetic structure is shown in Fig.~\ref{fig:SPXA_test}(h) and (i). The sign of $\zeta$ changes between the AB and BA regions of the moir\'e unit cell. As the peaks of $\zeta$ occur on the AB/BA regions, which form a hexagonal lattice on the moir\'e scale, this ordering is referred to as hexagonal anti-ferromagnetic order. This magnetic order is shown in Fig.~\ref{fig:SPXA_test}(h) and (i), and can be described by the analytical form
\begin{equation}
\zeta^{H}(\vb{r}) \approx \dfrac{\zeta_s}{6}\sum_{i  = 1}^6\sin(\textbf{G}_{i}\cdot \textbf{r}).
\end{equation}

This instability never appears to be a leading instability, but its critical Hubbard interaction was $\sim 5.4$~eV, which is only slightly higher than the leading instability. Therefore, we again found that constrained calculations were required to obtain mean-field values of its order parameter, as explained in Section \ref{sec:methods_atom}. In Fig.~\ref{fig:SPXA_test}(g) we show the mean-field quasi-particle band structure for $U = 5.94$~eV (this was the smallest $U$ which could stabilise the order with constrained calculations), where the different valleys have been coloured solid black and dotted grey. We find that this magnetic order does not create a gap at the K/K' point, despite it having a sublattice oscillation. This is because of the moir\'e-scale sine nature of its variation, which means it does not break C$_2$ on the moir\'e scale. It does cause the valleys to split at the K/K' points, however, resulting in one valley being pushed higher in energy and the other to lower energies. This is analogous to the effect of a perpendicular electric field has on the electronic structure. 


These calculations give some insight into the magnetic order from Hubbard interactions in TBG. However, we performed these well away from the magic angle and for Hubbard interaction parameters that are large ($U \approx 5.5$~eV for these calculations, but the value is thought to be $\sim4$~eV~\cite{GonzalezArraga2017}). Therefore, to understand the role of these instabilities close to the magic angle, we aim to include these magnetic states in the continuum model.


\subsection{Long Range Interactions in the Continuum Model}
\label{sec:results_cont}



\begin{figure*}[ht]
\centering
\includegraphics[width=\textwidth]{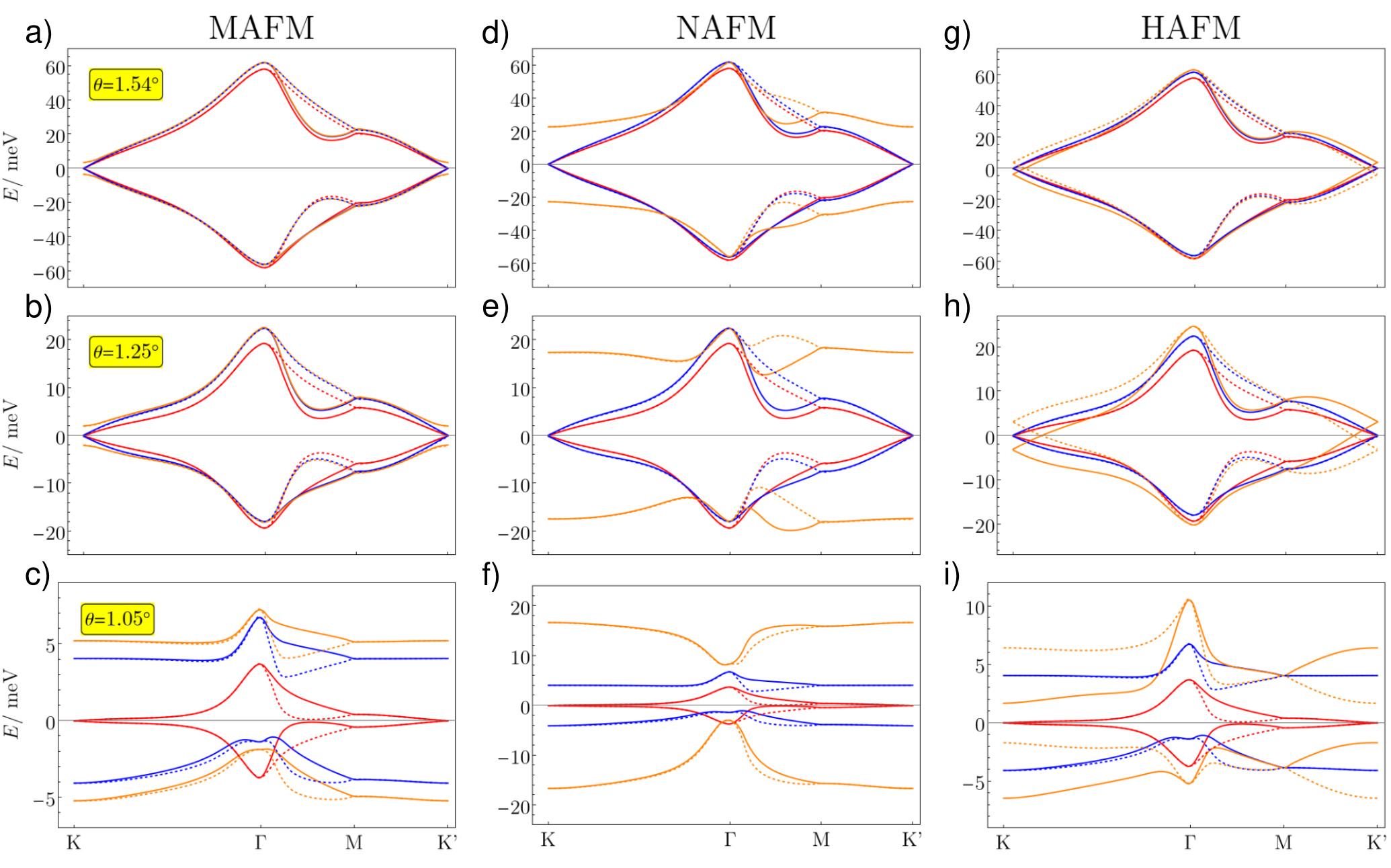}
\caption{Comparison of low energy band structure, along the path $\text{K} \rightarrow \Gamma \rightarrow \text{M}\rightarrow \text{K'}$ in the mBZ, of the non-interacting case in red, HF in blue and HF+MAFM (a, b, c), HF+NAFM (d, e, f), HF+HAFM (g, h, i) in orange, for $\theta = 1.54^{\circ}$ (a, d, g) $\theta=1.25^{\circ}$ (b, e, f) and $\theta = 1.05^{\circ}$ (c, f, i). Dashed lines correspond to valley flip.}
\label{fig:fig_main_f18}
\end{figure*}

In Fig.~\ref{fig:fig_main_f18} we show the Hartree-Fock quasi-particle band structure (solid blue line, for all subplots) for a number of twist angles at charge neutrality, in addition to the non-interacting band structure (solid red line, for all subplots). At 1.54$\degree$, Figs.~\ref{fig:fig_main_f18}(a), (d) and (g), we find that the Hartree-Fock potential slightly modifies the non-interacting band structure, indicating that this twist angle is too large for the formation of an insulating state. At a twist angle of 1.25$\degree$, Figs.~\ref{fig:fig_main_f18}(b), (e) and (h), the non-interacting bands are flat enough for the onset of a small gap due to the Hartree-Fock potential, significantly smaller than the bandwidth, at the Dirac cones K and K' points in the moir\'e Brillouin zone. Right at the magic angle of 1.05$\degree$, Figs.~\ref{fig:fig_main_f18}(c), (f) and (i), the Hartree-Fock interactions induce a large gap at the K and K' points, on a similar scale to the bandwidth. At half filling (charge neutrality) these insulating states are characterised by a broken sublattice symmetry and a preserved spin and valley symmetry with respect to the non-interacting picture. This implies that the mean-field ground state associated to the Hartree-Fock band structure is actually a linear combination of 4 states with different spin and valley indexes. These calculations are in good agreement with a large body of literature which investigates these long ranged interactions ...

\begin{table}[h]
\begin{tabular}{lccc}
\hline\hline
Order &$\quad$& $\abs{\delta_1}$ & $\abs{\delta_2}$\\
\hline
MAFM ($U=5.4$~eV) && $12.47$ & $12.54$\\
NAFM ($U=5.94$~eV) && $0$ & $237.09$\\
HAFM ($U=5.94$~eV) && $0$ & $211.89$\\
\hline \hline
\end{tabular}
\caption{Parameters $\abs{\delta_1}$ and $\abs{\delta_2}$ extracted form the atomistic Hubbard calculations for the different orderings considered in this work. All values are expressed in meV.}
\label{tab:parameterValues}
\end{table}

\subsection{Short Range Interactions in the Continuum Model}
\label{sec:results_cont_short}


In Section~\ref{sec:results_atom}, we described the leading anti-ferromagnetic instabilities obtained from RPA calculations of the atomistic model, performed mean-field Hubbard calculations of these, and found analytical forms which approximate the spin polarisation well, with the magnetic potential they create simply being $U\zeta/2$. Taking inspiration from how the Hartree potential is calculated in the atomistic model relative to the continuum model, a general expression for the potential induced by these anti-ferromagnetic instabilities would be
\begin{equation}
    U_{\delta}(\vb{r}) = \delta_{1} \tau_{z} + \dfrac{1}{6}\sum_{i=1}^6\delta_2\tau_{z} e^{\mathrm{i}\vb{G}_{i}\cdot\vb{r}},
\end{equation}
\noindent where $\delta_{1}$ corresponds to the constant sublattice polarization strength, $\delta_2(\vb{G}_{i})$ is the moir\'e modulated part of the potential, i.e. the weight associated to the expansion of the potential in the $i$-th BZ reciprocal lattice vector, and $\tau_z$ is the Pauli matrix. 

For the MAFM instability both $\delta_1$ and $\delta_2$ are real valued. In the case of NAFM order there is not a constant contribution to $\zeta^{N}$ so $\delta_{1}=0$, and $\delta_2$ is a real number. Similarly for the HAFM instability $\delta_{1}=0$ but $\delta_2$ is a purely imaginary number so the modulation is a sine function. Both MAFM and NAFM orderings are degenerate in the valley index but the HAFM is not and the modulated contribution to the potential must be complex conjugated when exchanging valleys. The value of the parameters $\delta_{1}$ and $\delta_{2}$ are obtained numerically within the self-consistent atomistic Hubbard calculation at a twist angle of 1.54$\degree$ and at charge neutrality, as described in Section~\ref{sec:results_atom}. Their values depending on the magnetic orders are summarized in Tab.~\ref{tab:parameterValues}. 

These parameters should be linear in the interaction strength and the
spin-polarised electron density, meaning an overall non-linear dependence on $U$, but we shall not seek self-consistent
solutions to this potential in the continuum model here. Instead, these potentials are included in the continuum model through
perturbation theory on the converged self-consistent
Hartree-Fock calculations, as explained in Section \ref{sec:methods_cont}. Note that the sublattice is polarised in the Hartree-Fock calculations, which means the overall sign of the potential can be chosen in two different ways, but we always choose the sign such that the sublattice polarisation matches, as this should increase any gaps, and further lower the energy. In the Appendix, we show the band structures in the continuum model (without the Hartree-Fock contribution) at 1.54$\degree$ with these perturbed potentials, and find good agreement with the atomistic calculations.

For twist angles smaller than 1.54$\degree$, the Hubbard interaction in the continuum model should scale as $\abs{\delta_{1/2}\left(\theta\right)} = \abs{\delta_{1/2}\left(1.54^{^{\circ}}\right)}\left(\frac{\theta}{1.54^\circ}\right)^{2}$,~\cite{EE}. However, extending the parameters to the magic angle in this way does not yield very physical results, as the gaps in the flat bands are much larger than is ever found in experiments, as shown in Fig.~\ref{fig:fig_f6}(c), (f) and (i). This is because to obtain a mean-field solution of the magnetic order at 1.54$\degree$ a $U = 5.4-6$~eV was required. However, a more physical value is $U = 4$~eV, or smaller~\cite{GonzalezArraga2017}. Moreover, the large interaction strength gives rise to large polarised spin densities, as seen in Section~\ref{sec:results_atom}, where the $\zeta$ values indicated that not just the flat band electrons were being polarised. Thus, we rescale these parameters to obtain more suitable estimates for these for the changes to the electronic structure. In Fig.~\ref{fig:fig_main_f18} we show the results with a scaling factor of $3$ in the main text, with other scaling factors being shown in the Appendix. As these perturbed calculations are not self-consistent, we cannot say what is the ground state. However, we can interpret the changes to the electronic structure, which provides information about where these interactions might be significant.






First we shall discuss MAFM order, as shown in Figs.~\ref{fig:fig_main_f18}(a), (b) and (c). At a twist angle of 1.54$\degree$, Fig.~\ref{fig:fig_main_f18}(a), we find a small gap at the K/K' points, which is more significant than the Hartree-Fock distortions. For a smaller twist angle of 1.25$\degree$, Fig.~\ref{fig:fig_main_f18}(b), a similar situation is found, the magnetic order opens a gap at the K/K' points while the Hartree-Fock potential is responsible for minor adjustments in the band structure. At the magic angle of 1.05$\degree$, Fig.~\ref{fig:fig_main_f18}(c), the situation totally changes, now the magnetic potential only slightly modifies the gap at the K/K' points, while the Hartree-Fock contribution dominates the deformations to the electronic structure. 

Next, we move on to describing the NAFM ordering, as seen in Fig.~\ref{fig:fig_main_f18}(d), (e) and (f). At the largest twist angle of 1.54$\degree$, Fig.~\ref{fig:fig_main_f18}(d), this magnetic order creates a significant gap at the K/K' points. This could be attributed to the non-self-consistent nature of the calculations, and form using large values of the parameters. This large band deformations persists at the smaller twist angles of 1.25$\degree$, Fig.~\ref{fig:fig_main_f18}(e), and 1.05$\degree$, Fig.~\ref{fig:fig_main_f18}(f). Even if smaller values of the parameters are utilised, this NAFM order induces large band deformations, lowering the energy of the occupied valence band. Therefore, it appears that this magnetic order can compete with the Hartree-Fock contribution.

Finally, we describe the effect of HAFM. As can be seen in Figs.~\ref{fig:fig_main_f18}(g), (h) and (i), the HAFM order does not create a gap at the K/K' points. Instead it causes the Dirac cones at K and K' to shift up and down, respectively, for the single spin and valley channel. At 1.54$\degree$, Fig.~\ref{fig:fig_main_f18}(g), this effect is almost imperceptible but stronger than the one induced by the Hartree-Fock interactions. For the smaller twist angle of 1.25$\degree$, Fig.~\ref{fig:fig_main_f18}(h) the situation is similar but the energy gap between K and K' points increases. While at the magic-angle, Fig.~\ref{fig:fig_main_f18}(i), it slightly contributes to reshaping the band structure, which in contrast is heavily affected by the Hartree-Fock contribution, so we can safely say that the HAFM is a secondary effect in this case.

\section{Discussion}

Overall, it appears that these magnetic potentials are more significant away from the magic angle, but at angles close enough that there could still be broken symmetry phases~\cite{PHD_6,Youngjoon2021}. For example, 1.25$\degree$ seems to be the most significantly affected by these Hubbard potentials relative to the Hartree-Fock contribution. At the magic angle, the Hartree-Fock contribution dominates, and at large twist angles the effects are small relative to the bandwidth, suggesting that these magnetic orders are not significant at these twist angles. This twist angle dependence could suggest why the predictions of Klebl \textit{et al.}~\cite{PHD_6}, in terms of the twist angle and doping dependence of magnetic states, agreed well with subsequent experiments~\cite{Youngjoon2021}. This could be because these Hubbard interactions are important close to the onset of broken symmetry phases, but close to the magic angle these Hubbard interactions are dominated by long-ranged Hartree-Fock interactions~\cite{EE}.

The NAFM order appears to effect the electronic structure most significantly, significantly lowering the eigenvalues of the occupied valence band, and therefore, it is a possible candidate for magnetic order in TBG. In contrast, the MAFM order effects the electronic structure more weakly. Finally, the HAFM appears to only effect the electronic structure slightly. This magnetic order should, however, couple to perpendicular electric fields~\cite{GonzalezArraga2017,Goodwin2022thesis}, which could make this ordering tendency more important. These perturbative calculations are interesting to be performed because they are a very natural explanation for the correlated insulating states in TBG~\cite{NAT_I}. From the Hartree-Fock calculations, a spin-valley degenerate insulating state is obtained. The atomistic Hubbard interaction, however, should break this symmetry and cause the onset of magnetic order. 



We have focused on TBG here, but many more moir\'e materials comprised of graphene exist~\cite{Cea20193D,Pierre2020}. Perhaps the most promising ones are where there is a $\pm\theta$ twist between each adjacent graphene layer~\cite{TBorNTB,khalaf2019magic,Zhu2020,Kruchkov2020,Fischer_TTLG,Phong2021trilayer,Shin2021,Alejandro2020}. These moir\'e graphene multilayers have been shown to host highly tunable superconducting phases~\cite{Park2021,Cao2021large,hao2021engineering,Turkel2022,Kim2022tri}, and as the number of layers increases, the superconducting phase occurs over wider and wider doping ranges~\cite{Zhang2022tri,Burg2022quad}. Fischer \textit{et al.}~\cite{Fischer_TTLG} has shown similar types of magnetic order occur in these systems, which means it will be possible to use the developed method. Another example of moir\'e structures is graphene twisted on a graphene multilayer, such as twisted mono-bilayer graphene~\cite{Shi2020tTLG,Yankowitz2020}. The magnetic structure of these systems was shown to be more complex by Goodwin \textit{et al.}~\cite{PHD_8}, which suggests the approach described here could be difficult to utilise. Finally, another class of moir\'e graphene multilayers is twisted bilayers composed of graphene multilayers, such as twisted double bilayer graphene~\cite{TCT,BIBI,PhysRevLett.123.197702,Crommie2021tDBLG,nematicity2020,cao2019electric}. Further investigation of this system would be of interest.




\section{Conclusion}

In summary, starting from atomistic methods, we studied several leading magnetic instabilities of charge neutral TBG at a large twist. These calculations permitted analytical forms for the magnetic ordering. These Hubbard potentials were investigated perturbatively from self-consistent Hartree-Fock calculations in the continuum model, allowing a comparison of long and short range exchange interactions. From these calculations, our take-home conclusions are:

\begin{enumerate}
    \item These atomistic Hubbard interactions break the spin-valley degeneracy of the insulating state at charge neutrality of TBG obtained from self-consistent Hubbard calculations. Therefore, these insulating states are likely to have some Mottness.
    \item These magnetic orders are most significant for intermediate twist angles between the magic-angle and angles where non-interacting physics is sufficient. At the magic-angle, the Hartree-Fock contribution dominates, and at large angles the bandwidth dominates. 
    \item Out of the studied magnetic orders, nodal anti-ferromagnetic order appears to be the most significant for changes to the electronic structure.
\end{enumerate}

It is hoped that these results further motivate inclusion of atomistic effects in the continuum model. Moreover, performing self-consistent magnetic calculations should also be possible, and investigating such ordering tendencies in other moir\'e graphene multilayers is possible now.

\section{Acknowledgments}
A.J.P., P.A.P. and F.G. acknowledge support from the Severo Ochoa programme for centres of excellence in R\&D (Grant No.\ SEV-2016-0686, Ministerio de Ciencia e Innovaci\'on, Spain); from the European Commission, within the Graphene Flagship, Core 3, grant number 881603 and from grants NMAT2D (Comunidad de Madrid, Spain) and SprQuMat (Ministerio de Ciencia e Innovaci\'on, Spain). ZG was supported through a studentship in the Centre for Doctoral Training on Theory and Simulation of Materials at Imperial College London funded by the EPSRC (EP/L015579/1). We acknowledge funding from EPSRC grant EP/S025324/1 and the Thomas Young Centre under grant number TYC-101. We acknowledge the Imperial College London Research Computing Service (DOI:10.14469/hpc/2232) for the computational resources used in carrying out this work. This  project  has  received  funding  from  the  European  Union’s  Horizon  2020  research  and  innovation programme under the Marie Sk\l{}odowska-Curie grant agreement No. 101067977. The Deutsche Forschungsgemeinschaft (DFG, German Research Foundation) is acknowledged for support through RTG 1995, within the Priority Program SPP 2244 “2DMP” and under Germany’s Excellence Strategy-Cluster of Excellence Matter and Light for Quantum Computing (ML4Q) EXC2004/1 - 390534769. We acknowledge support from the Max Planck-New York City Center for Non-Equilibrium Quantum Phenomena. Spin susceptibility calculations were performed with computing resources granted by RWTH Aachen University under projects rwth0496 and rwth0589.

\setcounter{equation}{0}
\setcounter{figure}{0}
\setcounter{table}{0}
\setcounter{section}{0}
\renewcommand{\theequation}{A\arabic{equation}}
\renewcommand{\thefigure}{A\arabic{figure}}
\renewcommand{\thesection}{\Alph{section}}

\appendix

\begin{figure*}[ht]
\centering
\includegraphics[width=\textwidth]{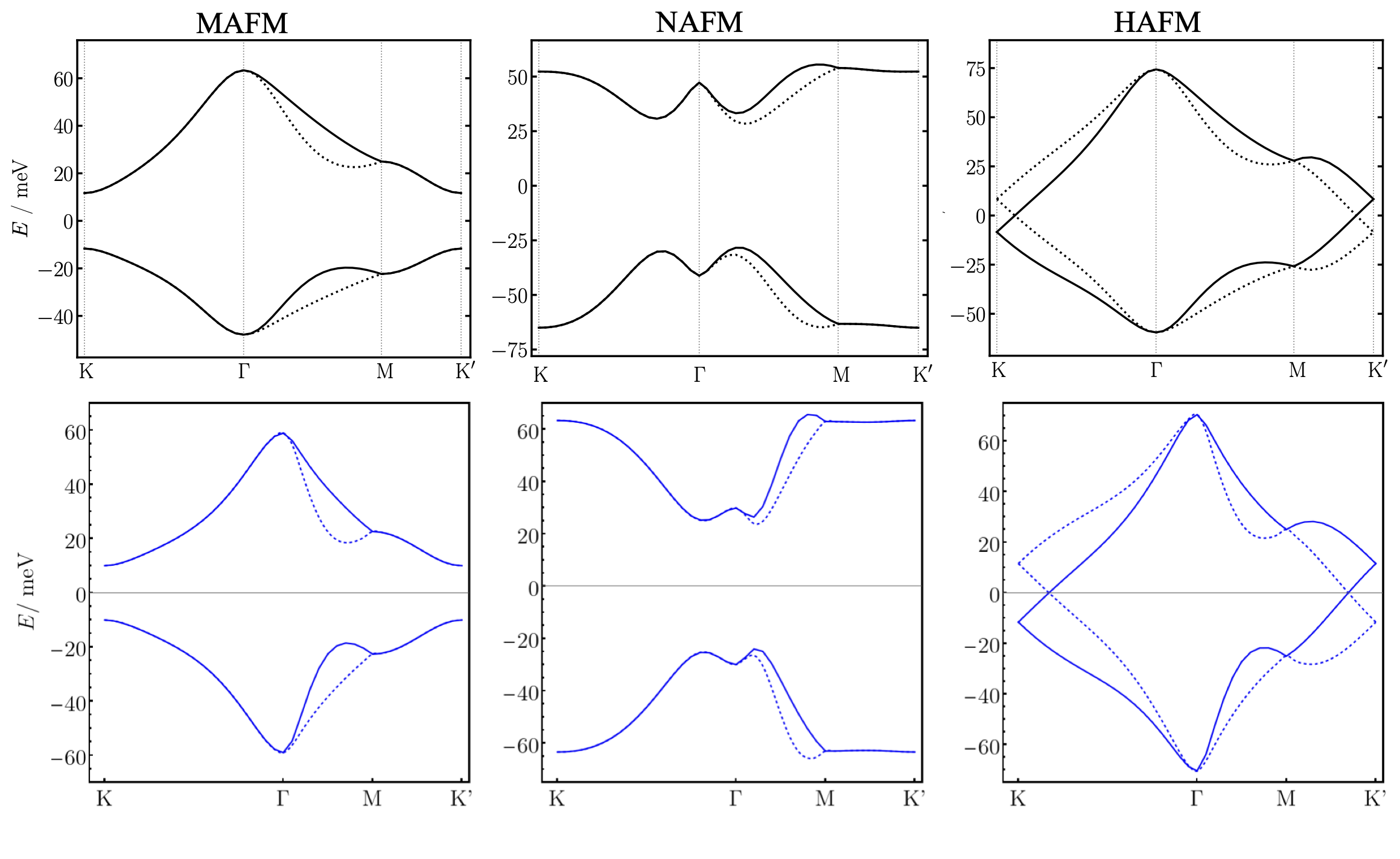}
\caption{Low energy band structure for $\theta = 1.54^{\circ}$, along the path $\text{K} \rightarrow \Gamma \rightarrow \text{M}\rightarrow \text{K'}$ in the mBZ. (a) for MAFM, (b) for NAFM, (c) for HAFM. Dashed lines correspond to valley flip.}
\label{fig:fig_fixFactor}
\end{figure*}

\begin{figure*}[ht]
\centering
\includegraphics[width=\textwidth]{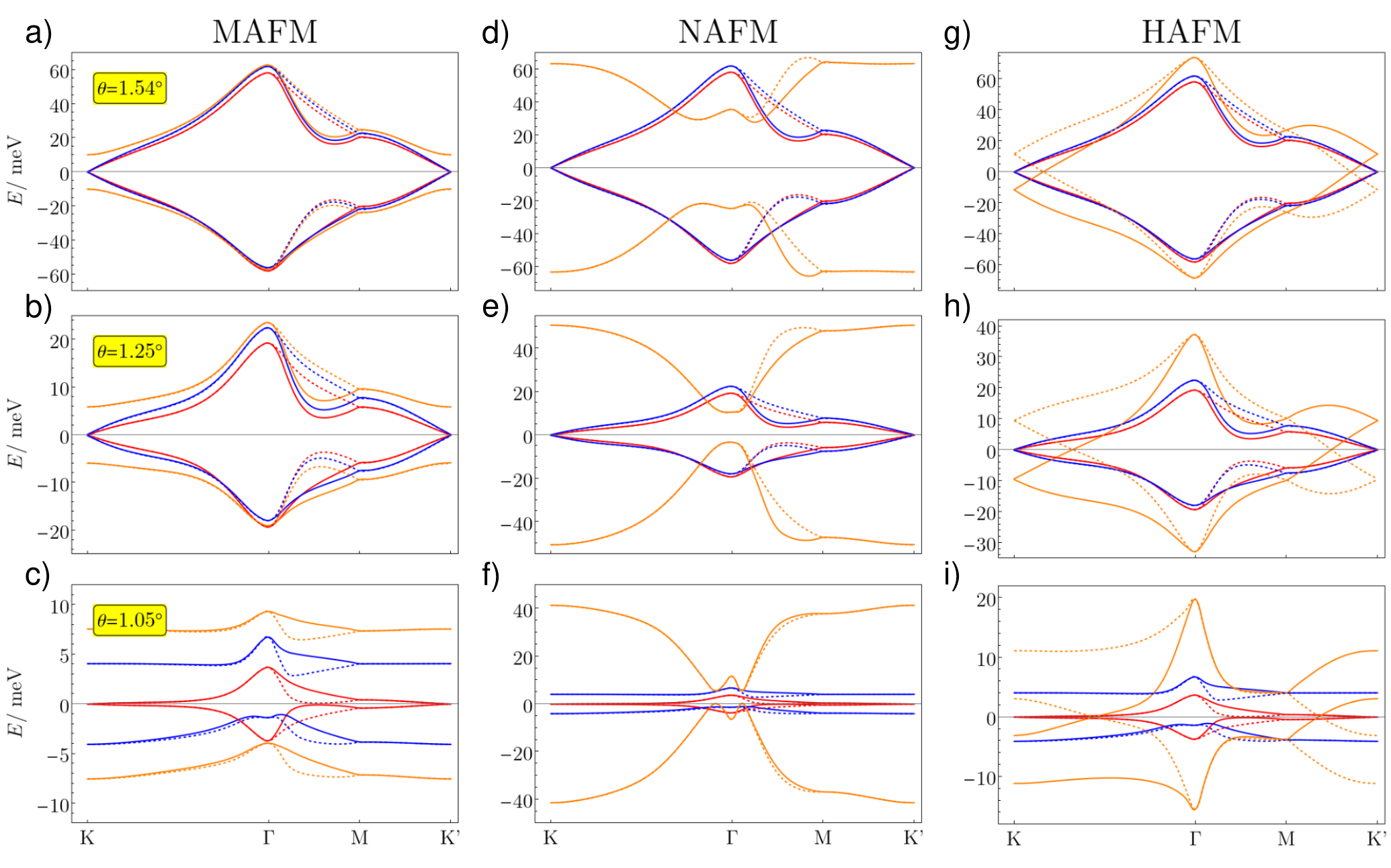}
\caption{Comparison of low energy band structure of the non-interacting case in red, HF in blue and HF+MAFM (a, b, c), HF+NAFM (d, e, f), HF+HAFM (g, h, i) in orange, for $\theta = 1.54^{\circ}$ (a, d, g) $\theta=1.25^{\circ}$ (b, e, f) and $\theta = 1.05^{\circ}$ (c, f, i). Dashed lines correspond to valley flip. This scenario correspond to a scaling factor of 1.}
\label{fig:fig_f6}
\end{figure*}

\begin{figure*}[ht]
\centering
\includegraphics[width=\textwidth]{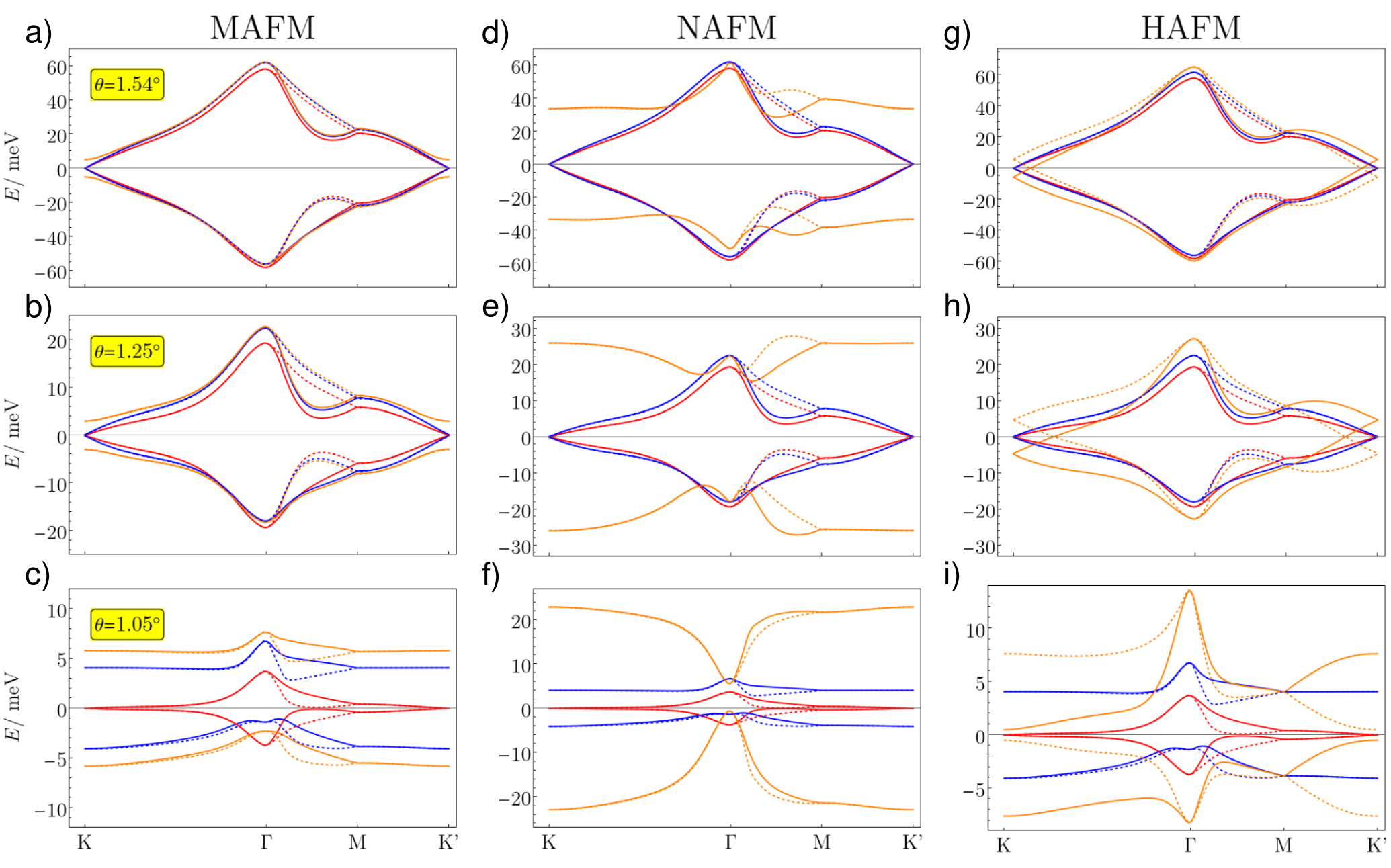}
\caption{Comparison of low energy band structure of the non-interacting case in red, HF in blue and HF+MAFM (a, b, c), HF+NAFM (d, e, f), HF+HAFM (g, h, i) in orange, for $\theta = 1.54^{\circ}$ (a, d, g) $\theta=1.25^{\circ}$ (b, e, f) and $\theta = 1.05^{\circ}$ (c, f, i). Dashed lines correspond to valley flip. This scenario correspond to a scaling factor of 2.}
\label{fig:fig_f12}
\end{figure*}


\begin{thebibliography}{118}%
\makeatletter
\providecommand \@ifxundefined [1]{%
 \@ifx{#1\undefined}
}%
\providecommand \@ifnum [1]{%
 \ifnum #1\expandafter \@firstoftwo
 \else \expandafter \@secondoftwo
 \fi
}%
\providecommand \@ifx [1]{%
 \ifx #1\expandafter \@firstoftwo
 \else \expandafter \@secondoftwo
 \fi
}%
\providecommand \natexlab [1]{#1}%
\providecommand \enquote  [1]{``#1''}%
\providecommand \bibnamefont  [1]{#1}%
\providecommand \bibfnamefont [1]{#1}%
\providecommand \citenamefont [1]{#1}%
\providecommand \href@noop [0]{\@secondoftwo}%
\providecommand \href [0]{\begingroup \@sanitize@url \@href}%
\providecommand \@href[1]{\@@startlink{#1}\@@href}%
\providecommand \@@href[1]{\endgroup#1\@@endlink}%
\providecommand \@sanitize@url [0]{\catcode `\\12\catcode `\$12\catcode
  `\&12\catcode `\#12\catcode `\^12\catcode `\_12\catcode `\%12\relax}%
\providecommand \@@startlink[1]{}%
\providecommand \@@endlink[0]{}%
\providecommand \url  [0]{\begingroup\@sanitize@url \@url }%
\providecommand \@url [1]{\endgroup\@href {#1}{\urlprefix }}%
\providecommand \urlprefix  [0]{URL }%
\providecommand \Eprint [0]{\href }%
\providecommand \doibase [0]{https://doi.org/}%
\providecommand \selectlanguage [0]{\@gobble}%
\providecommand \bibinfo  [0]{\@secondoftwo}%
\providecommand \bibfield  [0]{\@secondoftwo}%
\providecommand \translation [1]{[#1]}%
\providecommand \BibitemOpen [0]{}%
\providecommand \bibitemStop [0]{}%
\providecommand \bibitemNoStop [0]{.\EOS\space}%
\providecommand \EOS [0]{\spacefactor3000\relax}%
\providecommand \BibitemShut  [1]{\csname bibitem#1\endcsname}%
\let\auto@bib@innerbib\@empty
\bibitem [{\citenamefont {Carr}\ \emph
  {et~al.}(2020{\natexlab{a}})\citenamefont {Carr}, \citenamefont {Fang},\ and\
  \citenamefont {Kaxiras}}]{Carr2020NatRev}%
  \BibitemOpen
  \bibfield  {author} {\bibinfo {author} {\bibfnamefont {S.}~\bibnamefont
  {Carr}}, \bibinfo {author} {\bibfnamefont {S.}~\bibnamefont {Fang}},\ and\
  \bibinfo {author} {\bibfnamefont {E.}~\bibnamefont {Kaxiras}},\ }\bibfield
  {title} {\bibinfo {title} {Electronic-structure methods for twisted moir\'e
  layers},\ }\href {https://www.nature.com/articles/s41578-020-0214-0}
  {\bibfield  {journal} {\bibinfo  {journal} {Nat. Rev. Mater}\ }\textbf
  {\bibinfo {volume} {5}},\ \bibinfo {pages} {748} (\bibinfo {year}
  {2020}{\natexlab{a}})}\BibitemShut {NoStop}%
\bibitem [{\citenamefont {Carr}\ \emph {et~al.}(2017)\citenamefont {Carr},
  \citenamefont {Massatt}, \citenamefont {Fang}, \citenamefont {Cazeaux},
  \citenamefont {Luskin},\ and\ \citenamefont {Kaxiras}}]{TT}%
  \BibitemOpen
  \bibfield  {author} {\bibinfo {author} {\bibfnamefont {S.}~\bibnamefont
  {Carr}}, \bibinfo {author} {\bibfnamefont {D.}~\bibnamefont {Massatt}},
  \bibinfo {author} {\bibfnamefont {S.}~\bibnamefont {Fang}}, \bibinfo {author}
  {\bibfnamefont {P.}~\bibnamefont {Cazeaux}}, \bibinfo {author} {\bibfnamefont
  {M.}~\bibnamefont {Luskin}},\ and\ \bibinfo {author} {\bibfnamefont
  {E.}~\bibnamefont {Kaxiras}},\ }\bibfield  {title} {\bibinfo {title}
  {Twistronics: Manipulating the electronic properties of two-dimensional
  layered structures through their twist angle},\ }\href
  {https://link.aps.org/doi/10.1103/PhysRevB.95.075420} {\bibfield  {journal}
  {\bibinfo  {journal} {Phys. Rev. B}\ }\textbf {\bibinfo {volume} {95}},\
  \bibinfo {pages} {075420} (\bibinfo {year} {2017})}\BibitemShut {NoStop}%
\bibitem [{\citenamefont {Kennes}\ \emph {et~al.}(2021)\citenamefont {Kennes},
  \citenamefont {Claassen}, \citenamefont {Xian}, \citenamefont {Georges},
  \citenamefont {Millis}, \citenamefont {Hone}, \citenamefont {Dean},
  \citenamefont {Basov}, \citenamefont {Pasupathy},\ and\ \citenamefont
  {Rubio}}]{moiresim}%
  \BibitemOpen
  \bibfield  {author} {\bibinfo {author} {\bibfnamefont {D.~M.}\ \bibnamefont
  {Kennes}}, \bibinfo {author} {\bibfnamefont {M.}~\bibnamefont {Claassen}},
  \bibinfo {author} {\bibfnamefont {L.}~\bibnamefont {Xian}}, \bibinfo {author}
  {\bibfnamefont {A.}~\bibnamefont {Georges}}, \bibinfo {author} {\bibfnamefont
  {A.~J.}\ \bibnamefont {Millis}}, \bibinfo {author} {\bibfnamefont
  {J.}~\bibnamefont {Hone}}, \bibinfo {author} {\bibfnamefont {C.~R.}\
  \bibnamefont {Dean}}, \bibinfo {author} {\bibfnamefont {D.~N.}\ \bibnamefont
  {Basov}}, \bibinfo {author} {\bibfnamefont {A.}~\bibnamefont {Pasupathy}},\
  and\ \bibinfo {author} {\bibfnamefont {A.}~\bibnamefont {Rubio}},\ }\bibfield
   {title} {\bibinfo {title} {Moir\'e heterostructures: a condensed matter
  quantum simulator},\ }\href
  {https://www.nature.com/articles/s41567-020-01154-3} {\bibfield  {journal}
  {\bibinfo  {journal} {Nat. Phys.}\ }\textbf {\bibinfo {volume} {17}},\
  \bibinfo {pages} {155} (\bibinfo {year} {2021})}\BibitemShut {NoStop}%
\bibitem [{\citenamefont {\textit{et al.}}(2018)}]{NAT_I}%
  \BibitemOpen
  \bibfield  {author} {\bibinfo {author} {\bibfnamefont {Y.~C.}\ \bibnamefont
  {\textit{et al.}}},\ }\bibfield  {title} {\bibinfo {title} {Correlated
  insulator behaviour at half-filling in magic-angle graphene superlattices},\
  }\href {https://www.nature.com/articles/nature26154} {\bibfield  {journal}
  {\bibinfo  {journal} {Nature}\ }\textbf {\bibinfo {volume} {556}},\ \bibinfo
  {pages} {80} (\bibinfo {year} {2018})}\BibitemShut {NoStop}%
\bibitem [{\citenamefont {Cao}\ \emph {et~al.}(2018)\citenamefont {Cao},
  \citenamefont {Fatemi}, \citenamefont {Fang}, \citenamefont {Watanabe},
  \citenamefont {Taniguchi}, \citenamefont {Kaxiras},\ and\ \citenamefont
  {Jarillo-Herrero}}]{NAT_S}%
  \BibitemOpen
  \bibfield  {author} {\bibinfo {author} {\bibfnamefont {Y.}~\bibnamefont
  {Cao}}, \bibinfo {author} {\bibfnamefont {V.}~\bibnamefont {Fatemi}},
  \bibinfo {author} {\bibfnamefont {S.}~\bibnamefont {Fang}}, \bibinfo {author}
  {\bibfnamefont {K.}~\bibnamefont {Watanabe}}, \bibinfo {author}
  {\bibfnamefont {T.}~\bibnamefont {Taniguchi}}, \bibinfo {author}
  {\bibfnamefont {E.}~\bibnamefont {Kaxiras}},\ and\ \bibinfo {author}
  {\bibfnamefont {P.}~\bibnamefont {Jarillo-Herrero}},\ }\bibfield  {title}
  {\bibinfo {title} {Unconventional superconductivity in magic-angle graphene
  superlattices},\ }\href {https://www.nature.com/articles/nature26160}
  {\bibfield  {journal} {\bibinfo  {journal} {Nature}\ }\textbf {\bibinfo
  {volume} {556}},\ \bibinfo {pages} {43} (\bibinfo {year} {2018})}\BibitemShut
  {NoStop}%
\bibitem [{\citenamefont {Cao}\ \emph {et~al.}(2016)\citenamefont {Cao},
  \citenamefont {Luo}, \citenamefont {Fatemi}, \citenamefont {Fang},
  \citenamefont {Sanchez-Yamagishi}, \citenamefont {Watanabe}, \citenamefont
  {Taniguchi}, \citenamefont {Kaxiras},\ and\ \citenamefont
  {Jarillo-Herrero}}]{SLG}%
  \BibitemOpen
  \bibfield  {author} {\bibinfo {author} {\bibfnamefont {Y.}~\bibnamefont
  {Cao}}, \bibinfo {author} {\bibfnamefont {J.~Y.}\ \bibnamefont {Luo}},
  \bibinfo {author} {\bibfnamefont {V.}~\bibnamefont {Fatemi}}, \bibinfo
  {author} {\bibfnamefont {S.}~\bibnamefont {Fang}}, \bibinfo {author}
  {\bibfnamefont {J.~D.}\ \bibnamefont {Sanchez-Yamagishi}}, \bibinfo {author}
  {\bibfnamefont {K.}~\bibnamefont {Watanabe}}, \bibinfo {author}
  {\bibfnamefont {T.}~\bibnamefont {Taniguchi}}, \bibinfo {author}
  {\bibfnamefont {E.}~\bibnamefont {Kaxiras}},\ and\ \bibinfo {author}
  {\bibfnamefont {P.}~\bibnamefont {Jarillo-Herrero}},\ }\bibfield  {title}
  {\bibinfo {title} {Superlattice-induced insulating states and
  valley-protected orbits in twisted bilayer graphene},\ }\href
  {https://link.aps.org/doi/10.1103/PhysRevLett.117.116804} {\bibfield
  {journal} {\bibinfo  {journal} {Phys. Rev. Lett.}\ }\textbf {\bibinfo
  {volume} {117}},\ \bibinfo {pages} {116804} (\bibinfo {year}
  {2016})}\BibitemShut {NoStop}%
\bibitem [{\citenamefont {Balents}\ \emph {et~al.}(2020)\citenamefont
  {Balents}, \citenamefont {Dean}, \citenamefont {Efetov},\ and\ \citenamefont
  {Young}}]{Balents2020}%
  \BibitemOpen
  \bibfield  {author} {\bibinfo {author} {\bibfnamefont {L.}~\bibnamefont
  {Balents}}, \bibinfo {author} {\bibfnamefont {C.~R.}\ \bibnamefont {Dean}},
  \bibinfo {author} {\bibfnamefont {D.~K.}\ \bibnamefont {Efetov}},\ and\
  \bibinfo {author} {\bibfnamefont {A.~F.}\ \bibnamefont {Young}},\ }\bibfield
  {title} {\bibinfo {title} {Superconductivity and strong correlations in
  moir\'e flat bands},\ }\href
  {https://www.nature.com/articles/s41567-020-0906-9} {\bibfield  {journal}
  {\bibinfo  {journal} {Nat. Phys.}\ }\textbf {\bibinfo {volume} {16}},\
  \bibinfo {pages} {725} (\bibinfo {year} {2020})}\BibitemShut {NoStop}%
\bibitem [{\citenamefont {Andrei}\ and\ \citenamefont
  {MacDonald}(2020)}]{Andrei2020}%
  \BibitemOpen
  \bibfield  {author} {\bibinfo {author} {\bibfnamefont {E.~Y.}\ \bibnamefont
  {Andrei}}\ and\ \bibinfo {author} {\bibfnamefont {A.~H.}\ \bibnamefont
  {MacDonald}},\ }\bibfield  {title} {\bibinfo {title} {Graphene bilayers with
  a twist},\ }\href {https://www.nature.com/articles/s41563-020-00840-0}
  {\bibfield  {journal} {\bibinfo  {journal} {Nat. Mater.}\ }\textbf {\bibinfo
  {volume} {19}},\ \bibinfo {pages} {1265} (\bibinfo {year}
  {2020})}\BibitemShut {NoStop}%
\bibitem [{\citenamefont {Saito}\ \emph {et~al.}(2020)\citenamefont {Saito},
  \citenamefont {Ge}, \citenamefont {Watanabe}, \citenamefont {Taniguchi},\
  and\ \citenamefont {Young}}]{Saito2020}%
  \BibitemOpen
  \bibfield  {author} {\bibinfo {author} {\bibfnamefont {Y.}~\bibnamefont
  {Saito}}, \bibinfo {author} {\bibfnamefont {J.}~\bibnamefont {Ge}}, \bibinfo
  {author} {\bibfnamefont {K.}~\bibnamefont {Watanabe}}, \bibinfo {author}
  {\bibfnamefont {T.}~\bibnamefont {Taniguchi}},\ and\ \bibinfo {author}
  {\bibfnamefont {A.~F.}\ \bibnamefont {Young}},\ }\bibfield  {title} {\bibinfo
  {title} {Independent superconductors and correlated insulators in twisted
  bilayer graphene},\ }\href
  {https://www.nature.com/articles/s41567-020-0928-3} {\bibfield  {journal}
  {\bibinfo  {journal} {Nat. Phys.}\ }\textbf {\bibinfo {volume} {16}},\
  \bibinfo {pages} {926} (\bibinfo {year} {2020})}\BibitemShut {NoStop}%
\bibitem [{\citenamefont {Stepanov}\ \emph {et~al.}(2020)\citenamefont
  {Stepanov}, \citenamefont {Das}, \citenamefont {Lu}, \citenamefont
  {Fahimniya}, \citenamefont {Watanabe}, \citenamefont {Taniguchi},
  \citenamefont {Koppens}, \citenamefont {Lischner}, \citenamefont {Levitov},\
  and\ \citenamefont {Efetov}}]{Stepanov2020}%
  \BibitemOpen
  \bibfield  {author} {\bibinfo {author} {\bibfnamefont {P.}~\bibnamefont
  {Stepanov}}, \bibinfo {author} {\bibfnamefont {I.}~\bibnamefont {Das}},
  \bibinfo {author} {\bibfnamefont {X.}~\bibnamefont {Lu}}, \bibinfo {author}
  {\bibfnamefont {A.}~\bibnamefont {Fahimniya}}, \bibinfo {author}
  {\bibfnamefont {K.}~\bibnamefont {Watanabe}}, \bibinfo {author}
  {\bibfnamefont {T.}~\bibnamefont {Taniguchi}}, \bibinfo {author}
  {\bibfnamefont {F.~H.~L.}\ \bibnamefont {Koppens}}, \bibinfo {author}
  {\bibfnamefont {J.}~\bibnamefont {Lischner}}, \bibinfo {author}
  {\bibfnamefont {L.}~\bibnamefont {Levitov}},\ and\ \bibinfo {author}
  {\bibfnamefont {D.~K.}\ \bibnamefont {Efetov}},\ }\bibfield  {title}
  {\bibinfo {title} {Untying the insulating and superconducting orders in
  magic-angle graphene},\ }\href
  {https://www.nature.com/articles/s41586-020-2459-6} {\bibfield  {journal}
  {\bibinfo  {journal} {Nature}\ }\textbf {\bibinfo {volume} {583}},\ \bibinfo
  {pages} {375} (\bibinfo {year} {2020})}\BibitemShut {NoStop}%
\bibitem [{\citenamefont {Oh}\ \emph {et~al.}(2021)\citenamefont {Oh},
  \citenamefont {Nuckolls}, \citenamefont {Wong}, \citenamefont {Lee},
  \citenamefont {Liu}, \citenamefont {Watanabe}, \citenamefont {Taniguchi},\
  and\ \citenamefont {Yazdani}}]{Oh2021un}%
  \BibitemOpen
  \bibfield  {author} {\bibinfo {author} {\bibfnamefont {M.}~\bibnamefont
  {Oh}}, \bibinfo {author} {\bibfnamefont {K.~P.}\ \bibnamefont {Nuckolls}},
  \bibinfo {author} {\bibfnamefont {D.}~\bibnamefont {Wong}}, \bibinfo {author}
  {\bibfnamefont {R.~L.}\ \bibnamefont {Lee}}, \bibinfo {author} {\bibfnamefont
  {X.}~\bibnamefont {Liu}}, \bibinfo {author} {\bibfnamefont {K.}~\bibnamefont
  {Watanabe}}, \bibinfo {author} {\bibfnamefont {T.}~\bibnamefont
  {Taniguchi}},\ and\ \bibinfo {author} {\bibfnamefont {A.}~\bibnamefont
  {Yazdani}},\ }\bibfield  {title} {\bibinfo {title} {Evidence for
  unconventional superconductivity in twisted bilayer graphene},\ }\href
  {https://www.nature.com/articles/s41586-021-04121-x} {\bibfield  {journal}
  {\bibinfo  {journal} {Nature}\ }\textbf {\bibinfo {volume} {600}},\ \bibinfo
  {pages} {240} (\bibinfo {year} {2021})}\BibitemShut {NoStop}%
\bibitem [{\citenamefont {Cao}\ \emph {et~al.}(2019)\citenamefont {Cao},
  \citenamefont {Chowdhury}, \citenamefont {Rodan-Legrain}, \citenamefont
  {O.Rubies-Bigord\`a}, \citenamefont {Watanabe}, \citenamefont {Taniguchi},
  \citenamefont {Senthil},\ and\ \citenamefont {Jarillo-Herrero}}]{SMTBLG}%
  \BibitemOpen
  \bibfield  {author} {\bibinfo {author} {\bibfnamefont {Y.}~\bibnamefont
  {Cao}}, \bibinfo {author} {\bibfnamefont {D.}~\bibnamefont {Chowdhury}},
  \bibinfo {author} {\bibfnamefont {D.}~\bibnamefont {Rodan-Legrain}}, \bibinfo
  {author} {\bibnamefont {O.Rubies-Bigord\`a}}, \bibinfo {author}
  {\bibfnamefont {K.}~\bibnamefont {Watanabe}}, \bibinfo {author}
  {\bibfnamefont {T.}~\bibnamefont {Taniguchi}}, \bibinfo {author}
  {\bibfnamefont {T.}~\bibnamefont {Senthil}},\ and\ \bibinfo {author}
  {\bibfnamefont {P.}~\bibnamefont {Jarillo-Herrero}},\ }\bibfield  {title}
  {\bibinfo {title} {Strange metal in magic-angle graphene with near
  \uppercase{P}lanckian dissipation},\ }\href
  {https://link.aps.org/doi/10.1103/PhysRevLett.124.076801} {\bibfield
  {journal} {\bibinfo  {journal} {Phys. Rev. Lett.}\ }\textbf {\bibinfo
  {volume} {124}},\ \bibinfo {pages} {076801} (\bibinfo {year}
  {2019})}\BibitemShut {NoStop}%
\bibitem [{\citenamefont {Polshyn}\ \emph {et~al.}(2019)\citenamefont
  {Polshyn}, \citenamefont {Yankowitz}, \citenamefont {Chen}, \citenamefont
  {Zhang}, \citenamefont {Watanabe}, \citenamefont {Taniguchi}, \citenamefont
  {Dean},\ and\ \citenamefont {Young}}]{Polshyn2019}%
  \BibitemOpen
  \bibfield  {author} {\bibinfo {author} {\bibfnamefont {H.}~\bibnamefont
  {Polshyn}}, \bibinfo {author} {\bibfnamefont {M.}~\bibnamefont {Yankowitz}},
  \bibinfo {author} {\bibfnamefont {S.}~\bibnamefont {Chen}}, \bibinfo {author}
  {\bibfnamefont {Y.}~\bibnamefont {Zhang}}, \bibinfo {author} {\bibfnamefont
  {K.}~\bibnamefont {Watanabe}}, \bibinfo {author} {\bibfnamefont
  {T.}~\bibnamefont {Taniguchi}}, \bibinfo {author} {\bibfnamefont {C.~R.}\
  \bibnamefont {Dean}},\ and\ \bibinfo {author} {\bibfnamefont {A.~F.}\
  \bibnamefont {Young}},\ }\bibfield  {title} {\bibinfo {title} {Large
  linear-in-temperature resistivity in twisted bilayer graphene},\ }\href
  {https://www.nature.com/articles/s41567-019-0596-3} {\bibfield  {journal}
  {\bibinfo  {journal} {Nat. Phys.}\ }\textbf {\bibinfo {volume} {15}},\
  \bibinfo {pages} {1011} (\bibinfo {year} {2019})}\BibitemShut {NoStop}%
\bibitem [{\citenamefont {Cao}\ \emph {et~al.}(2021{\natexlab{a}})\citenamefont
  {Cao}, \citenamefont {Rodan-Legrain}, \citenamefont {Park}, \citenamefont
  {Yuan}, \citenamefont {Watanabe}, \citenamefont {Taniguchi}, \citenamefont
  {Fernandes}, \citenamefont {Fu},\ and\ \citenamefont
  {Jarillo-Herrero}}]{Cao2020}%
  \BibitemOpen
  \bibfield  {author} {\bibinfo {author} {\bibfnamefont {Y.}~\bibnamefont
  {Cao}}, \bibinfo {author} {\bibfnamefont {D.}~\bibnamefont {Rodan-Legrain}},
  \bibinfo {author} {\bibfnamefont {J.~M.}\ \bibnamefont {Park}}, \bibinfo
  {author} {\bibfnamefont {F.~N.}\ \bibnamefont {Yuan}}, \bibinfo {author}
  {\bibfnamefont {K.}~\bibnamefont {Watanabe}}, \bibinfo {author}
  {\bibfnamefont {T.}~\bibnamefont {Taniguchi}}, \bibinfo {author}
  {\bibfnamefont {R.~M.}\ \bibnamefont {Fernandes}}, \bibinfo {author}
  {\bibfnamefont {L.}~\bibnamefont {Fu}},\ and\ \bibinfo {author}
  {\bibfnamefont {P.}~\bibnamefont {Jarillo-Herrero}},\ }\bibfield  {title}
  {\bibinfo {title} {Nematicity and competing orders in superconducting
  magic-angle graphene},\ }\href
  {https://www.science.org/doi/10.1126/science.abc2836} {\bibfield  {journal}
  {\bibinfo  {journal} {Science}\ }\textbf {\bibinfo {volume} {372}},\ \bibinfo
  {pages} {264} (\bibinfo {year} {2021}{\natexlab{a}})}\BibitemShut {NoStop}%
\bibitem [{\citenamefont {\textit{et al.}}(2019{\natexlab{a}})}]{NAT_MEI}%
  \BibitemOpen
  \bibfield  {author} {\bibinfo {author} {\bibfnamefont {A.~K.}\ \bibnamefont
  {\textit{et al.}}},\ }\bibfield  {title} {\bibinfo {title} {Maximized
  electron interactions at the magic angle in twisted bilayer graphene},\
  }\href {https://www.nature.com/articles/s41586-019-1431-9} {\bibfield
  {journal} {\bibinfo  {journal} {Nature}\ }\textbf {\bibinfo {volume} {572}},\
  \bibinfo {pages} {95} (\bibinfo {year} {2019}{\natexlab{a}})}\BibitemShut
  {NoStop}%
\bibitem [{\citenamefont {Jiang}\ \emph {et~al.}(2019)\citenamefont {Jiang},
  \citenamefont {Lai}, \citenamefont {Watanabe}, \citenamefont {Taniguchi},
  \citenamefont {Haule}, \citenamefont {Mao},\ and\ \citenamefont
  {Andrei}}]{NAT_CO}%
  \BibitemOpen
  \bibfield  {author} {\bibinfo {author} {\bibfnamefont {Y.}~\bibnamefont
  {Jiang}}, \bibinfo {author} {\bibfnamefont {X.}~\bibnamefont {Lai}}, \bibinfo
  {author} {\bibfnamefont {K.}~\bibnamefont {Watanabe}}, \bibinfo {author}
  {\bibfnamefont {T.}~\bibnamefont {Taniguchi}}, \bibinfo {author}
  {\bibfnamefont {K.}~\bibnamefont {Haule}}, \bibinfo {author} {\bibfnamefont
  {J.}~\bibnamefont {Mao}},\ and\ \bibinfo {author} {\bibfnamefont {E.~Y.}\
  \bibnamefont {Andrei}},\ }\bibfield  {title} {\bibinfo {title} {Charge order
  and broken rotational symmetry in magic-angle twisted bilayer graphene},\
  }\href {https://www.nature.com/articles/s41586-019-1460-4} {\bibfield
  {journal} {\bibinfo  {journal} {Nature}\ }\textbf {\bibinfo {volume} {573}},\
  \bibinfo {pages} {91} (\bibinfo {year} {2019})}\BibitemShut {NoStop}%
\bibitem [{\citenamefont {\textit{et al.}}(2019{\natexlab{b}})}]{IEC}%
  \BibitemOpen
  \bibfield  {author} {\bibinfo {author} {\bibfnamefont {Y.~C.}\ \bibnamefont
  {\textit{et al.}}},\ }\bibfield  {title} {\bibinfo {title} {Electronic
  correlations in twisted bilayer graphene near the magic angle},\ }\href
  {https://www.nature.com/articles/s41567-019-0606-5} {\bibfield  {journal}
  {\bibinfo  {journal} {Nat. Phys.}\ }\textbf {\bibinfo {volume} {15}},\
  \bibinfo {pages} {1174} (\bibinfo {year} {2019}{\natexlab{b}})}\BibitemShut
  {NoStop}%
\bibitem [{\citenamefont {\textit{et al.}}(2020{\natexlab{a}})}]{Zondiner2020}%
  \BibitemOpen
  \bibfield  {author} {\bibinfo {author} {\bibfnamefont {U.~Z.}\ \bibnamefont
  {\textit{et al.}}},\ }\bibfield  {title} {\bibinfo {title} {Cascade of phase
  transitions and \uppercase{D}irac revivals in magic-angle graphene},\ }\href
  {https://www.nature.com/articles/s41586-020-2373-y} {\bibfield  {journal}
  {\bibinfo  {journal} {Nature}\ }\textbf {\bibinfo {volume} {582}},\ \bibinfo
  {pages} {203} (\bibinfo {year} {2020}{\natexlab{a}})}\BibitemShut {NoStop}%
\bibitem [{\citenamefont {Wong}\ \emph {et~al.}(2020)\citenamefont {Wong},
  \citenamefont {Nuckolls}, \citenamefont {Oh}, \citenamefont {Lian},
  \citenamefont {Xie}, \citenamefont {Jeon}, \citenamefont {Watanabe},
  \citenamefont {Taniguchi}, \citenamefont {Bernevig},\ and\ \citenamefont
  {Yazdani}}]{Wong2020}%
  \BibitemOpen
  \bibfield  {author} {\bibinfo {author} {\bibfnamefont {D.}~\bibnamefont
  {Wong}}, \bibinfo {author} {\bibfnamefont {K.~P.}\ \bibnamefont {Nuckolls}},
  \bibinfo {author} {\bibfnamefont {M.}~\bibnamefont {Oh}}, \bibinfo {author}
  {\bibfnamefont {B.}~\bibnamefont {Lian}}, \bibinfo {author} {\bibfnamefont
  {Y.}~\bibnamefont {Xie}}, \bibinfo {author} {\bibfnamefont {S.}~\bibnamefont
  {Jeon}}, \bibinfo {author} {\bibfnamefont {K.}~\bibnamefont {Watanabe}},
  \bibinfo {author} {\bibfnamefont {T.}~\bibnamefont {Taniguchi}}, \bibinfo
  {author} {\bibfnamefont {B.~A.}\ \bibnamefont {Bernevig}},\ and\ \bibinfo
  {author} {\bibfnamefont {A.}~\bibnamefont {Yazdani}},\ }\bibfield  {title}
  {\bibinfo {title} {Cascade of electronic transitions in magic-angle twisted
  bilayer graphene},\ }\href
  {https://www.nature.com/articles/s41586-020-2339-0} {\bibfield  {journal}
  {\bibinfo  {journal} {Nature}\ }\textbf {\bibinfo {volume} {582}},\ \bibinfo
  {pages} {198} (\bibinfo {year} {2020})}\BibitemShut {NoStop}%
\bibitem [{\citenamefont {\textit{et al.}}(2021{\natexlab{a}})}]{Rozen2021}%
  \BibitemOpen
  \bibfield  {author} {\bibinfo {author} {\bibfnamefont {A.~R.}\ \bibnamefont
  {\textit{et al.}}},\ }\bibfield  {title} {\bibinfo {title} {Entropic evidence
  for a \uppercase{P}omeranchuk effect in magic-angle graphene},\ }\href
  {https://www.nature.com/articles/s41586-021-03319-3} {\bibfield  {journal}
  {\bibinfo  {journal} {Nature}\ }\textbf {\bibinfo {volume} {592}},\ \bibinfo
  {pages} {214} (\bibinfo {year} {2021}{\natexlab{a}})}\BibitemShut {NoStop}%
\bibitem [{\citenamefont {Saito}\ \emph {et~al.}(2021)\citenamefont {Saito},
  \citenamefont {Yang}, \citenamefont {Ge}, \citenamefont {Liu}, \citenamefont
  {Taniguchi}, \citenamefont {Watanabe}, \citenamefont {Li}, \citenamefont
  {Berg},\ and\ \citenamefont {Young}}]{Saito2021}%
  \BibitemOpen
  \bibfield  {author} {\bibinfo {author} {\bibfnamefont {Y.}~\bibnamefont
  {Saito}}, \bibinfo {author} {\bibfnamefont {F.}~\bibnamefont {Yang}},
  \bibinfo {author} {\bibfnamefont {J.}~\bibnamefont {Ge}}, \bibinfo {author}
  {\bibfnamefont {X.}~\bibnamefont {Liu}}, \bibinfo {author} {\bibfnamefont
  {T.}~\bibnamefont {Taniguchi}}, \bibinfo {author} {\bibfnamefont
  {K.}~\bibnamefont {Watanabe}}, \bibinfo {author} {\bibfnamefont {J.~I.~A.}\
  \bibnamefont {Li}}, \bibinfo {author} {\bibfnamefont {E.}~\bibnamefont
  {Berg}},\ and\ \bibinfo {author} {\bibfnamefont {A.~F.}\ \bibnamefont
  {Young}},\ }\bibfield  {title} {\bibinfo {title} {Isospin
  \uppercase{P}omeranchuk effect in twisted bilayer graphene},\ }\href
  {https://www.nature.com/articles/s41586-021-03409-2} {\bibfield  {journal}
  {\bibinfo  {journal} {Nature}\ }\textbf {\bibinfo {volume} {592}},\ \bibinfo
  {pages} {220} (\bibinfo {year} {2021})}\BibitemShut {NoStop}%
\bibitem [{\citenamefont {Wu}\ \emph {et~al.}(2020)\citenamefont {Wu},
  \citenamefont {Zhang}, \citenamefont {Watanabe}, \citenamefont {Taniguchi},\
  and\ \citenamefont {Andrei}}]{Chern_Wu}%
  \BibitemOpen
  \bibfield  {author} {\bibinfo {author} {\bibfnamefont {S.}~\bibnamefont
  {Wu}}, \bibinfo {author} {\bibfnamefont {Z.}~\bibnamefont {Zhang}}, \bibinfo
  {author} {\bibfnamefont {K.}~\bibnamefont {Watanabe}}, \bibinfo {author}
  {\bibfnamefont {T.}~\bibnamefont {Taniguchi}},\ and\ \bibinfo {author}
  {\bibfnamefont {E.~Y.}\ \bibnamefont {Andrei}},\ }\bibfield  {title}
  {\bibinfo {title} {Chern insulators, van \uppercase{H}ove singularities and
  topological flat bands in magic-angle twisted bilayer graphene},\ }\href
  {https://www.nature.com/articles/s41563-020-00911-2} {\bibfield  {journal}
  {\bibinfo  {journal} {Nat. Mater.}\ }\textbf {\bibinfo {volume} {21}},\
  \bibinfo {pages} {488} (\bibinfo {year} {2020})}\BibitemShut {NoStop}%
\bibitem [{\citenamefont {Nuckolls}\ \emph {et~al.}(2020)\citenamefont
  {Nuckolls}, \citenamefont {Oh}, \citenamefont {Wong}, \citenamefont {Lian},
  \citenamefont {Watanabe}, \citenamefont {Taniguchi}, \citenamefont
  {Bernevig},\ and\ \citenamefont {Yazdani}}]{Chern_Nuckolls}%
  \BibitemOpen
  \bibfield  {author} {\bibinfo {author} {\bibfnamefont {K.~P.}\ \bibnamefont
  {Nuckolls}}, \bibinfo {author} {\bibfnamefont {M.}~\bibnamefont {Oh}},
  \bibinfo {author} {\bibfnamefont {D.}~\bibnamefont {Wong}}, \bibinfo {author}
  {\bibfnamefont {B.}~\bibnamefont {Lian}}, \bibinfo {author} {\bibfnamefont
  {K.}~\bibnamefont {Watanabe}}, \bibinfo {author} {\bibfnamefont
  {T.}~\bibnamefont {Taniguchi}}, \bibinfo {author} {\bibfnamefont {B.~A.}\
  \bibnamefont {Bernevig}},\ and\ \bibinfo {author} {\bibfnamefont
  {A.}~\bibnamefont {Yazdani}},\ }\bibfield  {title} {\bibinfo {title}
  {Strongly correlated \uppercase{C}hern insulators in magic-angle twisted
  bilayer graphene},\ }\href
  {https://www.nature.com/articles/s41586-020-3028-8} {\bibfield  {journal}
  {\bibinfo  {journal} {Nature}\ }\textbf {\bibinfo {volume} {588}},\ \bibinfo
  {pages} {610} (\bibinfo {year} {2020})}\BibitemShut {NoStop}%
\bibitem [{\citenamefont {Das}\ \emph {et~al.}(2021)\citenamefont {Das},
  \citenamefont {Lu}, \citenamefont {Herzog-Arbeitman}, \citenamefont {Song},
  \citenamefont {Watanabe}, \citenamefont {Taniguchi}, \citenamefont
  {Bernevig},\ and\ \citenamefont {Efetov}}]{Chern_Das}%
  \BibitemOpen
  \bibfield  {author} {\bibinfo {author} {\bibfnamefont {I.}~\bibnamefont
  {Das}}, \bibinfo {author} {\bibfnamefont {X.}~\bibnamefont {Lu}}, \bibinfo
  {author} {\bibfnamefont {J.}~\bibnamefont {Herzog-Arbeitman}}, \bibinfo
  {author} {\bibfnamefont {Z.-D.}\ \bibnamefont {Song}}, \bibinfo {author}
  {\bibfnamefont {K.}~\bibnamefont {Watanabe}}, \bibinfo {author}
  {\bibfnamefont {T.}~\bibnamefont {Taniguchi}}, \bibinfo {author}
  {\bibfnamefont {B.~A.}\ \bibnamefont {Bernevig}},\ and\ \bibinfo {author}
  {\bibfnamefont {D.~K.}\ \bibnamefont {Efetov}},\ }\bibfield  {title}
  {\bibinfo {title} {Symmetry broken \uppercase{C}hern insulators and ``magic
  series'' of \uppercase{R}ashbalike \uppercase{L}andau level crossings in
  magic angle bilayer graphene},\ }\href
  {https://www.nature.com/articles/s41567-021-01186-3} {\bibfield  {journal}
  {\bibinfo  {journal} {Nat. Phys.}\ }\textbf {\bibinfo {volume} {17}},\
  \bibinfo {pages} {710} (\bibinfo {year} {2021})}\BibitemShut {NoStop}%
\bibitem [{\citenamefont {\textit{et al.}}(2021{\natexlab{b}})}]{Xie2021frac}%
  \BibitemOpen
  \bibfield  {author} {\bibinfo {author} {\bibfnamefont {Y.~X.}\ \bibnamefont
  {\textit{et al.}}},\ }\bibfield  {title} {\bibinfo {title} {Fractional chern
  insulators in magic-angle twisted bilayer graphene},\ }\href
  {https://www.nature.com/articles/s41586-021-04002-3} {\bibfield  {journal}
  {\bibinfo  {journal} {Nature}\ }\textbf {\bibinfo {volume} {600}},\ \bibinfo
  {pages} {439} (\bibinfo {year} {2021}{\natexlab{b}})}\BibitemShut {NoStop}%
\bibitem [{\citenamefont {Choi}\ \emph {et~al.}(2021)\citenamefont {Choi},
  \citenamefont {Kim}, \citenamefont {Lewandowski}, \citenamefont {Peng},
  \citenamefont {Thomson}, \citenamefont {Polski}, \citenamefont {Zhang},
  \citenamefont {Watanabe}, \citenamefont {Taniguchi}, \citenamefont {Alicea},\
  and\ \citenamefont {Nadj-Perge}}]{Youngjoon2021}%
  \BibitemOpen
  \bibfield  {author} {\bibinfo {author} {\bibfnamefont {Y.}~\bibnamefont
  {Choi}}, \bibinfo {author} {\bibfnamefont {H.}~\bibnamefont {Kim}}, \bibinfo
  {author} {\bibfnamefont {C.}~\bibnamefont {Lewandowski}}, \bibinfo {author}
  {\bibfnamefont {Y.}~\bibnamefont {Peng}}, \bibinfo {author} {\bibfnamefont
  {A.}~\bibnamefont {Thomson}}, \bibinfo {author} {\bibfnamefont
  {R.}~\bibnamefont {Polski}}, \bibinfo {author} {\bibfnamefont
  {Y.}~\bibnamefont {Zhang}}, \bibinfo {author} {\bibfnamefont
  {K.}~\bibnamefont {Watanabe}}, \bibinfo {author} {\bibfnamefont
  {T.}~\bibnamefont {Taniguchi}}, \bibinfo {author} {\bibfnamefont
  {J.}~\bibnamefont {Alicea}},\ and\ \bibinfo {author} {\bibfnamefont
  {S.}~\bibnamefont {Nadj-Perge}},\ }\bibfield  {title} {\bibinfo {title}
  {Interaction-driven band flattening and correlated phases in twisted bilayer
  graphene},\ }\href {https://doi.org/10.1038/s41567-021-01359-0} {\bibfield
  {journal} {\bibinfo  {journal} {Nature Physics}\ }\textbf {\bibinfo {volume}
  {17}},\ \bibinfo {pages} {1375} (\bibinfo {year} {2021})}\BibitemShut
  {NoStop}%
\bibitem [{\citenamefont {Yankowitz}\ \emph {et~al.}(2019)\citenamefont
  {Yankowitz}, \citenamefont {Chen}, \citenamefont {Polshyn}, \citenamefont
  {Zhang}, \citenamefont {Watanabe}, \citenamefont {Taniguchi}, \citenamefont
  {Graf}, \citenamefont {Young},\ and\ \citenamefont {Dean}}]{TSTBLG}%
  \BibitemOpen
  \bibfield  {author} {\bibinfo {author} {\bibfnamefont {M.}~\bibnamefont
  {Yankowitz}}, \bibinfo {author} {\bibfnamefont {S.}~\bibnamefont {Chen}},
  \bibinfo {author} {\bibfnamefont {H.}~\bibnamefont {Polshyn}}, \bibinfo
  {author} {\bibfnamefont {Y.}~\bibnamefont {Zhang}}, \bibinfo {author}
  {\bibfnamefont {K.}~\bibnamefont {Watanabe}}, \bibinfo {author}
  {\bibfnamefont {T.}~\bibnamefont {Taniguchi}}, \bibinfo {author}
  {\bibfnamefont {D.}~\bibnamefont {Graf}}, \bibinfo {author} {\bibfnamefont
  {A.~F.}\ \bibnamefont {Young}},\ and\ \bibinfo {author} {\bibfnamefont
  {C.~R.}\ \bibnamefont {Dean}},\ }\bibfield  {title} {\bibinfo {title} {Tuning
  superconductivity in twisted bilayer graphene},\ }\href
  {https://www.science.org/doi/10.1126/science.aav1910} {\bibfield  {journal}
  {\bibinfo  {journal} {Science}\ }\textbf {\bibinfo {volume} {363}},\ \bibinfo
  {pages} {1059} (\bibinfo {year} {2019})}\BibitemShut {NoStop}%
\bibitem [{\citenamefont {Xie}\ \emph {et~al.}(2019)\citenamefont {Xie},
  \citenamefont {Lian}, \citenamefont {J\"{a}ck}, \citenamefont {Liu},
  \citenamefont {Chiu}, \citenamefont {Watanabe}, \citenamefont {Taniguchi},
  \citenamefont {Bernevig},\ and\ \citenamefont {Yazdani}}]{NAT_SS}%
  \BibitemOpen
  \bibfield  {author} {\bibinfo {author} {\bibfnamefont {Y.}~\bibnamefont
  {Xie}}, \bibinfo {author} {\bibfnamefont {B.}~\bibnamefont {Lian}}, \bibinfo
  {author} {\bibfnamefont {B.}~\bibnamefont {J\"{a}ck}}, \bibinfo {author}
  {\bibfnamefont {X.}~\bibnamefont {Liu}}, \bibinfo {author} {\bibfnamefont
  {C.-L.}\ \bibnamefont {Chiu}}, \bibinfo {author} {\bibfnamefont
  {K.}~\bibnamefont {Watanabe}}, \bibinfo {author} {\bibfnamefont
  {T.}~\bibnamefont {Taniguchi}}, \bibinfo {author} {\bibfnamefont {B.~A.}\
  \bibnamefont {Bernevig}},\ and\ \bibinfo {author} {\bibfnamefont
  {A.}~\bibnamefont {Yazdani}},\ }\bibfield  {title} {\bibinfo {title}
  {Spectroscopic signatures of many-body correlations in magic-angle twisted
  bilayer graphene},\ }\href
  {https://www.nature.com/articles/s41586-019-1422-x} {\bibfield  {journal}
  {\bibinfo  {journal} {Nature}\ }\textbf {\bibinfo {volume} {572}},\ \bibinfo
  {pages} {101} (\bibinfo {year} {2019})}\BibitemShut {NoStop}%
\bibitem [{\citenamefont {\textit{et al.}}(2019{\natexlab{c}})}]{SOM}%
  \BibitemOpen
  \bibfield  {author} {\bibinfo {author} {\bibfnamefont {X.~L.}\ \bibnamefont
  {\textit{et al.}}},\ }\bibfield  {title} {\bibinfo {title} {Superconductors,
  orbital magnets, and correlated states in magic angle bilayer graphene},\
  }\href {https://www.nature.com/articles/s41586-019-1695-0} {\bibfield
  {journal} {\bibinfo  {journal} {Nature}\ }\textbf {\bibinfo {volume} {574}},\
  \bibinfo {pages} {653} (\bibinfo {year} {2019}{\natexlab{c}})}\BibitemShut
  {NoStop}%
\bibitem [{\citenamefont {Sharpe}\ \emph {et~al.}(2019)\citenamefont {Sharpe},
  \citenamefont {Fox}, \citenamefont {Barnard}, \citenamefont {Finney},
  \citenamefont {Watanabe}, \citenamefont {Taniguchi}, \citenamefont
  {Kastner},\ and\ \citenamefont {Goldhaber-Gordon}}]{EFM}%
  \BibitemOpen
  \bibfield  {author} {\bibinfo {author} {\bibfnamefont {A.~L.}\ \bibnamefont
  {Sharpe}}, \bibinfo {author} {\bibfnamefont {E.~J.}\ \bibnamefont {Fox}},
  \bibinfo {author} {\bibfnamefont {A.~W.}\ \bibnamefont {Barnard}}, \bibinfo
  {author} {\bibfnamefont {J.}~\bibnamefont {Finney}}, \bibinfo {author}
  {\bibfnamefont {K.}~\bibnamefont {Watanabe}}, \bibinfo {author}
  {\bibfnamefont {T.}~\bibnamefont {Taniguchi}}, \bibinfo {author}
  {\bibfnamefont {M.~A.}\ \bibnamefont {Kastner}},\ and\ \bibinfo {author}
  {\bibfnamefont {D.}~\bibnamefont {Goldhaber-Gordon}},\ }\bibfield  {title}
  {\bibinfo {title} {Emergent ferromagnetism near three-quarters filling in
  twisted bilayer graphene},\ }\href
  {https://www.science.org/doi/10.1126/science.aaw3780} {\bibfield  {journal}
  {\bibinfo  {journal} {Science}\ }\textbf {\bibinfo {volume} {365}},\ \bibinfo
  {pages} {605} (\bibinfo {year} {2019})}\BibitemShut {NoStop}%
\bibitem [{\citenamefont {Serlin}\ \emph {et~al.}(2020)\citenamefont {Serlin},
  \citenamefont {Tschirhart}, \citenamefont {Polshyn}, \citenamefont {Zhang},
  \citenamefont {Zhu}, \citenamefont {Watanabe}, \citenamefont {Taniguchi},
  \citenamefont {Balents},\ and\ \citenamefont {Young}}]{Serlin2020}%
  \BibitemOpen
  \bibfield  {author} {\bibinfo {author} {\bibfnamefont {M.}~\bibnamefont
  {Serlin}}, \bibinfo {author} {\bibfnamefont {C.~L.}\ \bibnamefont
  {Tschirhart}}, \bibinfo {author} {\bibfnamefont {H.}~\bibnamefont {Polshyn}},
  \bibinfo {author} {\bibfnamefont {Y.}~\bibnamefont {Zhang}}, \bibinfo
  {author} {\bibfnamefont {J.}~\bibnamefont {Zhu}}, \bibinfo {author}
  {\bibfnamefont {K.}~\bibnamefont {Watanabe}}, \bibinfo {author}
  {\bibfnamefont {T.}~\bibnamefont {Taniguchi}}, \bibinfo {author}
  {\bibfnamefont {L.}~\bibnamefont {Balents}},\ and\ \bibinfo {author}
  {\bibfnamefont {A.~F.}\ \bibnamefont {Young}},\ }\bibfield  {title} {\bibinfo
  {title} {Intrinsic quantized anomalous \uppercase{H}all effect in a moir\'e
  heterostructure},\ }\href
  {https://www.science.org/doi/abs/10.1126/science.aay5533} {\bibfield
  {journal} {\bibinfo  {journal} {Science}\ }\textbf {\bibinfo {volume}
  {367}},\ \bibinfo {pages} {900} (\bibinfo {year} {2020})}\BibitemShut
  {NoStop}%
\bibitem [{\citenamefont {\textit{et al.}}(2019{\natexlab{d}})}]{Yoo2019}%
  \BibitemOpen
  \bibfield  {author} {\bibinfo {author} {\bibfnamefont {H.~Y.}\ \bibnamefont
  {\textit{et al.}}},\ }\bibfield  {title} {\bibinfo {title} {Atomic and
  electronic reconstruction at van der \uppercase{W}aals interface in twisted
  bilayer graphene},\ }\href
  {https://www.nature.com/articles/s41563-019-0346-z} {\bibfield  {journal}
  {\bibinfo  {journal} {Nat. Mater.}\ }\textbf {\bibinfo {volume} {18}},\
  \bibinfo {pages} {448} (\bibinfo {year} {2019}{\natexlab{d}})}\BibitemShut
  {NoStop}%
\bibitem [{\citenamefont {\textit{et al.}}(2020{\natexlab{b}})}]{Uri2020}%
  \BibitemOpen
  \bibfield  {author} {\bibinfo {author} {\bibfnamefont {A.~U.}\ \bibnamefont
  {\textit{et al.}}},\ }\bibfield  {title} {\bibinfo {title} {Mapping the
  twist-angle disorder and \uppercase{L}andau levels in magic-angle graphene},\
  }\href {https://www.nature.com/articles/s41586-020-2255-3} {\bibfield
  {journal} {\bibinfo  {journal} {Nature}\ }\textbf {\bibinfo {volume} {581}},\
  \bibinfo {pages} {47} (\bibinfo {year} {2020}{\natexlab{b}})}\BibitemShut
  {NoStop}%
\bibitem [{\citenamefont {Uchida}\ \emph {et~al.}(2014)\citenamefont {Uchida},
  \citenamefont {Furuya}, \citenamefont {Iwata},\ and\ \citenamefont
  {Oshiyama}}]{AC}%
  \BibitemOpen
  \bibfield  {author} {\bibinfo {author} {\bibfnamefont {K.}~\bibnamefont
  {Uchida}}, \bibinfo {author} {\bibfnamefont {S.}~\bibnamefont {Furuya}},
  \bibinfo {author} {\bibfnamefont {J.-I.}\ \bibnamefont {Iwata}},\ and\
  \bibinfo {author} {\bibfnamefont {A.}~\bibnamefont {Oshiyama}},\ }\bibfield
  {title} {\bibinfo {title} {Atomic corrugation and electron localization due
  to moir\'e patterns in twisted bilayer graphenes},\ }\href
  {https://doi.org/10.1103/physrevb.90.155451} {\bibfield  {journal} {\bibinfo
  {journal} {Phys. Rev. B}\ }\textbf {\bibinfo {volume} {90}},\ \bibinfo
  {pages} {155451} (\bibinfo {year} {2014})}\BibitemShut {NoStop}%
\bibitem [{\citenamefont {Lucignano}\ \emph {et~al.}(2019)\citenamefont
  {Lucignano}, \citenamefont {Alf\'e}, \citenamefont {Cataudella},
  \citenamefont {Ninno},\ and\ \citenamefont {Cantele}}]{CRAC}%
  \BibitemOpen
  \bibfield  {author} {\bibinfo {author} {\bibfnamefont {P.}~\bibnamefont
  {Lucignano}}, \bibinfo {author} {\bibfnamefont {D.}~\bibnamefont {Alf\'e}},
  \bibinfo {author} {\bibfnamefont {V.}~\bibnamefont {Cataudella}}, \bibinfo
  {author} {\bibfnamefont {D.}~\bibnamefont {Ninno}},\ and\ \bibinfo {author}
  {\bibfnamefont {G.}~\bibnamefont {Cantele}},\ }\bibfield  {title} {\bibinfo
  {title} {Crucial role of atomic corrugation on the flat bands and energy gaps
  of twisted bilayer graphene at the magic angle},\ }\href
  {https://link.aps.org/doi/10.1103/PhysRevB.99.195419} {\bibfield  {journal}
  {\bibinfo  {journal} {Phys. Rev. B}\ }\textbf {\bibinfo {volume} {99}},\
  \bibinfo {pages} {195419} (\bibinfo {year} {2019})}\BibitemShut {NoStop}%
\bibitem [{\citenamefont {de~Laissardi\`ere}\ \emph {et~al.}(2010)\citenamefont
  {de~Laissardi\`ere}, \citenamefont {Mayou},\ and\ \citenamefont
  {Magaud}}]{LDE}%
  \BibitemOpen
  \bibfield  {author} {\bibinfo {author} {\bibfnamefont {G.~T.}\ \bibnamefont
  {de~Laissardi\`ere}}, \bibinfo {author} {\bibfnamefont {D.}~\bibnamefont
  {Mayou}},\ and\ \bibinfo {author} {\bibfnamefont {L.}~\bibnamefont
  {Magaud}},\ }\bibfield  {title} {\bibinfo {title} {Localization of
  \uppercase{D}irac electrons in rotated graphene bilayers},\ }\href
  {https://pubs.acs.org/doi/10.1021/nl902948m} {\bibfield  {journal} {\bibinfo
  {journal} {Nano Lett.}\ }\textbf {\bibinfo {volume} {10}},\ \bibinfo {pages}
  {804} (\bibinfo {year} {2010})}\BibitemShut {NoStop}%
\bibitem [{\citenamefont {de~Laissardi\`ere}\ \emph {et~al.}(2012)\citenamefont
  {de~Laissardi\`ere}, \citenamefont {Mayou},\ and\ \citenamefont
  {Magaud}}]{NSCS}%
  \BibitemOpen
  \bibfield  {author} {\bibinfo {author} {\bibfnamefont {G.~T.}\ \bibnamefont
  {de~Laissardi\`ere}}, \bibinfo {author} {\bibfnamefont {D.}~\bibnamefont
  {Mayou}},\ and\ \bibinfo {author} {\bibfnamefont {L.}~\bibnamefont
  {Magaud}},\ }\bibfield  {title} {\bibinfo {title} {Numerical studies of
  confined states in rotated bilayers of graphene},\ }\href
  {https://link.aps.org/doi/10.1103/PhysRevB.86.125413} {\bibfield  {journal}
  {\bibinfo  {journal} {Phys. Rev. B}\ }\textbf {\bibinfo {volume} {86}},\
  \bibinfo {pages} {125413} (\bibinfo {year} {2012})}\BibitemShut {NoStop}%
\bibitem [{\citenamefont {Lopes~dos Santos}\ \emph {et~al.}(2007)\citenamefont
  {Lopes~dos Santos}, \citenamefont {Peres},\ and\ \citenamefont
  {Castro~Neto}}]{LopesDosSantos2007}%
  \BibitemOpen
  \bibfield  {author} {\bibinfo {author} {\bibfnamefont {J.~M.~B.}\
  \bibnamefont {Lopes~dos Santos}}, \bibinfo {author} {\bibfnamefont
  {N.~M.~R.}\ \bibnamefont {Peres}},\ and\ \bibinfo {author} {\bibfnamefont
  {A.~H.}\ \bibnamefont {Castro~Neto}},\ }\bibfield  {title} {\bibinfo {title}
  {Graphene bilayer with a twist: Electronic structure},\ }\href
  {https://doi.org/10.1103/PhysRevLett.99.256802} {\bibfield  {journal}
  {\bibinfo  {journal} {Phys. Rev. Lett.}\ }\textbf {\bibinfo {volume} {99}},\
  \bibinfo {pages} {256802} (\bibinfo {year} {2007})}\BibitemShut {NoStop}%
\bibitem [{\citenamefont {Lopes~dos Santos}\ \emph {et~al.}(2012)\citenamefont
  {Lopes~dos Santos}, \citenamefont {Peres},\ and\ \citenamefont
  {Castro~Neto}}]{LopesDosSantos2012}%
  \BibitemOpen
  \bibfield  {author} {\bibinfo {author} {\bibfnamefont {J.~M.~B.}\
  \bibnamefont {Lopes~dos Santos}}, \bibinfo {author} {\bibfnamefont
  {N.~M.~R.}\ \bibnamefont {Peres}},\ and\ \bibinfo {author} {\bibfnamefont
  {A.~H.}\ \bibnamefont {Castro~Neto}},\ }\bibfield  {title} {\bibinfo {title}
  {Continuum model of the twisted graphene bilayer},\ }\href
  {https://doi.org/10.1103/PhysRevB.86.155449} {\bibfield  {journal} {\bibinfo
  {journal} {Phys. Rev. B}\ }\textbf {\bibinfo {volume} {86}},\ \bibinfo
  {pages} {155449} (\bibinfo {year} {2012})}\BibitemShut {NoStop}%
\bibitem [{\citenamefont {Bistritzer}\ and\ \citenamefont
  {MacDonald}(2011)}]{Bistritzer2011}%
  \BibitemOpen
  \bibfield  {author} {\bibinfo {author} {\bibfnamefont {R.}~\bibnamefont
  {Bistritzer}}\ and\ \bibinfo {author} {\bibfnamefont {A.~H.}\ \bibnamefont
  {MacDonald}},\ }\bibfield  {title} {\bibinfo {title} {Moir{\'{e}} bands in
  twisted double-layer graphene},\ }\href
  {https://doi.org/10.1073/pnas.1108174108} {\bibfield  {journal} {\bibinfo
  {journal} {Proceedings of the National Academy of Sciences}\ }\textbf
  {\bibinfo {volume} {108}},\ \bibinfo {pages} {12233} (\bibinfo {year}
  {2011})}\BibitemShut {NoStop}%
\bibitem [{\citenamefont {Guinea}\ and\ \citenamefont {Walet}(2018)}]{EE}%
  \BibitemOpen
  \bibfield  {author} {\bibinfo {author} {\bibfnamefont {F.}~\bibnamefont
  {Guinea}}\ and\ \bibinfo {author} {\bibfnamefont {N.~R.}\ \bibnamefont
  {Walet}},\ }\bibfield  {title} {\bibinfo {title} {Electrostatic effects, band
  distortions, and superconductivity in twisted graphene bilayers},\ }\href
  {https://www.pnas.org/doi/10.1073/pnas.1810947115} {\bibfield  {journal}
  {\bibinfo  {journal} {PNAS}\ }\textbf {\bibinfo {volume} {115}},\ \bibinfo
  {pages} {13174} (\bibinfo {year} {2018})}\BibitemShut {NoStop}%
\bibitem [{\citenamefont {Xie}\ and\ \citenamefont
  {MacDonald}(2020)}]{Xie2020CorrelatedIns}%
  \BibitemOpen
  \bibfield  {author} {\bibinfo {author} {\bibfnamefont {M.}~\bibnamefont
  {Xie}}\ and\ \bibinfo {author} {\bibfnamefont {A.~H.}\ \bibnamefont
  {MacDonald}},\ }\bibfield  {title} {\bibinfo {title} {Nature of the
  correlated insulator states in twisted bilayer graphene},\ }\href
  {https://doi.org/10.1103/PhysRevLett.124.097601} {\bibfield  {journal}
  {\bibinfo  {journal} {Phys. Rev. Lett.}\ }\textbf {\bibinfo {volume} {124}},\
  \bibinfo {pages} {097601} (\bibinfo {year} {2020})}\BibitemShut {NoStop}%
\bibitem [{\citenamefont {Bultinck}\ \emph {et~al.}(2020)\citenamefont
  {Bultinck}, \citenamefont {Khalaf}, \citenamefont {Liu}, \citenamefont
  {Chatterjee}, \citenamefont {Vishwanath},\ and\ \citenamefont
  {Zaletel}}]{Bultinck2020}%
  \BibitemOpen
  \bibfield  {author} {\bibinfo {author} {\bibfnamefont {N.}~\bibnamefont
  {Bultinck}}, \bibinfo {author} {\bibfnamefont {E.}~\bibnamefont {Khalaf}},
  \bibinfo {author} {\bibfnamefont {S.}~\bibnamefont {Liu}}, \bibinfo {author}
  {\bibfnamefont {S.}~\bibnamefont {Chatterjee}}, \bibinfo {author}
  {\bibfnamefont {A.}~\bibnamefont {Vishwanath}},\ and\ \bibinfo {author}
  {\bibfnamefont {M.~P.}\ \bibnamefont {Zaletel}},\ }\bibfield  {title}
  {\bibinfo {title} {Ground state and hidden symmetry of magic-angle graphene
  at even integer filling},\ }\href
  {https://doi.org/10.1103/PhysRevX.10.031034} {\bibfield  {journal} {\bibinfo
  {journal} {Phys. Rev. X}\ }\textbf {\bibinfo {volume} {10}},\ \bibinfo
  {pages} {031034} (\bibinfo {year} {2020})}\BibitemShut {NoStop}%
\bibitem [{\citenamefont {Liu}\ and\ \citenamefont
  {Dai}(2021)}]{Liu2021Insulating}%
  \BibitemOpen
  \bibfield  {author} {\bibinfo {author} {\bibfnamefont {J.}~\bibnamefont
  {Liu}}\ and\ \bibinfo {author} {\bibfnamefont {X.}~\bibnamefont {Dai}},\
  }\bibfield  {title} {\bibinfo {title} {Theories for the correlated insulating
  states and quantum anomalous hall effect phenomena in twisted bilayer
  graphene},\ }\href {https://doi.org/10.1103/PhysRevB.103.035427} {\bibfield
  {journal} {\bibinfo  {journal} {Phys. Rev. B}\ }\textbf {\bibinfo {volume}
  {103}},\ \bibinfo {pages} {035427} (\bibinfo {year} {2021})}\BibitemShut
  {NoStop}%
\bibitem [{\citenamefont {Cea}\ \emph {et~al.}(2019{\natexlab{a}})\citenamefont
  {Cea}, \citenamefont {Walet},\ and\ \citenamefont {Guinea}}]{Cea2019Pinning}%
  \BibitemOpen
  \bibfield  {author} {\bibinfo {author} {\bibfnamefont {T.}~\bibnamefont
  {Cea}}, \bibinfo {author} {\bibfnamefont {N.~R.}\ \bibnamefont {Walet}},\
  and\ \bibinfo {author} {\bibfnamefont {F.}~\bibnamefont {Guinea}},\
  }\bibfield  {title} {\bibinfo {title} {Electronic band structure and pinning
  of fermi energy to van hove singularities in twisted bilayer graphene: A
  self-consistent approach},\ }\href
  {https://doi.org/10.1103/PhysRevB.100.205113} {\bibfield  {journal} {\bibinfo
   {journal} {Phys. Rev. B}\ }\textbf {\bibinfo {volume} {100}},\ \bibinfo
  {pages} {205113} (\bibinfo {year} {2019}{\natexlab{a}})}\BibitemShut
  {NoStop}%
\bibitem [{\citenamefont {Zhang}\ \emph {et~al.}(2020)\citenamefont {Zhang},
  \citenamefont {Jiang}, \citenamefont {Wang},\ and\ \citenamefont
  {Zhang}}]{Zhang2020InsulatingHF}%
  \BibitemOpen
  \bibfield  {author} {\bibinfo {author} {\bibfnamefont {Y.}~\bibnamefont
  {Zhang}}, \bibinfo {author} {\bibfnamefont {K.}~\bibnamefont {Jiang}},
  \bibinfo {author} {\bibfnamefont {Z.}~\bibnamefont {Wang}},\ and\ \bibinfo
  {author} {\bibfnamefont {F.}~\bibnamefont {Zhang}},\ }\bibfield  {title}
  {\bibinfo {title} {Correlated insulating phases of twisted bilayer graphene
  at commensurate filling fractions: A hartree-fock study},\ }\href
  {https://doi.org/10.1103/PhysRevB.102.035136} {\bibfield  {journal} {\bibinfo
   {journal} {Phys. Rev. B}\ }\textbf {\bibinfo {volume} {102}},\ \bibinfo
  {pages} {035136} (\bibinfo {year} {2020})}\BibitemShut {NoStop}%
\bibitem [{\citenamefont {Cea}\ \emph {et~al.}(2022)\citenamefont {Cea},
  \citenamefont {Pantaleón}, \citenamefont {Walet},\ and\ \citenamefont
  {Guinea}}]{CeaElectrostatic2022}%
  \BibitemOpen
  \bibfield  {author} {\bibinfo {author} {\bibfnamefont {T.}~\bibnamefont
  {Cea}}, \bibinfo {author} {\bibfnamefont {P.~A.}\ \bibnamefont {Pantaleón}},
  \bibinfo {author} {\bibfnamefont {N.~R.}\ \bibnamefont {Walet}},\ and\
  \bibinfo {author} {\bibfnamefont {F.}~\bibnamefont {Guinea}},\ }\bibfield
  {title} {\bibinfo {title} {Electrostatic interactions in twisted bilayer
  graphene},\ }\href
  {https://doi.org/https://doi.org/10.1016/j.nanoms.2021.10.001} {\bibfield
  {journal} {\bibinfo  {journal} {Nano Materials Science}\ }\textbf {\bibinfo
  {volume} {4}},\ \bibinfo {pages} {27} (\bibinfo {year} {2022})},\ \bibinfo
  {note} {special issue on Graphene and 2D Alternative Materials}\BibitemShut
  {NoStop}%
\bibitem [{\citenamefont {Cea}\ and\ \citenamefont
  {Guinea}(2021)}]{CeaCoulomb2021}%
  \BibitemOpen
  \bibfield  {author} {\bibinfo {author} {\bibfnamefont {T.}~\bibnamefont
  {Cea}}\ and\ \bibinfo {author} {\bibfnamefont {F.}~\bibnamefont {Guinea}},\
  }\bibfield  {title} {\bibinfo {title} {Coulomb interaction, phonons, and
  superconductivity in twisted bilayer graphene},\ }\href
  {https://doi.org/10.1073/pnas.2107874118} {\bibfield  {journal} {\bibinfo
  {journal} {Proceedings of the National Academy of Sciences}\ }\textbf
  {\bibinfo {volume} {118}},\ \bibinfo {pages} {e2107874118} (\bibinfo {year}
  {2021})}\BibitemShut {NoStop}%
\bibitem [{\citenamefont {Cea}\ and\ \citenamefont {Guinea}(2020)}]{Cea2020}%
  \BibitemOpen
  \bibfield  {author} {\bibinfo {author} {\bibfnamefont {T.}~\bibnamefont
  {Cea}}\ and\ \bibinfo {author} {\bibfnamefont {F.}~\bibnamefont {Guinea}},\
  }\bibfield  {title} {\bibinfo {title} {Band structure and insulating states
  driven by coulomb interaction in twisted bilayer graphene},\ }\href
  {https://journals.aps.org/prb/abstract/10.1103/PhysRevB.102.045107}
  {\bibfield  {journal} {\bibinfo  {journal} {Phys. Rev. B}\ }\textbf {\bibinfo
  {volume} {102}},\ \bibinfo {pages} {045107} (\bibinfo {year}
  {2020})}\BibitemShut {NoStop}%
\bibitem [{\citenamefont {Zou}\ \emph {et~al.}(2018)\citenamefont {Zou},
  \citenamefont {Po}, \citenamefont {Vishwanath},\ and\ \citenamefont
  {Senthil}}]{Zou2018}%
  \BibitemOpen
  \bibfield  {author} {\bibinfo {author} {\bibfnamefont {L.}~\bibnamefont
  {Zou}}, \bibinfo {author} {\bibfnamefont {H.~C.}\ \bibnamefont {Po}},
  \bibinfo {author} {\bibfnamefont {A.}~\bibnamefont {Vishwanath}},\ and\
  \bibinfo {author} {\bibfnamefont {T.}~\bibnamefont {Senthil}},\ }\bibfield
  {title} {\bibinfo {title} {Band structure of twisted bilayer graphene:
  Emergent symmetries, commensurate approximants, and wannier obstructions},\
  }\href {https://doi.org/10.1103/PhysRevB.98.085435} {\bibfield  {journal}
  {\bibinfo  {journal} {Phys. Rev. B}\ }\textbf {\bibinfo {volume} {98}},\
  \bibinfo {pages} {085435} (\bibinfo {year} {2018})}\BibitemShut {NoStop}%
\bibitem [{\citenamefont {Po}\ \emph {et~al.}(2018)\citenamefont {Po},
  \citenamefont {Zou}, \citenamefont {Vishwanath},\ and\ \citenamefont
  {Senthil}}]{Po2018}%
  \BibitemOpen
  \bibfield  {author} {\bibinfo {author} {\bibfnamefont {H.~C.}\ \bibnamefont
  {Po}}, \bibinfo {author} {\bibfnamefont {L.}~\bibnamefont {Zou}}, \bibinfo
  {author} {\bibfnamefont {A.}~\bibnamefont {Vishwanath}},\ and\ \bibinfo
  {author} {\bibfnamefont {T.}~\bibnamefont {Senthil}},\ }\bibfield  {title}
  {\bibinfo {title} {Origin of mott insulating behavior and superconductivity
  in twisted bilayer graphene},\ }\href
  {https://doi.org/10.1103/PhysRevX.8.031089} {\bibfield  {journal} {\bibinfo
  {journal} {Phys. Rev. X}\ }\textbf {\bibinfo {volume} {8}},\ \bibinfo {pages}
  {031089} (\bibinfo {year} {2018})}\BibitemShut {NoStop}%
\bibitem [{\citenamefont {Kang}\ and\ \citenamefont
  {Vafek}(2019)}]{Kang2019StrongCoupling}%
  \BibitemOpen
  \bibfield  {author} {\bibinfo {author} {\bibfnamefont {J.}~\bibnamefont
  {Kang}}\ and\ \bibinfo {author} {\bibfnamefont {O.}~\bibnamefont {Vafek}},\
  }\bibfield  {title} {\bibinfo {title} {Strong coupling phases of partially
  filled twisted bilayer graphene narrow bands},\ }\href
  {https://doi.org/10.1103/PhysRevLett.122.246401} {\bibfield  {journal}
  {\bibinfo  {journal} {Phys. Rev. Lett.}\ }\textbf {\bibinfo {volume} {122}},\
  \bibinfo {pages} {246401} (\bibinfo {year} {2019})}\BibitemShut {NoStop}%
\bibitem [{\citenamefont {Seo}\ \emph {et~al.}(2019)\citenamefont {Seo},
  \citenamefont {Kotov},\ and\ \citenamefont {Uchoa}}]{Seo2019FerroMott}%
  \BibitemOpen
  \bibfield  {author} {\bibinfo {author} {\bibfnamefont {K.}~\bibnamefont
  {Seo}}, \bibinfo {author} {\bibfnamefont {V.~N.}\ \bibnamefont {Kotov}},\
  and\ \bibinfo {author} {\bibfnamefont {B.}~\bibnamefont {Uchoa}},\ }\bibfield
   {title} {\bibinfo {title} {Ferromagnetic mott state in twisted graphene
  bilayers at the magic angle},\ }\href
  {https://doi.org/10.1103/PhysRevLett.122.246402} {\bibfield  {journal}
  {\bibinfo  {journal} {Phys. Rev. Lett.}\ }\textbf {\bibinfo {volume} {122}},\
  \bibinfo {pages} {246402} (\bibinfo {year} {2019})}\BibitemShut {NoStop}%
\bibitem [{\citenamefont {Calder\'on}\ and\ \citenamefont
  {Bascones}(2020)}]{Bascones2020}%
  \BibitemOpen
  \bibfield  {author} {\bibinfo {author} {\bibfnamefont {M.}~\bibnamefont
  {Calder\'on}}\ and\ \bibinfo {author} {\bibfnamefont {E.}~\bibnamefont
  {Bascones}},\ }\bibfield  {title} {\bibinfo {title} {Interactions in the
  8-orbital model for twisted bilayer graphene},\ }\href
  {https://link.aps.org/doi/10.1103/PhysRevB.102.155149} {\bibfield  {journal}
  {\bibinfo  {journal} {Phys. Rev. B}\ }\textbf {\bibinfo {volume} {102}},\
  \bibinfo {pages} {155149} (\bibinfo {year} {2020})}\BibitemShut {NoStop}%
\bibitem [{\citenamefont {Carr}\ \emph {et~al.}(2019)\citenamefont {Carr},
  \citenamefont {Fang}, \citenamefont {Po}, \citenamefont {Vishwanath},\ and\
  \citenamefont {Kaxiras}}]{Carr2019wannier}%
  \BibitemOpen
  \bibfield  {author} {\bibinfo {author} {\bibfnamefont {S.}~\bibnamefont
  {Carr}}, \bibinfo {author} {\bibfnamefont {S.}~\bibnamefont {Fang}}, \bibinfo
  {author} {\bibfnamefont {H.~C.}\ \bibnamefont {Po}}, \bibinfo {author}
  {\bibfnamefont {A.}~\bibnamefont {Vishwanath}},\ and\ \bibinfo {author}
  {\bibfnamefont {E.}~\bibnamefont {Kaxiras}},\ }\bibfield  {title} {\bibinfo
  {title} {Derivation of \uppercase{W}annier orbitals and minimal-basis
  tight-binding \uppercase{h}amiltonians for twisted bilayer graphene: a
  first-principles approach},\ }\href
  {https://link.aps.org/doi/10.1103/PhysRevResearch.1.033072} {\bibfield
  {journal} {\bibinfo  {journal} {Phys. Rev. Research}\ }\textbf {\bibinfo
  {volume} {1}},\ \bibinfo {pages} {033072} (\bibinfo {year}
  {2019})}\BibitemShut {NoStop}%
\bibitem [{\citenamefont {Goodwin}\ \emph
  {et~al.}(2019{\natexlab{a}})\citenamefont {Goodwin}, \citenamefont
  {Corsetti}, \citenamefont {Mostofi},\ and\ \citenamefont {Lischner}}]{PHD_1}%
  \BibitemOpen
  \bibfield  {author} {\bibinfo {author} {\bibfnamefont {Z.~A.~H.}\
  \bibnamefont {Goodwin}}, \bibinfo {author} {\bibfnamefont {F.}~\bibnamefont
  {Corsetti}}, \bibinfo {author} {\bibfnamefont {A.~A.}\ \bibnamefont
  {Mostofi}},\ and\ \bibinfo {author} {\bibfnamefont {J.}~\bibnamefont
  {Lischner}},\ }\bibfield  {title} {\bibinfo {title} {Twist-angle sensitivity
  of electron correlations in moir\'e graphene bilayers},\ }\href
  {https://link.aps.org/doi/10.1103/PhysRevB.100.121106} {\bibfield  {journal}
  {\bibinfo  {journal} {Phys. Rev. B}\ }\textbf {\bibinfo {volume} {100}},\
  \bibinfo {pages} {121106(R)} (\bibinfo {year}
  {2019}{\natexlab{a}})}\BibitemShut {NoStop}%
\bibitem [{\citenamefont {Goodwin}\ \emph
  {et~al.}(2019{\natexlab{b}})\citenamefont {Goodwin}, \citenamefont
  {Corsetti}, \citenamefont {Mostofi},\ and\ \citenamefont {Lischner}}]{PHD_2}%
  \BibitemOpen
  \bibfield  {author} {\bibinfo {author} {\bibfnamefont {Z.~A.~H.}\
  \bibnamefont {Goodwin}}, \bibinfo {author} {\bibfnamefont {F.}~\bibnamefont
  {Corsetti}}, \bibinfo {author} {\bibfnamefont {A.~A.}\ \bibnamefont
  {Mostofi}},\ and\ \bibinfo {author} {\bibfnamefont {J.}~\bibnamefont
  {Lischner}},\ }\bibfield  {title} {\bibinfo {title} {Attractive
  electron-electron interactions from internal screening in magic angle twisted
  bilayer graphene},\ }\href
  {https://link.aps.org/doi/10.1103/PhysRevB.100.235424} {\bibfield  {journal}
  {\bibinfo  {journal} {Phys. Rev. B}\ }\textbf {\bibinfo {volume} {100}},\
  \bibinfo {pages} {235424} (\bibinfo {year} {2019}{\natexlab{b}})}\BibitemShut
  {NoStop}%
\bibitem [{\citenamefont {Goodwin}\ \emph
  {et~al.}(2020{\natexlab{a}})\citenamefont {Goodwin}, \citenamefont {Vitale},
  \citenamefont {Corsetti}, \citenamefont {Efetov}, \citenamefont {Mostofi},\
  and\ \citenamefont {Lischner}}]{PHD_3}%
  \BibitemOpen
  \bibfield  {author} {\bibinfo {author} {\bibfnamefont {Z.~A.~H.}\
  \bibnamefont {Goodwin}}, \bibinfo {author} {\bibfnamefont {V.}~\bibnamefont
  {Vitale}}, \bibinfo {author} {\bibfnamefont {F.}~\bibnamefont {Corsetti}},
  \bibinfo {author} {\bibfnamefont {D.}~\bibnamefont {Efetov}}, \bibinfo
  {author} {\bibfnamefont {A.~A.}\ \bibnamefont {Mostofi}},\ and\ \bibinfo
  {author} {\bibfnamefont {J.}~\bibnamefont {Lischner}},\ }\bibfield  {title}
  {\bibinfo {title} {Critical role of device geometry for the phase diagram of
  twisted bilayer graphene},\ }\href
  {https://link.aps.org/doi/10.1103/PhysRevB.101.165110} {\bibfield  {journal}
  {\bibinfo  {journal} {Phys. Rev. B}\ }\textbf {\bibinfo {volume} {101}},\
  \bibinfo {pages} {165110} (\bibinfo {year} {2020}{\natexlab{a}})}\BibitemShut
  {NoStop}%
\bibitem [{\citenamefont {Kennes}\ \emph {et~al.}(2018)\citenamefont {Kennes},
  \citenamefont {Lischner},\ and\ \citenamefont {Karrasch}}]{Kennes2018}%
  \BibitemOpen
  \bibfield  {author} {\bibinfo {author} {\bibfnamefont {D.~M.}\ \bibnamefont
  {Kennes}}, \bibinfo {author} {\bibfnamefont {J.}~\bibnamefont {Lischner}},\
  and\ \bibinfo {author} {\bibfnamefont {C.}~\bibnamefont {Karrasch}},\
  }\bibfield  {title} {\bibinfo {title} {Strong correlations and
  $d+\mathit{id}$ superconductivity in twisted bilayer graphene},\ }\href
  {https://doi.org/10.1103/PhysRevB.98.241407} {\bibfield  {journal} {\bibinfo
  {journal} {Phys. Rev. B}\ }\textbf {\bibinfo {volume} {98}},\ \bibinfo
  {pages} {241407} (\bibinfo {year} {2018})}\BibitemShut {NoStop}%
\bibitem [{\citenamefont {Roy}\ and\ \citenamefont {Juri\ifmmode \check{c}\else
  \v{c}\fi{}i\ifmmode~\acute{c}\else
  \'{c}\fi{}}(2019)}]{Roy2019Unconventional}%
  \BibitemOpen
  \bibfield  {author} {\bibinfo {author} {\bibfnamefont {B.}~\bibnamefont
  {Roy}}\ and\ \bibinfo {author} {\bibfnamefont {V.}~\bibnamefont {Juri\ifmmode
  \check{c}\else \v{c}\fi{}i\ifmmode~\acute{c}\else \'{c}\fi{}}},\ }\bibfield
  {title} {\bibinfo {title} {Unconventional superconductivity in nearly flat
  bands in twisted bilayer graphene},\ }\href
  {https://doi.org/10.1103/PhysRevB.99.121407} {\bibfield  {journal} {\bibinfo
  {journal} {Phys. Rev. B}\ }\textbf {\bibinfo {volume} {99}},\ \bibinfo
  {pages} {121407} (\bibinfo {year} {2019})}\BibitemShut {NoStop}%
\bibitem [{\citenamefont {Hsu}\ \emph {et~al.}(2020)\citenamefont {Hsu},
  \citenamefont {Wu},\ and\ \citenamefont {Das~Sarma}}]{Hsu2020RGanalysis}%
  \BibitemOpen
  \bibfield  {author} {\bibinfo {author} {\bibfnamefont {Y.-T.}\ \bibnamefont
  {Hsu}}, \bibinfo {author} {\bibfnamefont {F.}~\bibnamefont {Wu}},\ and\
  \bibinfo {author} {\bibfnamefont {S.}~\bibnamefont {Das~Sarma}},\ }\bibfield
  {title} {\bibinfo {title} {Topological superconductivity, ferromagnetism, and
  valley-polarized phases in moir\'e systems: Renormalization group analysis
  for twisted double bilayer graphene},\ }\href
  {https://doi.org/10.1103/PhysRevB.102.085103} {\bibfield  {journal} {\bibinfo
   {journal} {Phys. Rev. B}\ }\textbf {\bibinfo {volume} {102}},\ \bibinfo
  {pages} {085103} (\bibinfo {year} {2020})}\BibitemShut {NoStop}%
\bibitem [{\citenamefont {Laksono}\ \emph {et~al.}(2018)\citenamefont
  {Laksono}, \citenamefont {Leaw}, \citenamefont {Reaves}, \citenamefont
  {Singh}, \citenamefont {Wang}, \citenamefont {Adam},\ and\ \citenamefont
  {Gu}}]{Laksono2018}%
  \BibitemOpen
  \bibfield  {author} {\bibinfo {author} {\bibfnamefont {E.}~\bibnamefont
  {Laksono}}, \bibinfo {author} {\bibfnamefont {J.~N.}\ \bibnamefont {Leaw}},
  \bibinfo {author} {\bibfnamefont {A.}~\bibnamefont {Reaves}}, \bibinfo
  {author} {\bibfnamefont {M.}~\bibnamefont {Singh}}, \bibinfo {author}
  {\bibfnamefont {X.}~\bibnamefont {Wang}}, \bibinfo {author} {\bibfnamefont
  {S.}~\bibnamefont {Adam}},\ and\ \bibinfo {author} {\bibfnamefont
  {X.}~\bibnamefont {Gu}},\ }\bibfield  {title} {\bibinfo {title} {Singlet
  superconductivity enhanced by charge order in nested twisted bilayer graphene
  fermi surfaces},\ }\href
  {https://doi.org/https://doi.org/10.1016/j.ssc.2018.07.013} {\bibfield
  {journal} {\bibinfo  {journal} {Solid State Communications}\ }\textbf
  {\bibinfo {volume} {282}},\ \bibinfo {pages} {38} (\bibinfo {year}
  {2018})}\BibitemShut {NoStop}%
\bibitem [{\citenamefont {Klebl}\ and\ \citenamefont
  {Honerkamp}(2019)}]{LK_CH}%
  \BibitemOpen
  \bibfield  {author} {\bibinfo {author} {\bibfnamefont {L.}~\bibnamefont
  {Klebl}}\ and\ \bibinfo {author} {\bibfnamefont {C.}~\bibnamefont
  {Honerkamp}},\ }\bibfield  {title} {\bibinfo {title} {Inherited and
  flatband-induced ordering in twisted graphene bilayers},\ }\href
  {https://link.aps.org/doi/10.1103/PhysRevB.100.155145} {\bibfield  {journal}
  {\bibinfo  {journal} {Phys. Rev. B}\ }\textbf {\bibinfo {volume} {100}},\
  \bibinfo {pages} {155145} (\bibinfo {year} {2019})}\BibitemShut {NoStop}%
\bibitem [{\citenamefont {Klebl}\ \emph {et~al.}(2021)\citenamefont {Klebl},
  \citenamefont {Goodwin}, \citenamefont {Mostofi}, \citenamefont {Kennes},\
  and\ \citenamefont {Lischner}}]{PHD_6}%
  \BibitemOpen
  \bibfield  {author} {\bibinfo {author} {\bibfnamefont {L.}~\bibnamefont
  {Klebl}}, \bibinfo {author} {\bibfnamefont {Z.~A.~H.}\ \bibnamefont
  {Goodwin}}, \bibinfo {author} {\bibfnamefont {A.~A.}\ \bibnamefont
  {Mostofi}}, \bibinfo {author} {\bibfnamefont {D.~M.}\ \bibnamefont
  {Kennes}},\ and\ \bibinfo {author} {\bibfnamefont {J.}~\bibnamefont
  {Lischner}},\ }\bibfield  {title} {\bibinfo {title} {Importance of
  long-ranged electron-electron interactions for the magnetic phase diagram of
  twisted bilayer graphene},\ }\href
  {https://link.aps.org/doi/10.1103/PhysRevB.103.195127} {\bibfield  {journal}
  {\bibinfo  {journal} {Phys. Rev. B}\ }\textbf {\bibinfo {volume} {103}},\
  \bibinfo {pages} {195127} (\bibinfo {year} {2021})}\BibitemShut {NoStop}%
\bibitem [{\citenamefont {Fischer}\ \emph
  {et~al.}(2021{\natexlab{a}})\citenamefont {Fischer}, \citenamefont {Klebl},
  \citenamefont {Honerkamp},\ and\ \citenamefont {Kennes}}]{AF2020}%
  \BibitemOpen
  \bibfield  {author} {\bibinfo {author} {\bibfnamefont {A.}~\bibnamefont
  {Fischer}}, \bibinfo {author} {\bibfnamefont {L.}~\bibnamefont {Klebl}},
  \bibinfo {author} {\bibfnamefont {C.}~\bibnamefont {Honerkamp}},\ and\
  \bibinfo {author} {\bibfnamefont {D.~M.}\ \bibnamefont {Kennes}},\ }\bibfield
   {title} {\bibinfo {title} {Spin-fluctuation-induced pairing in twisted
  bilayer graphene},\ }\href
  {https://link.aps.org/doi/10.1103/PhysRevB.103.L041103} {\bibfield  {journal}
  {\bibinfo  {journal} {Phys. Rev. B}\ }\textbf {\bibinfo {volume} {103}},\
  \bibinfo {pages} {L041103} (\bibinfo {year}
  {2021}{\natexlab{a}})}\BibitemShut {NoStop}%
\bibitem [{\citenamefont {Rademaker}\ \emph {et~al.}(2019)\citenamefont
  {Rademaker}, \citenamefont {Abanin},\ and\ \citenamefont
  {Mellado}}]{Rademaker2019}%
  \BibitemOpen
  \bibfield  {author} {\bibinfo {author} {\bibfnamefont {L.}~\bibnamefont
  {Rademaker}}, \bibinfo {author} {\bibfnamefont {D.~A.}\ \bibnamefont
  {Abanin}},\ and\ \bibinfo {author} {\bibfnamefont {P.}~\bibnamefont
  {Mellado}},\ }\bibfield  {title} {\bibinfo {title} {Charge smoothening and
  band flattening due to \uppercase{H}artree corrections in twisted bilayer
  graphene},\ }\href {https://link.aps.org/doi/10.1103/PhysRevB.100.205114}
  {\bibfield  {journal} {\bibinfo  {journal} {Phys. Rev. B}\ }\textbf {\bibinfo
  {volume} {100}},\ \bibinfo {pages} {205114} (\bibinfo {year}
  {2019})}\BibitemShut {NoStop}%
\bibitem [{\citenamefont {Goodwin}\ \emph
  {et~al.}(2020{\natexlab{b}})\citenamefont {Goodwin}, \citenamefont {Vitale},
  \citenamefont {Liang}, \citenamefont {Mostofi},\ and\ \citenamefont
  {Lischner}}]{PHD_4}%
  \BibitemOpen
  \bibfield  {author} {\bibinfo {author} {\bibfnamefont {Z.~A.~H.}\
  \bibnamefont {Goodwin}}, \bibinfo {author} {\bibfnamefont {V.}~\bibnamefont
  {Vitale}}, \bibinfo {author} {\bibfnamefont {X.}~\bibnamefont {Liang}},
  \bibinfo {author} {\bibfnamefont {A.~A.}\ \bibnamefont {Mostofi}},\ and\
  \bibinfo {author} {\bibfnamefont {J.}~\bibnamefont {Lischner}},\ }\bibfield
  {title} {\bibinfo {title} {Hartree theory calculations of quasiparticle
  properties in twisted bilayer graphene},\ }\href
  {https://iopscience.iop.org/article/10.1088/2516-1075/ab9f94} {\bibfield
  {journal} {\bibinfo  {journal} {Electron. Struct.}\ }\textbf {\bibinfo
  {volume} {2}},\ \bibinfo {pages} {034001} (\bibinfo {year}
  {2020}{\natexlab{b}})}\BibitemShut {NoStop}%
\bibitem [{\citenamefont {Cheung}\ \emph {et~al.}(2021)\citenamefont {Cheung},
  \citenamefont {Goodwin}, \citenamefont {Vitale}, \citenamefont {Lischner},\
  and\ \citenamefont {Mostofi}}]{PHD_9}%
  \BibitemOpen
  \bibfield  {author} {\bibinfo {author} {\bibfnamefont {C.~S.}\ \bibnamefont
  {Cheung}}, \bibinfo {author} {\bibfnamefont {Z.~A.~H.}\ \bibnamefont
  {Goodwin}}, \bibinfo {author} {\bibfnamefont {V.}~\bibnamefont {Vitale}},
  \bibinfo {author} {\bibfnamefont {J.}~\bibnamefont {Lischner}},\ and\
  \bibinfo {author} {\bibfnamefont {A.~A.}\ \bibnamefont {Mostofi}},\
  }\bibfield  {title} {\bibinfo {title} {Hartree theory and crystal field of
  twisted double bilayer graphene near the magic-angle},\ }\href
  {https://iopscience.iop.org/article/10.1088/2516-1075/ac5eaa} {\bibfield
  {journal} {\bibinfo  {journal} {Electron. Struct.}\ }\textbf {\bibinfo
  {volume} {4}},\ \bibinfo {pages} {025001} (\bibinfo {year}
  {2021})}\BibitemShut {NoStop}%
\bibitem [{\citenamefont {Vahedi}\ \emph {et~al.}(2021)\citenamefont {Vahedi},
  \citenamefont {Peters}, \citenamefont {Missaoui}, \citenamefont {Honecker},\
  and\ \citenamefont {de~Laissardi\'ere}}]{Vahedi2021}%
  \BibitemOpen
  \bibfield  {author} {\bibinfo {author} {\bibfnamefont {J.}~\bibnamefont
  {Vahedi}}, \bibinfo {author} {\bibfnamefont {R.}~\bibnamefont {Peters}},
  \bibinfo {author} {\bibfnamefont {A.}~\bibnamefont {Missaoui}}, \bibinfo
  {author} {\bibfnamefont {A.}~\bibnamefont {Honecker}},\ and\ \bibinfo
  {author} {\bibfnamefont {G.~T.}\ \bibnamefont {de~Laissardi\'ere}},\
  }\bibfield  {title} {\bibinfo {title} {Magnetism of magic-angle twisted
  bilayer graphene},\ }\href {https://scipost.org/SciPostPhys.11.4.083/pdf}
  {\bibfield  {journal} {\bibinfo  {journal} {SciPost Phys.}\ }\textbf
  {\bibinfo {volume} {11}},\ \bibinfo {pages} {083} (\bibinfo {year}
  {2021})}\BibitemShut {NoStop}%
\bibitem [{\citenamefont {Gonz\'alez}\ and\ \citenamefont
  {Stauber}(2021)}]{Stauber2021}%
  \BibitemOpen
  \bibfield  {author} {\bibinfo {author} {\bibfnamefont {J.}~\bibnamefont
  {Gonz\'alez}}\ and\ \bibinfo {author} {\bibfnamefont {T.}~\bibnamefont
  {Stauber}},\ }\bibfield  {title} {\bibinfo {title} {Magnetic phases from
  competing \uppercase{H}ubbard and extended \uppercase{C}oulomb interactions
  in twisted bilayer graphene},\ }\href
  {https://journals.aps.org/prb/abstract/10.1103/PhysRevB.104.115110}
  {\bibfield  {journal} {\bibinfo  {journal} {Phys. Rev. B}\ }\textbf {\bibinfo
  {volume} {104}},\ \bibinfo {pages} {115110} (\bibinfo {year}
  {2021})}\BibitemShut {NoStop}%
\bibitem [{\citenamefont {Gonz\'alez}\ and\ \citenamefont
  {Stauber}(2020)}]{Gonzalex2020SymmetryBreaking}%
  \BibitemOpen
  \bibfield  {author} {\bibinfo {author} {\bibfnamefont {J.}~\bibnamefont
  {Gonz\'alez}}\ and\ \bibinfo {author} {\bibfnamefont {T.}~\bibnamefont
  {Stauber}},\ }\bibfield  {title} {\bibinfo {title} {Time-reversal symmetry
  breaking versus chiral symmetry breaking in twisted bilayer graphene},\
  }\href {https://doi.org/10.1103/PhysRevB.102.081118} {\bibfield  {journal}
  {\bibinfo  {journal} {Phys. Rev. B}\ }\textbf {\bibinfo {volume} {102}},\
  \bibinfo {pages} {081118} (\bibinfo {year} {2020})}\BibitemShut {NoStop}%
\bibitem [{\citenamefont {Sboychakov}\ \emph {et~al.}(2019)\citenamefont
  {Sboychakov}, \citenamefont {Rozhkov}, \citenamefont {Rakhmanov},\ and\
  \citenamefont {Nori}}]{Sboychakov2019ManyBodyEffects}%
  \BibitemOpen
  \bibfield  {author} {\bibinfo {author} {\bibfnamefont {A.~O.}\ \bibnamefont
  {Sboychakov}}, \bibinfo {author} {\bibfnamefont {A.~V.}\ \bibnamefont
  {Rozhkov}}, \bibinfo {author} {\bibfnamefont {A.~L.}\ \bibnamefont
  {Rakhmanov}},\ and\ \bibinfo {author} {\bibfnamefont {F.}~\bibnamefont
  {Nori}},\ }\bibfield  {title} {\bibinfo {title} {Many-body effects in twisted
  bilayer graphene at low twist angles},\ }\href
  {https://doi.org/10.1103/PhysRevB.100.045111} {\bibfield  {journal} {\bibinfo
   {journal} {Phys. Rev. B}\ }\textbf {\bibinfo {volume} {100}},\ \bibinfo
  {pages} {045111} (\bibinfo {year} {2019})}\BibitemShut {NoStop}%
\bibitem [{\citenamefont {Sboychakov}\ \emph {et~al.}(2020)\citenamefont
  {Sboychakov}, \citenamefont {Rozhkov}, \citenamefont {Rakhmanov},\ and\
  \citenamefont {Nori}}]{Sboychakov2020}%
  \BibitemOpen
  \bibfield  {author} {\bibinfo {author} {\bibfnamefont {A.~O.}\ \bibnamefont
  {Sboychakov}}, \bibinfo {author} {\bibfnamefont {A.~V.}\ \bibnamefont
  {Rozhkov}}, \bibinfo {author} {\bibfnamefont {A.~L.}\ \bibnamefont
  {Rakhmanov}},\ and\ \bibinfo {author} {\bibfnamefont {F.}~\bibnamefont
  {Nori}},\ }\bibfield  {title} {\bibinfo {title} {Spin density wave and
  electron nematicity in magic-angle twisted bilayer graphene},\ }\href
  {https://link.aps.org/doi/10.1103/PhysRevB.102.155142} {\bibfield  {journal}
  {\bibinfo  {journal} {Phys. Rev. B}\ }\textbf {\bibinfo {volume} {102}},\
  \bibinfo {pages} {155142} (\bibinfo {year} {2020})}\BibitemShut {NoStop}%
\bibitem [{\citenamefont {Chichinadze}\ \emph {et~al.}(2020)\citenamefont
  {Chichinadze}, \citenamefont {Classen},\ and\ \citenamefont
  {Chubukov}}]{Chichinadze2020NematicSupercond}%
  \BibitemOpen
  \bibfield  {author} {\bibinfo {author} {\bibfnamefont {D.~V.}\ \bibnamefont
  {Chichinadze}}, \bibinfo {author} {\bibfnamefont {L.}~\bibnamefont
  {Classen}},\ and\ \bibinfo {author} {\bibfnamefont {A.~V.}\ \bibnamefont
  {Chubukov}},\ }\bibfield  {title} {\bibinfo {title} {Nematic
  superconductivity in twisted bilayer graphene},\ }\href
  {https://doi.org/10.1103/PhysRevB.101.224513} {\bibfield  {journal} {\bibinfo
   {journal} {Phys. Rev. B}\ }\textbf {\bibinfo {volume} {101}},\ \bibinfo
  {pages} {224513} (\bibinfo {year} {2020})}\BibitemShut {NoStop}%
\bibitem [{\citenamefont {Gonzalez-Arraga}\ \emph {et~al.}(2017)\citenamefont
  {Gonzalez-Arraga}, \citenamefont {Lado}, \citenamefont {Guinea},\ and\
  \citenamefont {San-Jose}}]{GonzalezArraga2017}%
  \BibitemOpen
  \bibfield  {author} {\bibinfo {author} {\bibfnamefont {L.~A.}\ \bibnamefont
  {Gonzalez-Arraga}}, \bibinfo {author} {\bibfnamefont {J.~L.}\ \bibnamefont
  {Lado}}, \bibinfo {author} {\bibfnamefont {F.}~\bibnamefont {Guinea}},\ and\
  \bibinfo {author} {\bibfnamefont {P.}~\bibnamefont {San-Jose}},\ }\bibfield
  {title} {\bibinfo {title} {Electrically controllable magnetism in twisted
  bilayer graphene},\ }\href {https://doi.org/10.1103/PhysRevLett.119.107201}
  {\bibfield  {journal} {\bibinfo  {journal} {Phys. Rev. Lett.}\ }\textbf
  {\bibinfo {volume} {119}},\ \bibinfo {pages} {107201} (\bibinfo {year}
  {2017})}\BibitemShut {NoStop}%
\bibitem [{\citenamefont {Ramires}\ and\ \citenamefont
  {Lado}(2019)}]{Ramires2019}%
  \BibitemOpen
  \bibfield  {author} {\bibinfo {author} {\bibfnamefont {A.}~\bibnamefont
  {Ramires}}\ and\ \bibinfo {author} {\bibfnamefont {J.~L.}\ \bibnamefont
  {Lado}},\ }\bibfield  {title} {\bibinfo {title} {Impurity-induced triple
  point fermions in twisted bilayer graphene},\ }\href
  {https://link.aps.org/doi/10.1103/PhysRevB.99.245118} {\bibfield  {journal}
  {\bibinfo  {journal} {Phys. Rev. B}\ }\textbf {\bibinfo {volume} {99}},\
  \bibinfo {pages} {245118} (\bibinfo {year} {2019})}\BibitemShut {NoStop}%
\bibitem [{\citenamefont {Wolf}\ \emph {et~al.}(2019)\citenamefont {Wolf},
  \citenamefont {Lado}, \citenamefont {Blatter},\ and\ \citenamefont
  {Zilberberg}}]{Wolf2019}%
  \BibitemOpen
  \bibfield  {author} {\bibinfo {author} {\bibfnamefont {T.~M.~R.}\
  \bibnamefont {Wolf}}, \bibinfo {author} {\bibfnamefont {J.~L.}\ \bibnamefont
  {Lado}}, \bibinfo {author} {\bibfnamefont {G.}~\bibnamefont {Blatter}},\ and\
  \bibinfo {author} {\bibfnamefont {O.}~\bibnamefont {Zilberberg}},\ }\bibfield
   {title} {\bibinfo {title} {Electrically tunable flat bands and magnetism in
  twisted bilayer graphene},\ }\href
  {https://link.aps.org/doi/10.1103/PhysRevLett.123.096802} {\bibfield
  {journal} {\bibinfo  {journal} {Phys. Rev. Lett.}\ }\textbf {\bibinfo
  {volume} {123}},\ \bibinfo {pages} {096802} (\bibinfo {year}
  {2019})}\BibitemShut {NoStop}%
\bibitem [{\citenamefont {Carr}\ \emph {et~al.}(2018)\citenamefont {Carr},
  \citenamefont {Fang}, \citenamefont {Jarillo-Herrero},\ and\ \citenamefont
  {Kaxiras}}]{PDTBLG}%
  \BibitemOpen
  \bibfield  {author} {\bibinfo {author} {\bibfnamefont {S.}~\bibnamefont
  {Carr}}, \bibinfo {author} {\bibfnamefont {S.}~\bibnamefont {Fang}}, \bibinfo
  {author} {\bibfnamefont {P.}~\bibnamefont {Jarillo-Herrero}},\ and\ \bibinfo
  {author} {\bibfnamefont {E.}~\bibnamefont {Kaxiras}},\ }\bibfield  {title}
  {\bibinfo {title} {Pressure dependence of the magic twist angle in graphene
  superlattices},\ }\href {https://link.aps.org/doi/10.1103/PhysRevB.98.085144}
  {\bibfield  {journal} {\bibinfo  {journal} {Phys. Rev. B}\ }\textbf {\bibinfo
  {volume} {98}},\ \bibinfo {pages} {085144} (\bibinfo {year}
  {2018})}\BibitemShut {NoStop}%
\bibitem [{\citenamefont {Lopez-Bezanilla}(2019)}]{Bezanilla2019}%
  \BibitemOpen
  \bibfield  {author} {\bibinfo {author} {\bibfnamefont {A.}~\bibnamefont
  {Lopez-Bezanilla}},\ }\bibfield  {title} {\bibinfo {title} {Emergence of
  flat-band magnetism and half-metallicity in twisted bilayer graphene},\
  }\href {https://link.aps.org/doi/10.1103/PhysRevMaterials.3.054003}
  {\bibfield  {journal} {\bibinfo  {journal} {Phys. Rev. Materials}\ }\textbf
  {\bibinfo {volume} {3}},\ \bibinfo {pages} {054003} (\bibinfo {year}
  {2019})}\BibitemShut {NoStop}%
\bibitem [{\citenamefont {Chen}\ \emph {et~al.}(2022)\citenamefont {Chen},
  \citenamefont {Liu}, \citenamefont {Fry},\ and\ \citenamefont
  {Cheng}}]{Chen2022first}%
  \BibitemOpen
  \bibfield  {author} {\bibinfo {author} {\bibfnamefont {X.}~\bibnamefont
  {Chen}}, \bibinfo {author} {\bibfnamefont {S.}~\bibnamefont {Liu}}, \bibinfo
  {author} {\bibfnamefont {J.~N.}\ \bibnamefont {Fry}},\ and\ \bibinfo {author}
  {\bibfnamefont {H.-P.}\ \bibnamefont {Cheng}},\ }\bibfield  {title} {\bibinfo
  {title} {First-principles calculation of gate-tunable ferromagnetism in
  magic-angle twisted bilayer graphene under pressure},\ }\href
  {https://iopscience.iop.org/article/10.1088/1361-648X/ac7e9a} {\bibfield
  {journal} {\bibinfo  {journal} {J. Phys.: Condens. Matter.}\ }\textbf
  {\bibinfo {volume} {34}},\ \bibinfo {pages} {385501} (\bibinfo {year}
  {2022})}\BibitemShut {NoStop}%
\bibitem [{\citenamefont {Nam}\ and\ \citenamefont {Koshino}(2017)}]{LREBM}%
  \BibitemOpen
  \bibfield  {author} {\bibinfo {author} {\bibfnamefont {N.~N.~T.}\
  \bibnamefont {Nam}}\ and\ \bibinfo {author} {\bibfnamefont {M.}~\bibnamefont
  {Koshino}},\ }\bibfield  {title} {\bibinfo {title} {Lattice relaxation and
  energy band modulation in twisted bilayer graphene},\ }\href
  {https://doi.org/10.1103/physrevb.96.075311} {\bibfield  {journal} {\bibinfo
  {journal} {Phys. Rev. B}\ }\textbf {\bibinfo {volume} {96}},\ \bibinfo
  {pages} {075311} (\bibinfo {year} {2017})}\BibitemShut {NoStop}%
\bibitem [{\citenamefont {Jain}\ \emph {et~al.}(2017)\citenamefont {Jain},
  \citenamefont {Juri\u{c}i\'c},\ and\ \citenamefont {Barkema}}]{STBBG}%
  \BibitemOpen
  \bibfield  {author} {\bibinfo {author} {\bibfnamefont {S.~K.}\ \bibnamefont
  {Jain}}, \bibinfo {author} {\bibfnamefont {V.}~\bibnamefont
  {Juri\u{c}i\'c}},\ and\ \bibinfo {author} {\bibfnamefont {G.~T.}\
  \bibnamefont {Barkema}},\ }\bibfield  {title} {\bibinfo {title} {Structure of
  twisted and buckled bilayer graphene},\ }\href
  {https://doi.org/10.1088/2053-1583/4/1/015018} {\bibfield  {journal}
  {\bibinfo  {journal} {2D Mater.}\ }\textbf {\bibinfo {volume} {4}},\ \bibinfo
  {pages} {015018} (\bibinfo {year} {2017})}\BibitemShut {NoStop}%
\bibitem [{\citenamefont {Gargiulo}\ and\ \citenamefont
  {Yazyev}(2018)}]{SETLA}%
  \BibitemOpen
  \bibfield  {author} {\bibinfo {author} {\bibfnamefont {F.}~\bibnamefont
  {Gargiulo}}\ and\ \bibinfo {author} {\bibfnamefont {O.~V.}\ \bibnamefont
  {Yazyev}},\ }\bibfield  {title} {\bibinfo {title} {Structural and electronic
  transformation in low-angle twisted bilayer graphene},\ }\href
  {https://doi.org/10.1088/2053-1583/aa9640} {\bibfield  {journal} {\bibinfo
  {journal} {2D Mater.}\ }\textbf {\bibinfo {volume} {5}},\ \bibinfo {pages}
  {015019} (\bibinfo {year} {2018})}\BibitemShut {NoStop}%
\bibitem [{\citenamefont {Guinea}\ and\ \citenamefont {Walet}(2019)}]{LDLE}%
  \BibitemOpen
  \bibfield  {author} {\bibinfo {author} {\bibfnamefont {F.}~\bibnamefont
  {Guinea}}\ and\ \bibinfo {author} {\bibfnamefont {N.~R.}\ \bibnamefont
  {Walet}},\ }\bibfield  {title} {\bibinfo {title} {Continuum models for
  twisted bilayer graphene: Effect of lattice deformation and hopping
  parameters},\ }\bibfield  {journal} {\bibinfo  {journal} {Physical Review B}\
  }\textbf {\bibinfo {volume} {99}},\ \href
  {https://doi.org/10.1103/physrevb.99.205134} {10.1103/physrevb.99.205134}
  (\bibinfo {year} {2019})\BibitemShut {NoStop}%
\bibitem [{\citenamefont {Liang}\ \emph {et~al.}(2020)\citenamefont {Liang},
  \citenamefont {Goodwin}, \citenamefont {Vitale}, \citenamefont {Corsetti},
  \citenamefont {Mostofi},\ and\ \citenamefont {Lischner}}]{Xia2020}%
  \BibitemOpen
  \bibfield  {author} {\bibinfo {author} {\bibfnamefont {X.}~\bibnamefont
  {Liang}}, \bibinfo {author} {\bibfnamefont {Z.~A.~H.}\ \bibnamefont
  {Goodwin}}, \bibinfo {author} {\bibfnamefont {V.}~\bibnamefont {Vitale}},
  \bibinfo {author} {\bibfnamefont {F.}~\bibnamefont {Corsetti}}, \bibinfo
  {author} {\bibfnamefont {A.}~\bibnamefont {Mostofi}},\ and\ \bibinfo {author}
  {\bibfnamefont {J.}~\bibnamefont {Lischner}},\ }\bibfield  {title} {\bibinfo
  {title} {Effect of bilayer stacking on the atomic and electronic structure of
  twisted double bilayer graphene},\ }\href
  {https://doi.org/10.1103/physrevb.102.155146} {\bibfield  {journal} {\bibinfo
   {journal} {Phys. Rev. B}\ }\textbf {\bibinfo {volume} {102}},\ \bibinfo
  {pages} {155146} (\bibinfo {year} {2020})}\BibitemShut {NoStop}%
\bibitem [{\citenamefont {Plimpton}(1995)}]{LAMMPS}%
  \BibitemOpen
  \bibfield  {author} {\bibinfo {author} {\bibfnamefont {S.}~\bibnamefont
  {Plimpton}},\ }\bibfield  {title} {\bibinfo {title} {Fast parallel algorithms
  for short-range molecular dynamics},\ }\href
  {https://doi.org/10.1006/jcph.1995.1039} {\bibfield  {journal} {\bibinfo
  {journal} {J. Comp. Phys.}\ }\textbf {\bibinfo {volume} {117}},\ \bibinfo
  {pages} {1} (\bibinfo {year} {1995})}\BibitemShut {NoStop}%
\bibitem [{\citenamefont {O'Connor}\ \emph {et~al.}(2015)\citenamefont
  {O'Connor}, \citenamefont {Andzelm},\ and\ \citenamefont {Robbins}}]{AIREBO}%
  \BibitemOpen
  \bibfield  {author} {\bibinfo {author} {\bibfnamefont {T.~C.}\ \bibnamefont
  {O'Connor}}, \bibinfo {author} {\bibfnamefont {J.}~\bibnamefont {Andzelm}},\
  and\ \bibinfo {author} {\bibfnamefont {M.~O.}\ \bibnamefont {Robbins}},\
  }\bibfield  {title} {\bibinfo {title} {Airebo-m: A reactive model for
  hydrocarbons at extreme pressures},\ }\href
  {https://doi.org/10.1063/1.4905549} {\bibfield  {journal} {\bibinfo
  {journal} {J. Chem. Phys.}\ }\textbf {\bibinfo {volume} {142}},\ \bibinfo
  {pages} {024903} (\bibinfo {year} {2015})}\BibitemShut {NoStop}%
\bibitem [{\citenamefont {Kolmogorov}\ and\ \citenamefont {Crespi}(2005)}]{KC}%
  \BibitemOpen
  \bibfield  {author} {\bibinfo {author} {\bibfnamefont {A.~N.}\ \bibnamefont
  {Kolmogorov}}\ and\ \bibinfo {author} {\bibfnamefont {V.~H.}\ \bibnamefont
  {Crespi}},\ }\bibfield  {title} {\bibinfo {title} {Registry-dependent
  interlayer potential for graphitic systems},\ }\href
  {https://doi.org/10.1103/physrevb.71.235415} {\bibfield  {journal} {\bibinfo
  {journal} {Phys. Rev. B}\ }\textbf {\bibinfo {volume} {71}},\ \bibinfo
  {pages} {235415} (\bibinfo {year} {2005})}\BibitemShut {NoStop}%
\bibitem [{\citenamefont {Slater}\ and\ \citenamefont {Koster}(1954)}]{SK}%
  \BibitemOpen
  \bibfield  {author} {\bibinfo {author} {\bibfnamefont {J.~C.}\ \bibnamefont
  {Slater}}\ and\ \bibinfo {author} {\bibfnamefont {G.~F.}\ \bibnamefont
  {Koster}},\ }\bibfield  {title} {\bibinfo {title} {Simplified
  \uppercase{LCAO} method for the periodic potential problem},\ }\href
  {https://doi.org/10.1103/physrev.94.1498} {\bibfield  {journal} {\bibinfo
  {journal} {Phys. Rev.}\ }\textbf {\bibinfo {volume} {94}},\ \bibinfo {pages}
  {1498} (\bibinfo {year} {1954})}\BibitemShut {NoStop}%
\bibitem [{\citenamefont {Angeli}\ \emph {et~al.}(2018)\citenamefont {Angeli},
  \citenamefont {Mandelli}, \citenamefont {Valli}, \citenamefont {Amaricci},
  \citenamefont {Capone}, \citenamefont {Tosatti},\ and\ \citenamefont
  {Fabrizio}}]{EDS}%
  \BibitemOpen
  \bibfield  {author} {\bibinfo {author} {\bibfnamefont {M.}~\bibnamefont
  {Angeli}}, \bibinfo {author} {\bibfnamefont {D.}~\bibnamefont {Mandelli}},
  \bibinfo {author} {\bibfnamefont {A.}~\bibnamefont {Valli}}, \bibinfo
  {author} {\bibfnamefont {A.}~\bibnamefont {Amaricci}}, \bibinfo {author}
  {\bibfnamefont {M.}~\bibnamefont {Capone}}, \bibinfo {author} {\bibfnamefont
  {E.}~\bibnamefont {Tosatti}},\ and\ \bibinfo {author} {\bibfnamefont
  {M.}~\bibnamefont {Fabrizio}},\ }\bibfield  {title} {\bibinfo {title}
  {Emergent \uppercase{D}$_6$ symmetry in fully-relaxed magic-angle twisted
  bilayer graphene},\ }\href {https://doi.org/10.1103/physrevb.98.235137}
  {\bibfield  {journal} {\bibinfo  {journal} {Phys. Rev. B}\ }\textbf {\bibinfo
  {volume} {98}},\ \bibinfo {pages} {235137} (\bibinfo {year}
  {2018})}\BibitemShut {NoStop}%
\bibitem [{\citenamefont {Goodwin}(2022)}]{Goodwin2022thesis}%
  \BibitemOpen
  \bibfield  {author} {\bibinfo {author} {\bibfnamefont {Z.~A.~H.}\
  \bibnamefont {Goodwin}},\ }\href {https://doi.org/10.25560/96089} {\bibinfo
  {title} {Theory and simulation of moir\'e graphene multilayers}} (\bibinfo
  {year} {2022})\BibitemShut {NoStop}%
\bibitem [{\citenamefont {Koshino}\ \emph {et~al.}(2018)\citenamefont
  {Koshino}, \citenamefont {Yuan}, \citenamefont {Koretsune}, \citenamefont
  {Ochi}, \citenamefont {Kuroki},\ and\ \citenamefont {Fu}}]{Koshino2018}%
  \BibitemOpen
  \bibfield  {author} {\bibinfo {author} {\bibfnamefont {M.}~\bibnamefont
  {Koshino}}, \bibinfo {author} {\bibfnamefont {N.~F.~Q.}\ \bibnamefont
  {Yuan}}, \bibinfo {author} {\bibfnamefont {T.}~\bibnamefont {Koretsune}},
  \bibinfo {author} {\bibfnamefont {M.}~\bibnamefont {Ochi}}, \bibinfo {author}
  {\bibfnamefont {K.}~\bibnamefont {Kuroki}},\ and\ \bibinfo {author}
  {\bibfnamefont {L.}~\bibnamefont {Fu}},\ }\bibfield  {title} {\bibinfo
  {title} {Maximally localized wannier orbitals and the extended hubbard model
  for twisted bilayer graphene},\ }\href
  {https://doi.org/10.1103/PhysRevX.8.031087} {\bibfield  {journal} {\bibinfo
  {journal} {Phys. Rev. X}\ }\textbf {\bibinfo {volume} {8}},\ \bibinfo {pages}
  {031087} (\bibinfo {year} {2018})}\BibitemShut {NoStop}%
\bibitem [{\citenamefont {Lopez-Bezanilla}\ and\ \citenamefont
  {Lado}(2020)}]{Alejandro2020}%
  \BibitemOpen
  \bibfield  {author} {\bibinfo {author} {\bibfnamefont {A.}~\bibnamefont
  {Lopez-Bezanilla}}\ and\ \bibinfo {author} {\bibfnamefont {J.~L.}\
  \bibnamefont {Lado}},\ }\bibfield  {title} {\bibinfo {title} {Electrical band
  flattening, valley flux, and superconductivity in twisted trilayer
  graphene},\ }\href
  {https://link.aps.org/doi/10.1103/PhysRevResearch.2.033357} {\bibfield
  {journal} {\bibinfo  {journal} {Phys. Rev. Research}\ }\textbf {\bibinfo
  {volume} {2}},\ \bibinfo {pages} {033357} (\bibinfo {year}
  {2020})}\BibitemShut {NoStop}%
\bibitem [{\citenamefont {Cea}\ \emph {et~al.}(2019{\natexlab{b}})\citenamefont
  {Cea}, \citenamefont {Walet},\ and\ \citenamefont {Guinea}}]{Cea20193D}%
  \BibitemOpen
  \bibfield  {author} {\bibinfo {author} {\bibfnamefont {T.}~\bibnamefont
  {Cea}}, \bibinfo {author} {\bibfnamefont {N.~R.}\ \bibnamefont {Walet}},\
  and\ \bibinfo {author} {\bibfnamefont {F.}~\bibnamefont {Guinea}},\
  }\bibfield  {title} {\bibinfo {title} {Twists and the electronic structure of
  graphitic materials},\ }\href
  {https://pubs.acs.org/doi/10.1021/acs.nanolett.9b03335} {\bibfield  {journal}
  {\bibinfo  {journal} {Nano Lett.}\ }\textbf {\bibinfo {volume} {19}},\
  \bibinfo {pages} {8683} (\bibinfo {year} {2019}{\natexlab{b}})}\BibitemShut
  {NoStop}%
\bibitem [{\citenamefont {Pantale\'on}\ \emph {et~al.}(2021)\citenamefont
  {Pantale\'on}, \citenamefont {Cea}, \citenamefont {Brown}, \citenamefont
  {Walet},\ and\ \citenamefont {Guinea}}]{Pierre2020}%
  \BibitemOpen
  \bibfield  {author} {\bibinfo {author} {\bibfnamefont {P.~A.}\ \bibnamefont
  {Pantale\'on}}, \bibinfo {author} {\bibfnamefont {T.}~\bibnamefont {Cea}},
  \bibinfo {author} {\bibfnamefont {R.}~\bibnamefont {Brown}}, \bibinfo
  {author} {\bibfnamefont {N.~R.}\ \bibnamefont {Walet}},\ and\ \bibinfo
  {author} {\bibfnamefont {F.}~\bibnamefont {Guinea}},\ }\bibfield  {title}
  {\bibinfo {title} {Narrow bands and electrostatic interactions in graphene
  stacks},\ }\href
  {https://iopscience.iop.org/article/10.1088/2053-1583/ac1b6d/meta} {\bibfield
   {journal} {\bibinfo  {journal} {2D Mater.}\ }\textbf {\bibinfo {volume}
  {8}},\ \bibinfo {pages} {044006} (\bibinfo {year} {2021})}\BibitemShut
  {NoStop}%
\bibitem [{\citenamefont {Ledwith}\ \emph {et~al.}(2021)\citenamefont
  {Ledwith}, \citenamefont {Khalaf}, \citenamefont {Zhu}, \citenamefont {Carr},
  \citenamefont {Kaxiras},\ and\ \citenamefont {Vishwanath}}]{TBorNTB}%
  \BibitemOpen
  \bibfield  {author} {\bibinfo {author} {\bibfnamefont {P.~J.}\ \bibnamefont
  {Ledwith}}, \bibinfo {author} {\bibfnamefont {E.}~\bibnamefont {Khalaf}},
  \bibinfo {author} {\bibfnamefont {Z.}~\bibnamefont {Zhu}}, \bibinfo {author}
  {\bibfnamefont {S.}~\bibnamefont {Carr}}, \bibinfo {author} {\bibfnamefont
  {E.}~\bibnamefont {Kaxiras}},\ and\ \bibinfo {author} {\bibfnamefont
  {A.}~\bibnamefont {Vishwanath}},\ }\bibfield  {title} {\bibinfo {title} {Tb
  or not tb? contrasting properties of twisted bilayer graphene and the
  alternating twist n-layer structures (n=3,4,5...)},\ }\href
  {https://arxiv.org/abs/2111.11060} {\bibfield  {journal} {\bibinfo  {journal}
  {arXiv:2111.11060}\ } (\bibinfo {year} {2021})}\BibitemShut {NoStop}%
\bibitem [{\citenamefont {Khalaf}\ \emph {et~al.}(2019)\citenamefont {Khalaf},
  \citenamefont {J.Kruchkov}, \citenamefont {Tarnopolsky},\ and\ \citenamefont
  {Vishwanath}}]{khalaf2019magic}%
  \BibitemOpen
  \bibfield  {author} {\bibinfo {author} {\bibfnamefont {E.}~\bibnamefont
  {Khalaf}}, \bibinfo {author} {\bibfnamefont {A.}~\bibnamefont {J.Kruchkov}},
  \bibinfo {author} {\bibfnamefont {G.}~\bibnamefont {Tarnopolsky}},\ and\
  \bibinfo {author} {\bibfnamefont {A.}~\bibnamefont {Vishwanath}},\ }\bibfield
   {title} {\bibinfo {title} {Magic angle hierarchy in twisted graphene
  multilayers},\ }\href {https://link.aps.org/doi/10.1103/PhysRevB.100.085109}
  {\bibfield  {journal} {\bibinfo  {journal} {Phys. Rev. B}\ }\textbf {\bibinfo
  {volume} {100}},\ \bibinfo {pages} {085109} (\bibinfo {year}
  {2019})}\BibitemShut {NoStop}%
\bibitem [{\citenamefont {Zhu}\ \emph {et~al.}(2020)\citenamefont {Zhu},
  \citenamefont {Carr}, \citenamefont {Massatt}, \citenamefont {Luskin},\ and\
  \citenamefont {Kaxiras}}]{Zhu2020}%
  \BibitemOpen
  \bibfield  {author} {\bibinfo {author} {\bibfnamefont {Z.}~\bibnamefont
  {Zhu}}, \bibinfo {author} {\bibfnamefont {S.}~\bibnamefont {Carr}}, \bibinfo
  {author} {\bibfnamefont {D.}~\bibnamefont {Massatt}}, \bibinfo {author}
  {\bibfnamefont {M.}~\bibnamefont {Luskin}},\ and\ \bibinfo {author}
  {\bibfnamefont {E.}~\bibnamefont {Kaxiras}},\ }\bibfield  {title} {\bibinfo
  {title} {Twisted trilayer graphene: A precisely tunable platform for
  correlated electrons},\ }\href
  {https://link.aps.org/doi/10.1103/PhysRevLett.125.116404} {\bibfield
  {journal} {\bibinfo  {journal} {Phys. Rev. Lett.}\ }\textbf {\bibinfo
  {volume} {125}},\ \bibinfo {pages} {116404} (\bibinfo {year}
  {2020})}\BibitemShut {NoStop}%
\bibitem [{\citenamefont {Carr}\ \emph
  {et~al.}(2020{\natexlab{b}})\citenamefont {Carr}, \citenamefont {Li},
  \citenamefont {Zhu}, \citenamefont {Kaxiras}, \citenamefont {Sachdev},\ and\
  \citenamefont {Kruchkov}}]{Kruchkov2020}%
  \BibitemOpen
  \bibfield  {author} {\bibinfo {author} {\bibfnamefont {S.}~\bibnamefont
  {Carr}}, \bibinfo {author} {\bibfnamefont {C.}~\bibnamefont {Li}}, \bibinfo
  {author} {\bibfnamefont {Z.}~\bibnamefont {Zhu}}, \bibinfo {author}
  {\bibfnamefont {E.}~\bibnamefont {Kaxiras}}, \bibinfo {author} {\bibfnamefont
  {S.}~\bibnamefont {Sachdev}},\ and\ \bibinfo {author} {\bibfnamefont
  {A.}~\bibnamefont {Kruchkov}},\ }\bibfield  {title} {\bibinfo {title}
  {Ultraheavy and \uppercase{U}ltrarelativistic dirac quasiparticles in
  sandwiched graphenes},\ }\href
  {https://pubs.acs.org/doi/abs/10.1021/acs.nanolett.9b04979} {\bibfield
  {journal} {\bibinfo  {journal} {Nano Lett.}\ }\textbf {\bibinfo {volume}
  {20}},\ \bibinfo {pages} {3030} (\bibinfo {year}
  {2020}{\natexlab{b}})}\BibitemShut {NoStop}%
\bibitem [{\citenamefont {Fischer}\ \emph
  {et~al.}(2021{\natexlab{b}})\citenamefont {Fischer}, \citenamefont {Goodwin},
  \citenamefont {Mostofi}, \citenamefont {Lischner}, \citenamefont {Kennes},\
  and\ \citenamefont {Klebl}}]{Fischer_TTLG}%
  \BibitemOpen
  \bibfield  {author} {\bibinfo {author} {\bibfnamefont {A.}~\bibnamefont
  {Fischer}}, \bibinfo {author} {\bibfnamefont {Z.~H.}\ \bibnamefont
  {Goodwin}}, \bibinfo {author} {\bibfnamefont {A.~A.}\ \bibnamefont
  {Mostofi}}, \bibinfo {author} {\bibfnamefont {J.}~\bibnamefont {Lischner}},
  \bibinfo {author} {\bibfnamefont {D.~M.}\ \bibnamefont {Kennes}},\ and\
  \bibinfo {author} {\bibfnamefont {L.}~\bibnamefont {Klebl}},\ }\bibfield
  {title} {\bibinfo {title} {Unconventional superconductivity in magic-angle
  twisted trilayer graphene},\ }\href
  {https://www.nature.com/articles/s41535-021-00410-w} {\bibfield  {journal}
  {\bibinfo  {journal} {npj Quantum Mater.}\ }\textbf {\bibinfo {volume} {7}},\
  \bibinfo {pages} {5} (\bibinfo {year} {2021}{\natexlab{b}})}\BibitemShut
  {NoStop}%
\bibitem [{\citenamefont {o~Phong}\ \emph {et~al.}(2021)\citenamefont
  {o~Phong}, \citenamefont {Pantale\'on}, \citenamefont {Cea},\ and\
  \citenamefont {Guinea}}]{Phong2021trilayer}%
  \BibitemOpen
  \bibfield  {author} {\bibinfo {author} {\bibfnamefont {V.}~\bibnamefont
  {o~Phong}}, \bibinfo {author} {\bibfnamefont {P.~A.}\ \bibnamefont
  {Pantale\'on}}, \bibinfo {author} {\bibfnamefont {T.}~\bibnamefont {Cea}},\
  and\ \bibinfo {author} {\bibfnamefont {F.}~\bibnamefont {Guinea}},\
  }\bibfield  {title} {\bibinfo {title} {Band structure and superconductivity
  in twisted trilayer graphene},\ }\href
  {https://link.aps.org/doi/10.1103/PhysRevB.104.L121116} {\bibfield  {journal}
  {\bibinfo  {journal} {Phys. Rev. B}\ }\textbf {\bibinfo {volume} {104}},\
  \bibinfo {pages} {L121116} (\bibinfo {year} {2021})}\BibitemShut {NoStop}%
\bibitem [{\citenamefont {Shin}\ \emph {et~al.}(2021)\citenamefont {Shin},
  \citenamefont {Chittari},\ and\ \citenamefont {Jung}}]{Shin2021}%
  \BibitemOpen
  \bibfield  {author} {\bibinfo {author} {\bibfnamefont {J.}~\bibnamefont
  {Shin}}, \bibinfo {author} {\bibfnamefont {B.~L.}\ \bibnamefont {Chittari}},\
  and\ \bibinfo {author} {\bibfnamefont {J.}~\bibnamefont {Jung}},\ }\bibfield
  {title} {\bibinfo {title} {Stacking and gate tunable topological at bands,
  gaps and anisotropic strip patterns in twisted trilayer graphene},\ }\href
  {https://link.aps.org/doi/10.1103/PhysRevB.104.045413} {\bibfield  {journal}
  {\bibinfo  {journal} {Phys. Rev. B}\ }\textbf {\bibinfo {volume} {104}},\
  \bibinfo {pages} {045413} (\bibinfo {year} {2021})}\BibitemShut {NoStop}%
\bibitem [{\citenamefont {Park}\ \emph {et~al.}(2021)\citenamefont {Park},
  \citenamefont {Cao}, \citenamefont {Watanabe}, \citenamefont {Taniguchi},\
  and\ \citenamefont {Jarillo-Herrero}}]{Park2021}%
  \BibitemOpen
  \bibfield  {author} {\bibinfo {author} {\bibfnamefont {J.~M.}\ \bibnamefont
  {Park}}, \bibinfo {author} {\bibfnamefont {Y.}~\bibnamefont {Cao}}, \bibinfo
  {author} {\bibfnamefont {K.}~\bibnamefont {Watanabe}}, \bibinfo {author}
  {\bibfnamefont {T.}~\bibnamefont {Taniguchi}},\ and\ \bibinfo {author}
  {\bibfnamefont {P.}~\bibnamefont {Jarillo-Herrero}},\ }\bibfield  {title}
  {\bibinfo {title} {Tunable strongly coupled superconductivity in magic-angle
  twisted trilayer graphene},\ }\href
  {https://www.nature.com/articles/s41586-021-03192-0} {\bibfield  {journal}
  {\bibinfo  {journal} {Nature}\ }\textbf {\bibinfo {volume} {590}},\ \bibinfo
  {pages} {249} (\bibinfo {year} {2021})}\BibitemShut {NoStop}%
\bibitem [{\citenamefont {Cao}\ \emph {et~al.}(2021{\natexlab{b}})\citenamefont
  {Cao}, \citenamefont {Park}, \citenamefont {Watanabe}, \citenamefont
  {Taniguchi},\ and\ \citenamefont {Jarillo-Herrero}}]{Cao2021large}%
  \BibitemOpen
  \bibfield  {author} {\bibinfo {author} {\bibfnamefont {Y.}~\bibnamefont
  {Cao}}, \bibinfo {author} {\bibfnamefont {J.~M.}\ \bibnamefont {Park}},
  \bibinfo {author} {\bibfnamefont {K.}~\bibnamefont {Watanabe}}, \bibinfo
  {author} {\bibfnamefont {T.}~\bibnamefont {Taniguchi}},\ and\ \bibinfo
  {author} {\bibfnamefont {P.}~\bibnamefont {Jarillo-Herrero}},\ }\bibfield
  {title} {\bibinfo {title} {Large \uppercase{P}auli limit violation and
  reentrant superconductivity in magic-angle twisted trilayer graphene},\
  }\href {https://www.nature.com/articles/s41586-021-03685-y} {\bibfield
  {journal} {\bibinfo  {journal} {Nature}\ }\textbf {\bibinfo {volume} {595}},\
  \bibinfo {pages} {526} (\bibinfo {year} {2021}{\natexlab{b}})}\BibitemShut
  {NoStop}%
\bibitem [{\citenamefont {Hao}\ \emph {et~al.}(2021)\citenamefont {Hao},
  \citenamefont {Zimmerman}, \citenamefont {Ledwith}, \citenamefont {Khalaf},
  \citenamefont {Najafabadi}, \citenamefont {Watanabe}, \citenamefont
  {Taniguchi}, \citenamefont {Vishwanath},\ and\ \citenamefont
  {Kim}}]{hao2021engineering}%
  \BibitemOpen
  \bibfield  {author} {\bibinfo {author} {\bibfnamefont {Z.}~\bibnamefont
  {Hao}}, \bibinfo {author} {\bibfnamefont {A.~M.}\ \bibnamefont {Zimmerman}},
  \bibinfo {author} {\bibfnamefont {P.}~\bibnamefont {Ledwith}}, \bibinfo
  {author} {\bibfnamefont {E.}~\bibnamefont {Khalaf}}, \bibinfo {author}
  {\bibfnamefont {D.~H.}\ \bibnamefont {Najafabadi}}, \bibinfo {author}
  {\bibfnamefont {K.}~\bibnamefont {Watanabe}}, \bibinfo {author}
  {\bibfnamefont {T.}~\bibnamefont {Taniguchi}}, \bibinfo {author}
  {\bibfnamefont {A.}~\bibnamefont {Vishwanath}},\ and\ \bibinfo {author}
  {\bibfnamefont {P.}~\bibnamefont {Kim}},\ }\bibfield  {title} {\bibinfo
  {title} {Electric field tunable superconductivity in alternating twist
  magic-angle trilayer graphene},\ }\href
  {https://www.science.org/doi/10.1126/science.abg0399} {\bibfield  {journal}
  {\bibinfo  {journal} {Science}\ }\textbf {\bibinfo {volume} {371}},\ \bibinfo
  {pages} {1133} (\bibinfo {year} {2021})}\BibitemShut {NoStop}%
\bibitem [{\citenamefont {Turkel}\ \emph {et~al.}(2022)\citenamefont {Turkel},
  \citenamefont {Swann}, \citenamefont {Zhu}, \citenamefont {Christos},
  \citenamefont {Watanabe}, \citenamefont {Taniguchi}, \citenamefont {Sachdev},
  \citenamefont {Scheurer}, \citenamefont {Kaxiras}, \citenamefont {Dean},\
  and\ \citenamefont {Pasupathy}}]{Turkel2022}%
  \BibitemOpen
  \bibfield  {author} {\bibinfo {author} {\bibfnamefont {S.}~\bibnamefont
  {Turkel}}, \bibinfo {author} {\bibfnamefont {J.}~\bibnamefont {Swann}},
  \bibinfo {author} {\bibfnamefont {Z.}~\bibnamefont {Zhu}}, \bibinfo {author}
  {\bibfnamefont {M.}~\bibnamefont {Christos}}, \bibinfo {author}
  {\bibfnamefont {K.}~\bibnamefont {Watanabe}}, \bibinfo {author}
  {\bibfnamefont {T.}~\bibnamefont {Taniguchi}}, \bibinfo {author}
  {\bibfnamefont {S.}~\bibnamefont {Sachdev}}, \bibinfo {author} {\bibfnamefont
  {M.~S.}\ \bibnamefont {Scheurer}}, \bibinfo {author} {\bibfnamefont
  {E.}~\bibnamefont {Kaxiras}}, \bibinfo {author} {\bibfnamefont {C.~R.}\
  \bibnamefont {Dean}},\ and\ \bibinfo {author} {\bibfnamefont {A.~N.}\
  \bibnamefont {Pasupathy}},\ }\bibfield  {title} {\bibinfo {title} {Orderly
  disorder in magic-angle twisted trilayer graphene},\ }\href
  {https://www.science.org/doi/10.1126/science.abk1895} {\bibfield  {journal}
  {\bibinfo  {journal} {Science}\ }\textbf {\bibinfo {volume} {376}},\ \bibinfo
  {pages} {193–199} (\bibinfo {year} {2022})}\BibitemShut {NoStop}%
\bibitem [{\citenamefont {Kim}\ \emph {et~al.}(2022)\citenamefont {Kim},
  \citenamefont {Choi}, \citenamefont {Lewandowski}, \citenamefont {Thomson},
  \citenamefont {Zhang}, \citenamefont {Polski}, \citenamefont {Watanabe},
  \citenamefont {Taniguchi}, \citenamefont {Alicea},\ and\ \citenamefont
  {Nadj-Perge}}]{Kim2022tri}%
  \BibitemOpen
  \bibfield  {author} {\bibinfo {author} {\bibfnamefont {H.}~\bibnamefont
  {Kim}}, \bibinfo {author} {\bibfnamefont {Y.}~\bibnamefont {Choi}}, \bibinfo
  {author} {\bibfnamefont {C.}~\bibnamefont {Lewandowski}}, \bibinfo {author}
  {\bibfnamefont {A.}~\bibnamefont {Thomson}}, \bibinfo {author} {\bibfnamefont
  {Y.}~\bibnamefont {Zhang}}, \bibinfo {author} {\bibfnamefont
  {R.}~\bibnamefont {Polski}}, \bibinfo {author} {\bibfnamefont
  {K.}~\bibnamefont {Watanabe}}, \bibinfo {author} {\bibfnamefont
  {T.}~\bibnamefont {Taniguchi}}, \bibinfo {author} {\bibfnamefont
  {J.}~\bibnamefont {Alicea}},\ and\ \bibinfo {author} {\bibfnamefont
  {S.}~\bibnamefont {Nadj-Perge}},\ }\bibfield  {title} {\bibinfo {title}
  {Evidence for unconventional superconductivity in twisted trilayer
  graphene},\ }\href {https://www.nature.com/articles/s41586-022-04715-z}
  {\bibfield  {journal} {\bibinfo  {journal} {Nature}\ }\textbf {\bibinfo
  {volume} {606}},\ \bibinfo {pages} {494–500} (\bibinfo {year}
  {2022})}\BibitemShut {NoStop}%
\bibitem [{\citenamefont {Zhang}\ \emph {et~al.}(2022)\citenamefont {Zhang},
  \citenamefont {Polski}, \citenamefont {Lewandowski}, \citenamefont {Thomson},
  \citenamefont {Peng}, \citenamefont {Choi}, \citenamefont {Kim},
  \citenamefont {Watanabe}, \citenamefont {Taniguchi}, \citenamefont {Alicea},
  \citenamefont {von Oppen}, \citenamefont {Refael},\ and\ \citenamefont
  {Nadj-Perge}}]{Zhang2022tri}%
  \BibitemOpen
  \bibfield  {author} {\bibinfo {author} {\bibfnamefont {Y.}~\bibnamefont
  {Zhang}}, \bibinfo {author} {\bibfnamefont {R.}~\bibnamefont {Polski}},
  \bibinfo {author} {\bibfnamefont {C.}~\bibnamefont {Lewandowski}}, \bibinfo
  {author} {\bibfnamefont {A.}~\bibnamefont {Thomson}}, \bibinfo {author}
  {\bibfnamefont {Y.}~\bibnamefont {Peng}}, \bibinfo {author} {\bibfnamefont
  {Y.}~\bibnamefont {Choi}}, \bibinfo {author} {\bibfnamefont {H.}~\bibnamefont
  {Kim}}, \bibinfo {author} {\bibfnamefont {K.}~\bibnamefont {Watanabe}},
  \bibinfo {author} {\bibfnamefont {T.}~\bibnamefont {Taniguchi}}, \bibinfo
  {author} {\bibfnamefont {J.}~\bibnamefont {Alicea}}, \bibinfo {author}
  {\bibfnamefont {F.}~\bibnamefont {von Oppen}}, \bibinfo {author}
  {\bibfnamefont {G.}~\bibnamefont {Refael}},\ and\ \bibinfo {author}
  {\bibfnamefont {S.}~\bibnamefont {Nadj-Perge}},\ }\bibfield  {title}
  {\bibinfo {title} {Promotion of superconductivity in magic-anglegraphene
  multilayers},\ }\href
  {https://www.science.org/doi/epdf/10.1126/science.abn8585} {\bibfield
  {journal} {\bibinfo  {journal} {Science}\ }\textbf {\bibinfo {volume}
  {377}},\ \bibinfo {pages} {1538–1543} (\bibinfo {year} {2022})}\BibitemShut
  {NoStop}%
\bibitem [{\citenamefont {Burg}\ \emph {et~al.}(2022)\citenamefont {Burg},
  \citenamefont {Khalaf}, \citenamefont {Wang}, \citenamefont {Watanabe},
  \citenamefont {Taniguchi},\ and\ \citenamefont {Tutuc}}]{Burg2022quad}%
  \BibitemOpen
  \bibfield  {author} {\bibinfo {author} {\bibfnamefont {G.~W.}\ \bibnamefont
  {Burg}}, \bibinfo {author} {\bibfnamefont {E.}~\bibnamefont {Khalaf}},
  \bibinfo {author} {\bibfnamefont {Y.}~\bibnamefont {Wang}}, \bibinfo {author}
  {\bibfnamefont {K.}~\bibnamefont {Watanabe}}, \bibinfo {author}
  {\bibfnamefont {T.}~\bibnamefont {Taniguchi}},\ and\ \bibinfo {author}
  {\bibfnamefont {E.}~\bibnamefont {Tutuc}},\ }\bibfield  {title} {\bibinfo
  {title} {Emergence of correlations in alternating twist quadrilayer
  graphene},\ }\href {https://doi.org/10.1038/s41563-022-01286-2} {\bibfield
  {journal} {\bibinfo  {journal} {Nat. Mater}\ }\textbf {\bibinfo {volume}
  {21}},\ \bibinfo {pages} {884–889} (\bibinfo {year} {2022})}\BibitemShut
  {NoStop}%
\bibitem [{\citenamefont {\textit{et al.}}(2021{\natexlab{c}})}]{Shi2020tTLG}%
  \BibitemOpen
  \bibfield  {author} {\bibinfo {author} {\bibfnamefont {Y.~S.}\ \bibnamefont
  {\textit{et al.}}},\ }\bibfield  {title} {\bibinfo {title} {Tunable van
  \uppercase{H}ove singularities and correlated states in twisted trilayer
  graphene},\ }\href {https://www.nature.com/articles/s41567-021-01172-9}
  {\bibfield  {journal} {\bibinfo  {journal} {Nat. Phys.}\ }\textbf {\bibinfo
  {volume} {17}},\ \bibinfo {pages} {619} (\bibinfo {year}
  {2021}{\natexlab{c}})}\BibitemShut {NoStop}%
\bibitem [{\citenamefont {\textit{et
  al.}}(2021{\natexlab{d}})}]{Yankowitz2020}%
  \BibitemOpen
  \bibfield  {author} {\bibinfo {author} {\bibfnamefont {S.~C.}\ \bibnamefont
  {\textit{et al.}}},\ }\bibfield  {title} {\bibinfo {title} {Electrically
  tunable correlated and topological states in twisted monolayer--bilayer
  graphene},\ }\href {https://www.nature.com/articles/s41567-020-01062-6}
  {\bibfield  {journal} {\bibinfo  {journal} {Nat. Phys.}\ }\textbf {\bibinfo
  {volume} {17}},\ \bibinfo {pages} {374} (\bibinfo {year}
  {2021}{\natexlab{d}})}\BibitemShut {NoStop}%
\bibitem [{\citenamefont {Goodwin}\ \emph {et~al.}(2021)\citenamefont
  {Goodwin}, \citenamefont {Klebl}, \citenamefont {Vitale}, \citenamefont
  {Liang}, \citenamefont {Gogtay}, \citenamefont {van Gorp}, \citenamefont
  {Kennes}, \citenamefont {Mostofi},\ and\ \citenamefont {Lischner}}]{PHD_8}%
  \BibitemOpen
  \bibfield  {author} {\bibinfo {author} {\bibfnamefont {Z.~A.~H.}\
  \bibnamefont {Goodwin}}, \bibinfo {author} {\bibfnamefont {L.}~\bibnamefont
  {Klebl}}, \bibinfo {author} {\bibfnamefont {V.}~\bibnamefont {Vitale}},
  \bibinfo {author} {\bibfnamefont {X.}~\bibnamefont {Liang}}, \bibinfo
  {author} {\bibfnamefont {V.}~\bibnamefont {Gogtay}}, \bibinfo {author}
  {\bibfnamefont {X.}~\bibnamefont {van Gorp}}, \bibinfo {author}
  {\bibfnamefont {D.~M.}\ \bibnamefont {Kennes}}, \bibinfo {author}
  {\bibfnamefont {A.~A.}\ \bibnamefont {Mostofi}},\ and\ \bibinfo {author}
  {\bibfnamefont {J.}~\bibnamefont {Lischner}},\ }\bibfield  {title} {\bibinfo
  {title} {Flat bands, electron interactions, and magnetic order in magic-angle
  mono-trilayer graphene},\ }\href
  {https://doi.org/10.1103/physrevmaterials.5.084008} {\bibfield  {journal}
  {\bibinfo  {journal} {Phys. Rev. Materials}\ }\textbf {\bibinfo {volume}
  {5}},\ \bibinfo {pages} {084008} (\bibinfo {year} {2021})}\BibitemShut
  {NoStop}%
\bibitem [{\citenamefont {\textit{et al.}}(2020{\natexlab{c}})}]{TCT}%
  \BibitemOpen
  \bibfield  {author} {\bibinfo {author} {\bibfnamefont {C.~S.}\ \bibnamefont
  {\textit{et al.}}},\ }\bibfield  {title} {\bibinfo {title} {Correlated states
  in twisted double bilayer graphene},\ }\href
  {https://www.nature.com/articles/s41567-020-0825-9} {\bibfield  {journal}
  {\bibinfo  {journal} {Nat. Phys.}\ }\textbf {\bibinfo {volume} {16}},\
  \bibinfo {pages} {520} (\bibinfo {year} {2020}{\natexlab{c}})}\BibitemShut
  {NoStop}%
\bibitem [{\citenamefont {Liu}\ \emph {et~al.}(2020)\citenamefont {Liu},
  \citenamefont {Hao}, \citenamefont {Khalaf}, \citenamefont {Lee},
  \citenamefont {Watanabe}, \citenamefont {Taniguchi}, \citenamefont
  {Vishwanath},\ and\ \citenamefont {Kim}}]{BIBI}%
  \BibitemOpen
  \bibfield  {author} {\bibinfo {author} {\bibfnamefont {X.}~\bibnamefont
  {Liu}}, \bibinfo {author} {\bibfnamefont {Z.}~\bibnamefont {Hao}}, \bibinfo
  {author} {\bibfnamefont {E.}~\bibnamefont {Khalaf}}, \bibinfo {author}
  {\bibfnamefont {J.~Y.}\ \bibnamefont {Lee}}, \bibinfo {author} {\bibfnamefont
  {K.}~\bibnamefont {Watanabe}}, \bibinfo {author} {\bibfnamefont
  {T.}~\bibnamefont {Taniguchi}}, \bibinfo {author} {\bibfnamefont
  {A.}~\bibnamefont {Vishwanath}},\ and\ \bibinfo {author} {\bibfnamefont
  {P.}~\bibnamefont {Kim}},\ }\bibfield  {title} {\bibinfo {title} {Tunable
  spin-polarized correlated states in twisted double bilayer graphene},\ }\href
  {https://www.nature.com/articles/s41586-020-2458-7} {\bibfield  {journal}
  {\bibinfo  {journal} {Nature}\ }\textbf {\bibinfo {volume} {583}},\ \bibinfo
  {pages} {221} (\bibinfo {year} {2020})}\BibitemShut {NoStop}%
\bibitem [{\citenamefont {G.~W}\ \emph {et~al.}(2019)\citenamefont {G.~W},
  \citenamefont {Zhu}, \citenamefont {Taniguchi}, \citenamefont {Watanabe},
  \citenamefont {MacDonald},\ and\ \citenamefont
  {Tutuc}}]{PhysRevLett.123.197702}%
  \BibitemOpen
  \bibfield  {author} {\bibinfo {author} {\bibfnamefont {B.}~\bibnamefont
  {G.~W}}, \bibinfo {author} {\bibfnamefont {J.}~\bibnamefont {Zhu}}, \bibinfo
  {author} {\bibfnamefont {T.}~\bibnamefont {Taniguchi}}, \bibinfo {author}
  {\bibfnamefont {K.}~\bibnamefont {Watanabe}}, \bibinfo {author}
  {\bibfnamefont {A.~H.}\ \bibnamefont {MacDonald}},\ and\ \bibinfo {author}
  {\bibfnamefont {E.}~\bibnamefont {Tutuc}},\ }\bibfield  {title} {\bibinfo
  {title} {Correlated insulating states in twisted double bilayer graphene},\
  }\href {https://link.aps.org/doi/10.1103/PhysRevLett.123.197702} {\bibfield
  {journal} {\bibinfo  {journal} {Phys. Rev. Lett.}\ }\textbf {\bibinfo
  {volume} {123}},\ \bibinfo {pages} {197702} (\bibinfo {year}
  {2019})}\BibitemShut {NoStop}%
\bibitem [{\citenamefont {\textit{et
  al.}}(2021{\natexlab{e}})}]{Crommie2021tDBLG}%
  \BibitemOpen
  \bibfield  {author} {\bibinfo {author} {\bibfnamefont {C.~Z.}\ \bibnamefont
  {\textit{et al.}}},\ }\bibfield  {title} {\bibinfo {title} {Visualizing
  delocalized correlated electronic states in twisted double bilayer
  graphene},\ }\href {https://www.nature.com/articles/s41467-021-22711-1}
  {\bibfield  {journal} {\bibinfo  {journal} {Nat. Commun.}\ }\textbf {\bibinfo
  {volume} {12}},\ \bibinfo {pages} {2516} (\bibinfo {year}
  {2021}{\natexlab{e}})}\BibitemShut {NoStop}%
\bibitem [{\citenamefont {Rubio-Verd{\'{u}}}\ \emph {et~al.}(2021)\citenamefont
  {Rubio-Verd{\'{u}}}, \citenamefont {Turkel}, \citenamefont {Song},
  \citenamefont {Klebl}, \citenamefont {Samajdar}, \citenamefont {Scheurer},
  \citenamefont {Venderbos}, \citenamefont {Watanabe}, \citenamefont
  {Taniguchi}, \citenamefont {Ochoa}, \citenamefont {Xian}, \citenamefont
  {Kennes}, \citenamefont {Fernandes}, \citenamefont {Rubio},\ and\
  \citenamefont {Pasupathy}}]{nematicity2020}%
  \BibitemOpen
  \bibfield  {author} {\bibinfo {author} {\bibfnamefont {C.}~\bibnamefont
  {Rubio-Verd{\'{u}}}}, \bibinfo {author} {\bibfnamefont {S.}~\bibnamefont
  {Turkel}}, \bibinfo {author} {\bibfnamefont {Y.}~\bibnamefont {Song}},
  \bibinfo {author} {\bibfnamefont {L.}~\bibnamefont {Klebl}}, \bibinfo
  {author} {\bibfnamefont {R.}~\bibnamefont {Samajdar}}, \bibinfo {author}
  {\bibfnamefont {M.~S.}\ \bibnamefont {Scheurer}}, \bibinfo {author}
  {\bibfnamefont {J.~W.~F.}\ \bibnamefont {Venderbos}}, \bibinfo {author}
  {\bibfnamefont {K.}~\bibnamefont {Watanabe}}, \bibinfo {author}
  {\bibfnamefont {T.}~\bibnamefont {Taniguchi}}, \bibinfo {author}
  {\bibfnamefont {H.}~\bibnamefont {Ochoa}}, \bibinfo {author} {\bibfnamefont
  {L.}~\bibnamefont {Xian}}, \bibinfo {author} {\bibfnamefont {D.~M.}\
  \bibnamefont {Kennes}}, \bibinfo {author} {\bibfnamefont {R.~M.}\
  \bibnamefont {Fernandes}}, \bibinfo {author} {\bibfnamefont
  {{\'{A}}.}~\bibnamefont {Rubio}},\ and\ \bibinfo {author} {\bibfnamefont
  {A.~N.}\ \bibnamefont {Pasupathy}},\ }\bibfield  {title} {\bibinfo {title}
  {Moir{\'{e}} nematic phase in twisted double bilayer graphene},\ }\href
  {https://doi.org/10.1038/s41567-021-01438-2} {\bibfield  {journal} {\bibinfo
  {journal} {Nature Physics}\ }\textbf {\bibinfo {volume} {18}},\ \bibinfo
  {pages} {196} (\bibinfo {year} {2021})}\BibitemShut {NoStop}%
\bibitem [{\citenamefont {Cao}\ \emph {et~al.}(2020)\citenamefont {Cao},
  \citenamefont {Rodan-Legrain}, \citenamefont {Rubies-Bigorda}, \citenamefont
  {Park}, \citenamefont {Watanabe}, \citenamefont {Taniguchi},\ and\
  \citenamefont {Jarillo-Herrero}}]{cao2019electric}%
  \BibitemOpen
  \bibfield  {author} {\bibinfo {author} {\bibfnamefont {Y.}~\bibnamefont
  {Cao}}, \bibinfo {author} {\bibfnamefont {D.}~\bibnamefont {Rodan-Legrain}},
  \bibinfo {author} {\bibfnamefont {O.}~\bibnamefont {Rubies-Bigorda}},
  \bibinfo {author} {\bibfnamefont {J.~M.}\ \bibnamefont {Park}}, \bibinfo
  {author} {\bibfnamefont {K.}~\bibnamefont {Watanabe}}, \bibinfo {author}
  {\bibfnamefont {T.}~\bibnamefont {Taniguchi}},\ and\ \bibinfo {author}
  {\bibfnamefont {P.}~\bibnamefont {Jarillo-Herrero}},\ }\bibfield  {title}
  {\bibinfo {title} {Tunable correlated states and spin-polarized phases in
  twisted bilayer--bilayer graphene},\ }\href
  {https://doi.org/10.1038/s41586-020-2260-6} {\bibfield  {journal} {\bibinfo
  {journal} {Nature}\ }\textbf {\bibinfo {volume} {583}},\ \bibinfo {pages}
  {215} (\bibinfo {year} {2020})}\BibitemShut {NoStop}%
\end{thebibliography}
\end{document}